\documentclass[onecolumn,a4paper]{quantumarticle}
\pdfoutput=1
\usepackage[utf8]{inputenc}
\usepackage[T1]{fontenc}
\usepackage{amsmath,amssymb}
\usepackage{graphicx}
\usepackage{float}

\usepackage{booktabs}
\usepackage{tabularx}
\usepackage{array}
\usepackage{enumitem}
\usepackage{microtype}
\usepackage{caption}
\usepackage[dvipsnames]{xcolor}
\usepackage{cite}
\usepackage[colorlinks=true,allcolors=Blue,breaklinks=true]{hyperref}
\usepackage{pgfplots}
\pgfplotsset{compat=1.18}
\usepgfplotslibrary{fillbetween}
\usepgfplotslibrary{groupplots}
\usetikzlibrary{arrows.meta,positioning,backgrounds,fit,calc}
\definecolor{oiBlue}{HTML}{0072B2}
\definecolor{oiOrange}{HTML}{E69F00}
\definecolor{oiGreen}{HTML}{009E73}
\definecolor{oiVerm}{HTML}{D55E00}
\definecolor{oiPurple}{HTML}{8256B4}
\definecolor{oiSky}{HTML}{56B4E9}
\definecolor{oiGrey}{HTML}{8A8A8A}
\definecolor{oiInk}{HTML}{222222}
\pgfplotsset{
  paperplot/.style={
    axis line style={oiInk!75, line width=0.5pt},
    tick label style={font=\small}, label style={font=\small},
    title style={font=\small, align=center},
    grid=major, grid style={oiInk!12, line width=0.4pt},
    tick align=outside, tickpos=left, axis on top=false,
    legend style={font=\footnotesize, draw=none, fill=none, row sep=1pt},
    legend cell align=left,
  }
}

\newcommand{\ent}[1]{\raisebox{0.1ex}{\textcircled{\scriptsize #1}}}
\title{Machine learning for sample-based quantum diagonalization: a review of generative configuration recovery and the classical-simulability frontier}
\author{Nicol\'as Bonilla Vargas}
\affiliation{M.Sc.\ Physics, Universidad Nacional de Colombia; M.Sc.\ Applied Data Science \& AI, SRH University Heidelberg / Munich; Daita AI}
\email{ngbonillav@unal.edu.co}
\date{18 September 2026}

\begin{document}
\maketitle
\begin{abstract}
Sample-based quantum diagonalization (SQD), equivalently quantum-selected configuration interaction (QSCI), has become a centre of gravity of pre-fault-tolerant quantum chemistry: a quantum processor samples electronic configurations, and the Hamiltonian is diagonalized classically in the resulting subspace. Its accuracy is governed entirely by \emph{which} configurations enter it --- a machine-learning selection problem, made acute by a coupon-collector bottleneck. We review the ecosystem of generative and learned selectors, organized by what each generates and the signal it exploits, and identify one gap: no reward-proportional generative-flow-network proposer has been built for tail discovery. On the field's central question --- whether the quantum sampler beats classical selected CI --- the negative verdict is not ours to claim: priority belongs to Reinholdt \emph{et al.} [J.\ Chem.\ Theory Comput.\ \textbf{21}, 6811 (2025)], and polynomial-time classical energy estimation of the flagship single-layer circuits has since reinforced it. What we add is scope and apparatus. We state that verdict at the precision a falsifiable claim requires --- it concerns \emph{reproducible}, \emph{same-active-space} comparisons on \emph{molecular electronic structure} --- and weigh the live claims outside those qualifiers. We show that $\alpha$-string weights are not invariant under rotations inside degenerate orbital shells, so determinant counts are undefined until the orbital gauge is declared. We distil a ten-element benchmarking standard, apply it to our own deposit, and report what it returned --- five defects at the audit date, and more while preparing this version. FCI-exact experiments confirm one prediction and refute another: held to our own standard, the single generative advantage we find keeps no consistent sign along the dissociation coordinate at device-calibrated noise and reverses under a symmetric readout model, so it bounds a regime rather than announcing one. Against a noise-matched classical recovery loop the proposer does win on N$_2$, by $2.4$~mHa, a margin five seeds cannot resolve; against the classical selector that needs no sampler at all it loses by more than a factor of two. This review addresses quantum-algorithms researchers working on subspace and sampling methods, and machine-learning researchers entering electronic structure.
\end{abstract}

\section{Introduction}\label{sec:1}

The determination of the ground-state energy of a many-electron Hamiltonian is among the oldest and most consequential problems in the physical sciences, underlying our predictive understanding of chemical reactivity, catalysis, and correlated materials. Its difficulty is structural: the dimension of the full configuration-interaction (FCI) space grows combinatorially with the number of electrons and orbitals, and the exact problem is, in the worst case, QMA-complete \cite{kempe2006}. The last half-century of quantum chemistry can be read as a sequence of increasingly sophisticated strategies for evading this exponential wall without paying its full cost --- coupled-cluster expansions, the density-matrix renormalization group (DMRG), auxiliary-field quantum Monte Carlo (AFQMC), and, most directly relevant here, \emph{selected} configuration interaction (sCI), which builds a compact variational subspace by retaining only the determinants that matter and correcting the remainder perturbatively \cite{huron1973,holmes2016,sharma2017,tubman2016}.

The advent of programmable quantum processors reframed the same problem. If the exponential difficulty of electronic structure is a quantum difficulty, perhaps a quantum device can help. The first decade of this hope was dominated by the variational quantum eigensolver (VQE) \cite{peruzzo2014}, in which a parameterized circuit is optimized to minimize an energy expectation value. That programme has, for the purpose of demonstrating a scaling advantage, largely run its course: barren plateaus render deep expressive circuits untrainable \cite{mcclean2018,larocca2025}, the measurement cost of estimating chemical-accuracy energies is prohibitive, and --- most damningly --- the structural conditions that let a variational circuit \emph{avoid} barren plateaus tend also to render it classically simulable \cite{cerezo2025}. Into the vacuum left by VQE stepped a conceptually different, and strikingly effective, idea.

\textbf{Sample-based quantum diagonalization.} Rather than optimize a circuit to \emph{be} the ground state, one uses a shallow, noise-tolerant circuit only to \emph{sample} the electronic configurations (computational-basis bitstrings, i.e.\ Slater determinants) that carry appreciable ground-state weight, and then diagonalizes the Hamiltonian classically in the sampled subspace on high-performance hardware. Introduced as quantum-selected configuration interaction (QSCI) by Kanno and co-workers \cite{kanno2023} and scaled to quantum-centric supercomputing as SQD by an IBM--RIKEN collaboration \cite{robledomoreno2025}, the method inherits the variational guarantee of sCI, sidesteps the optimization and measurement pathologies of VQE, and is remarkably robust to hardware noise through a self-consistent configuration-recovery step that repairs symmetry-broken samples. Within two years it has been demonstrated on iron--sulfur clusters at 77 qubits with the entire Fugaku supercomputer in the loop \cite{robledomoreno2025,shirakawa2025}, extended to excited states, embeddings, and periodic systems, and pushed, through fragment-based workflows, to protein--ligand complexes of 11,608 and 12,635 atoms \cite{merz2026}. SQD is, today, the de facto default of pre-fault-tolerant quantum chemistry.

\textbf{The machine-learning entry point.} The accuracy of the SQD loop is determined entirely by which configurations populate the subspace, and this is where the method is most vulnerable and most interesting. The relevant determinants are heavy-tailed in amplitude; discovering the rare, chemically decisive ones by repeated sampling is a coupon-collector problem, in which the marginal cost of each newly discovered configuration grows sharply \cite{reinholdt2025}; the mapping has since been made analytic, yielding a closed expression and a scalable lower-bound estimator for the shot count needed to recover a target determinant set, validated in simulation and on a 127-qubit device \cite{khamadja2025}. The configuration-recovery step, a probabilistic repair of noise-corrupted samples toward an estimated occupation profile, is a statistical denoising problem. Both are natural targets for machine learning, and the field has responded with striking speed. In the two years since QSCI, generative and learned methods have been proposed for every stage of the loop: restricted Boltzmann machines that learn and resample the distribution of dominant determinants \cite{patra2025,patra2026}; autoregressive and transformer neural quantum states that propose determinants inside the sample--diagonalize--update loop \cite{chang2026,solanki2026,shang2025}; neural-network classifiers and learning-to-rank models that predict determinant importance \cite{zeni2026,zhanghaar2025,nie2026}; and generative circuit design that learns the sampling ansatz itself \cite{kemmoku2026}. This ecosystem has, to our knowledge, not yet been reviewed. The standing review of the broader family, Motta and co-workers on quantum subspace methods \cite{motta2024}, is the reference account of the projection framework; the one other review naming QSCI among its topics, Jiang and co-workers on quantum algorithms for sampling tasks in chemistry, treats it as one entry in a Monte-Carlo-organized sweep \cite{jiang2025}. Both predate the 2025 scaling demonstrations and cover none of the learned samplers, recovery models, or generative circuit designs surveyed here. The sampling bottleneck alone has had one survey-length treatment, the unrefereed master's thesis cited above \cite{khamadja2025}. What is missing is a critical account of the learning layer.

\textbf{The question the field turns on.} Beneath the methodological activity lies a single, contested, and increasingly uncomfortable question: does any of this actually beat classical selected configuration interaction? SQD \emph{is}, after all, a quantum-sampled sCI, and its classical competitors --- semistochastic heat-bath CI (SHCI), the perturbatively-selected CIPSI, adaptive sampling CI (ASCI), DMRG, and AFQMC --- are mature, fast, and run on a laptop. The critical literature has grown in tandem with the method. Reinholdt and co-workers showed that, even granting SQD a perfect noiseless sampler, its expansions are an order of magnitude less compact than heat-bath CI on the very systems used to showcase it \cite{reinholdt2025}; Gaberle and Jattana found that for strongly correlated lattice models the required configuration count grows exponentially \emph{even with an optimal sampler} \cite{gaberle2026}; Vaquero-Sabater and co-workers demonstrated that random determinants passed through the classical recovery loop reproduce the accuracy, implying that the classical recovery, not the quantum sampler, is doing the work \cite{vaquerosabater2026}; and, most strikingly, Belagali and co-workers showed in 2026 that the single-layer unitary cluster Jastrow circuit SQD samples is classically simulable in polynomial time, independently of the hardware locality constraints, reproducing the flagship 77-qubit experiment on a laptop at \emph{lower} energy --- at the energy (weak-simulation) level, not by out-sampling the device \cite{belagali2026}. The pro-quantum case remains alive but narrow: worst-case instantaneous-quantum-polynomial (IQP) hardness of the sampling circuits \cite{hafid2025}, and a contested 49-qubit advantage claim on sparse-ground-state Hamiltonians \cite{kirby2026}.

\textbf{This review.} We take the position that a field cannot mature while its central claim is unsettled, and that the right place to stand is neither uncritical enthusiasm nor dismissal, but a rigorous map of what is known. We make five contributions. First (\S\ref{sec:2}--\S\ref{sec:3}), we give a structured account of the SQD/QSCI method family and the configuration-recovery bottleneck that makes it a machine-learning problem. Second (\S\ref{sec:4}), we provide the first taxonomy of the ML and generative methods proposed for determinant and subspace selection, organized by the object they generate and the importance signal they exploit, and we identify the conspicuous methodological gap: that generative flow networks, purpose-built for diverse reward-proportional sampling over exponential discrete spaces, have not yet been applied to this problem. Third (\S\ref{sec:5}), we confront the quantum-advantage question directly. The negative answer is not ours: Reinholdt and co-workers reached and published it in 2025, on the same two systems \cite{reinholdt2025}, and the 2026 dequantization results have since converged on it independently. What we contribute there is scope and apparatus: we assemble the classical baselines, the benzene-benchmark accuracy standard, the critique literature, and the 2026 dequantization results into a single verdict stated at the precision a falsifiable claim requires --- scoped to \emph{reproducible}, \emph{same-active-space} comparisons on \emph{molecular electronic structure} --- and we weigh the live claims that lie outside those qualifiers \cite{kirby2026,weaving2025,kemmoku2026} on the same axis rather than leaving them outside the argument. Fourth (\S\ref{sec:6}), we propose a set of benchmarking standards that we contend any future claim in this field must meet. Fifth (\S\ref{sec:7}), we argue that the negative result is a compass, identifying where generative and quantum methods might retain a defensible or, in one case, provable advantage: robustness under \emph{elevated} noise (which a controlled test finds to be generic rather than a quantum edge, which has no consistent sign at the device-calibrated scale itself, and which reverses under a symmetric readout model), genuinely multireference chemistry (an open prediction we probe but do not confirm), and the distinct quantum-data-native task of learning from experiments. Throughout, we distinguish sharply between what is proven, what is demonstrated, and what is hoped.

\medskip\noindent\textbf{Abbreviations.} {\footnotesize SQD/QSCI (sample-based quantum diagonalization / quantum-selected configuration interaction); sCI (selected configuration interaction); SHCI (semistochastic heat-bath CI); CIPSI (perturbatively selected CI); ASCI (adaptive sampling CI); FCI (full configuration interaction); DMRG (density-matrix renormalization group); AFQMC (auxiliary-field quantum Monte Carlo); LUCJ (local unitary cluster Jastrow); S-CORE (self-consistent configuration recovery); VQE (variational quantum eigensolver); NQS (neural quantum state); RBM (restricted Boltzmann machine); GQE (generative quantum eigensolver); GFlowNet (generative flow network); EN (Epstein--Nesbet); PT (perturbation theory).}

\section{Sample-based quantum diagonalization: the method and its family}\label{sec:2}

\emph{Scope and method.} This review surveys the sample-based quantum diagonalization literature indexed on arXiv (quant-ph, physics.chem-ph) and in the major chemistry and quantum-information journals, together with QunaSys's continuously updated bibliography portal \emph{Awesome-QSCI}, from the introduction of QSCI in early 2023 to September 2026. We include the SQD/QSCI method family, the machine-learning and generative methods proposed for determinant or subspace selection, the classical selected-configuration-interaction competitors that constitute the benchmark, and the critique and dequantization literature; we exclude general variational-quantum-eigensolver work except where it bears directly on the sampling step. Because the field is moving on a scale of weeks and much of the primary literature is preprint, we date every claim explicitly and flag its standing --- established, demonstrated, or merely hoped.

\subsection{The core loop}\label{sec:2-1}

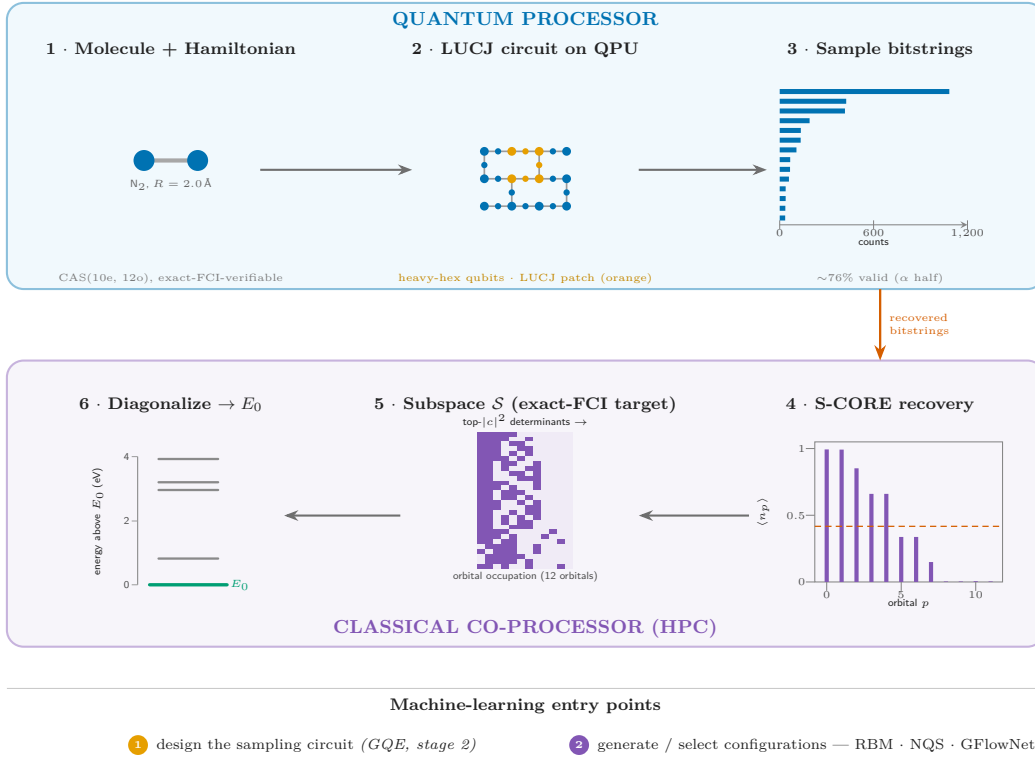
\begin{figure*}[tbp]
\centering
\resizebox{\linewidth}{!}{%
\begin{tikzpicture}[font=\sffamily,every node/.style={inner sep=0pt}]
\fill[oiSky!8,rounded corners=8pt] (0,4.7) rectangle (17.4,9.5);
\draw[oiBlue!45,line width=1pt,rounded corners=8pt] (0,4.7) rectangle (17.4,9.5);
\fill[oiPurple!6,rounded corners=8pt] (0,-1.3) rectangle (17.4,3.5);
\draw[oiPurple!45,line width=1pt,rounded corners=8pt] (0,-1.3) rectangle (17.4,3.5);
\node[anchor=north,font=\small\bfseries,text=oiBlue] at (8.70,9.35) {QUANTUM PROCESSOR};
\node[anchor=south,font=\small\bfseries,text=oiPurple] at (8.70,-1.15) {CLASSICAL CO-PROCESSOR (HPC)};
\node[anchor=north,font=\footnotesize\bfseries,text=oiInk] at (2.75,8.85) {1 $\cdot$ Molecule + Hamiltonian};
\node[anchor=north,font=\footnotesize\bfseries,text=oiInk] at (8.70,8.85) {2 $\cdot$ LUCJ circuit on QPU};
\node[anchor=north,font=\footnotesize\bfseries,text=oiInk] at (14.65,8.85) {3 $\cdot$ Sample bitstrings};
\node[anchor=center] at (2.75,6.7) {\begin{tikzpicture}\draw[oiInk!40,line width=2pt] (-0.45,0)--(0.45,0);\fill[oiBlue] (-0.45,0) circle (5pt);\fill[oiBlue] (0.45,0) circle (5pt);\node[font=\tiny,text=oiInk!70,anchor=north] at (0,-0.28){N$_2$, $R=2.0$\,\AA};\end{tikzpicture}};
\node[anchor=north,font=\tiny,text=oiInk!60] at (2.75,4.98) {CAS(10e, 12o), exact-FCI-verifiable};
\node[anchor=center] at (8.70,6.55) {\begin{tikzpicture}\draw[oiInk!45,line width=0.8pt] (0.00,0.00)--(1.38,0.00);\draw[oiInk!45,line width=0.8pt] (0.00,0.46)--(1.38,0.46);\draw[oiInk!45,line width=0.8pt] (0.00,0.92)--(1.38,0.92);\draw[oiInk!45,line width=0.8pt] (0.00,0.92)--(0.00,0.46);\draw[oiInk!45,line width=0.8pt] (0.92,0.92)--(0.92,0.46);\draw[oiInk!45,line width=0.8pt] (0.46,0.46)--(0.46,0.00);\draw[oiInk!45,line width=0.8pt] (1.38,0.46)--(1.38,0.00);\fill[oiBlue] (0.00,0.00) circle (2.1pt);\fill[oiBlue] (0.46,0.00) circle (2.1pt);\fill[oiBlue] (0.92,0.00) circle (2.1pt);\fill[oiBlue] (1.38,0.00) circle (2.1pt);\fill[oiBlue] (0.00,0.46) circle (2.1pt);\fill[oiOrange] (0.46,0.46) circle (2.1pt);\fill[oiOrange] (0.92,0.46) circle (2.1pt);\fill[oiBlue] (1.38,0.46) circle (2.1pt);\fill[oiBlue] (0.00,0.92) circle (2.1pt);\fill[oiOrange] (0.46,0.92) circle (2.1pt);\fill[oiOrange] (0.92,0.92) circle (2.1pt);\fill[oiBlue] (1.38,0.92) circle (2.1pt);\fill[oiBlue] (0.23,0.00) circle (1.5pt);\fill[oiBlue] (0.69,0.00) circle (1.5pt);\fill[oiBlue] (1.15,0.00) circle (1.5pt);\fill[oiBlue] (0.23,0.46) circle (1.5pt);\fill[oiOrange] (0.69,0.46) circle (1.5pt);\fill[oiBlue] (1.15,0.46) circle (1.5pt);\fill[oiBlue] (0.23,0.92) circle (1.5pt);\fill[oiOrange] (0.69,0.92) circle (1.5pt);\fill[oiBlue] (1.15,0.92) circle (1.5pt);\fill[oiBlue] (0.00,0.69) circle (1.5pt);\fill[oiOrange] (0.92,0.69) circle (1.5pt);\fill[oiBlue] (0.46,0.23) circle (1.5pt);\fill[oiBlue] (1.38,0.23) circle (1.5pt);\end{tikzpicture}};
\node[anchor=north,font=\tiny,text=oiOrange!85!black] at (8.70,4.98) {heavy-hex qubits $\cdot$ LUCJ patch (orange)};
\node[anchor=center] at (14.65,6.75) {\begin{tikzpicture}\begin{axis}[width=3.15cm,height=2.35cm,scale only axis,grid=none,axis line style={oiInk!55,line width=0.5pt},tick align=outside,tick label style={font=\tiny,text=oiInk!80},label style={font=\tiny,text=oiInk},tickpos=left,every axis plot/.append style={line width=0.9pt},xbar,y dir=reverse,bar width=2.4pt,xmin=0,xmax=1200,ymin=-0.7,ymax=13.7,axis y line=none,axis x line=bottom,xlabel={counts},xtick distance=600,enlarge y limits=false]\addplot[fill=oiBlue,draw=none] table[x=count,y=rank]{wf_hist.dat};\end{axis}\end{tikzpicture}};
\node[anchor=north,font=\tiny,text=oiInk!60,align=center] at (14.65,4.98) {$\sim$76\% valid ($\alpha$ half)};
\draw[-{Stealth[length=2.4mm]},oiInk!65,line width=1pt] (4.25,6.7)--(6.80,6.7);
\draw[-{Stealth[length=2.4mm]},oiInk!65,line width=1pt] (10.60,6.7)--(12.75,6.7);
\node[anchor=north,font=\footnotesize\bfseries,text=oiInk] at (2.75,2.9) {6 $\cdot$ Diagonalize $\rightarrow E_0$};
\node[anchor=north,font=\footnotesize\bfseries,text=oiInk] at (8.70,2.9) {5 $\cdot$ Subspace $\mathcal{S}$ (exact-FCI target)};
\node[anchor=north,font=\footnotesize\bfseries,text=oiInk] at (14.65,2.9) {4 $\cdot$ S-CORE recovery};
\node[anchor=center] at (14.65,0.75) {\begin{tikzpicture}\begin{axis}[width=3.15cm,height=2.35cm,scale only axis,grid=none,axis line style={oiInk!55,line width=0.5pt},tick align=outside,tick label style={font=\tiny,text=oiInk!80},label style={font=\tiny,text=oiInk},tickpos=left,every axis plot/.append style={line width=0.9pt},ybar,bar width=2.0pt,xmin=-0.8,xmax=11.8,ymin=0,ymax=1.05,xlabel={orbital $p$},ylabel={$\langle n_p\rangle$},ylabel style={yshift=8pt},xtick={0,5,10},ytick={0,0.5,1}]\addplot[fill=oiPurple,draw=none] table[x=orb,y=occ]{wf_occ.dat};\draw[oiVerm,densely dashed,line width=0.6pt] (axis cs:-0.8,0.417)--(axis cs:11.8,0.417);\end{axis}\end{tikzpicture}};
\node[anchor=north] at (8.70,2.60) {\begin{tikzpicture}\node[anchor=north,font=\tiny,text=oiInk] at (0.81,0.30) {top-$|c|^2$ determinants $\rightarrow$};\fill[oiPurple] (0.000,0.000) rectangle (0.123,-0.070);\fill[oiPurple] (0.135,0.000) rectangle (0.258,-0.070);\fill[oiPurple] (0.270,0.000) rectangle (0.393,-0.070);\fill[oiPurple] (0.405,0.000) rectangle (0.528,-0.070);\fill[oiPurple] (0.540,0.000) rectangle (0.663,-0.070);\fill[oiPurple!12] (0.675,0.000) rectangle (0.798,-0.070);\fill[oiPurple!12] (0.810,0.000) rectangle (0.933,-0.070);\fill[oiPurple!12] (0.945,0.000) rectangle (1.068,-0.070);\fill[oiPurple!12] (1.080,0.000) rectangle (1.203,-0.070);\fill[oiPurple!12] (1.215,0.000) rectangle (1.338,-0.070);\fill[oiPurple!12] (1.350,0.000) rectangle (1.473,-0.070);\fill[oiPurple!12] (1.485,0.000) rectangle (1.608,-0.070);\fill[oiPurple] (0.000,-0.082) rectangle (0.123,-0.152);\fill[oiPurple] (0.135,-0.082) rectangle (0.258,-0.152);\fill[oiPurple] (0.270,-0.082) rectangle (0.393,-0.152);\fill[oiPurple] (0.405,-0.082) rectangle (0.528,-0.152);\fill[oiPurple!12] (0.540,-0.082) rectangle (0.663,-0.152);\fill[oiPurple!12] (0.675,-0.082) rectangle (0.798,-0.152);\fill[oiPurple] (0.810,-0.082) rectangle (0.933,-0.152);\fill[oiPurple!12] (0.945,-0.082) rectangle (1.068,-0.152);\fill[oiPurple!12] (1.080,-0.082) rectangle (1.203,-0.152);\fill[oiPurple!12] (1.215,-0.082) rectangle (1.338,-0.152);\fill[oiPurple!12] (1.350,-0.082) rectangle (1.473,-0.152);\fill[oiPurple!12] (1.485,-0.082) rectangle (1.608,-0.152);\fill[oiPurple] (0.000,-0.164) rectangle (0.123,-0.234);\fill[oiPurple] (0.135,-0.164) rectangle (0.258,-0.234);\fill[oiPurple] (0.270,-0.164) rectangle (0.393,-0.234);\fill[oiPurple!12] (0.405,-0.164) rectangle (0.528,-0.234);\fill[oiPurple] (0.540,-0.164) rectangle (0.663,-0.234);\fill[oiPurple] (0.675,-0.164) rectangle (0.798,-0.234);\fill[oiPurple!12] (0.810,-0.164) rectangle (0.933,-0.234);\fill[oiPurple!12] (0.945,-0.164) rectangle (1.068,-0.234);\fill[oiPurple!12] (1.080,-0.164) rectangle (1.203,-0.234);\fill[oiPurple!12] (1.215,-0.164) rectangle (1.338,-0.234);\fill[oiPurple!12] (1.350,-0.164) rectangle (1.473,-0.234);\fill[oiPurple!12] (1.485,-0.164) rectangle (1.608,-0.234);\fill[oiPurple] (0.000,-0.246) rectangle (0.123,-0.316);\fill[oiPurple] (0.135,-0.246) rectangle (0.258,-0.316);\fill[oiPurple] (0.270,-0.246) rectangle (0.393,-0.316);\fill[oiPurple!12] (0.405,-0.246) rectangle (0.528,-0.316);\fill[oiPurple!12] (0.540,-0.246) rectangle (0.663,-0.316);\fill[oiPurple] (0.675,-0.246) rectangle (0.798,-0.316);\fill[oiPurple] (0.810,-0.246) rectangle (0.933,-0.316);\fill[oiPurple!12] (0.945,-0.246) rectangle (1.068,-0.316);\fill[oiPurple!12] (1.080,-0.246) rectangle (1.203,-0.316);\fill[oiPurple!12] (1.215,-0.246) rectangle (1.338,-0.316);\fill[oiPurple!12] (1.350,-0.246) rectangle (1.473,-0.316);\fill[oiPurple!12] (1.485,-0.246) rectangle (1.608,-0.316);\fill[oiPurple] (0.000,-0.328) rectangle (0.123,-0.398);\fill[oiPurple] (0.135,-0.328) rectangle (0.258,-0.398);\fill[oiPurple] (0.270,-0.328) rectangle (0.393,-0.398);\fill[oiPurple!12] (0.405,-0.328) rectangle (0.528,-0.398);\fill[oiPurple] (0.540,-0.328) rectangle (0.663,-0.398);\fill[oiPurple!12] (0.675,-0.328) rectangle (0.798,-0.398);\fill[oiPurple] (0.810,-0.328) rectangle (0.933,-0.398);\fill[oiPurple!12] (0.945,-0.328) rectangle (1.068,-0.398);\fill[oiPurple!12] (1.080,-0.328) rectangle (1.203,-0.398);\fill[oiPurple!12] (1.215,-0.328) rectangle (1.338,-0.398);\fill[oiPurple!12] (1.350,-0.328) rectangle (1.473,-0.398);\fill[oiPurple!12] (1.485,-0.328) rectangle (1.608,-0.398);\fill[oiPurple] (0.000,-0.410) rectangle (0.123,-0.480);\fill[oiPurple] (0.135,-0.410) rectangle (0.258,-0.480);\fill[oiPurple] (0.270,-0.410) rectangle (0.393,-0.480);\fill[oiPurple] (0.405,-0.410) rectangle (0.528,-0.480);\fill[oiPurple!12] (0.540,-0.410) rectangle (0.663,-0.480);\fill[oiPurple] (0.675,-0.410) rectangle (0.798,-0.480);\fill[oiPurple!12] (0.810,-0.410) rectangle (0.933,-0.480);\fill[oiPurple!12] (0.945,-0.410) rectangle (1.068,-0.480);\fill[oiPurple!12] (1.080,-0.410) rectangle (1.203,-0.480);\fill[oiPurple!12] (1.215,-0.410) rectangle (1.338,-0.480);\fill[oiPurple!12] (1.350,-0.410) rectangle (1.473,-0.480);\fill[oiPurple!12] (1.485,-0.410) rectangle (1.608,-0.480);\fill[oiPurple] (0.000,-0.492) rectangle (0.123,-0.562);\fill[oiPurple] (0.135,-0.492) rectangle (0.258,-0.562);\fill[oiPurple!12] (0.270,-0.492) rectangle (0.393,-0.562);\fill[oiPurple] (0.405,-0.492) rectangle (0.528,-0.562);\fill[oiPurple] (0.540,-0.492) rectangle (0.663,-0.562);\fill[oiPurple!12] (0.675,-0.492) rectangle (0.798,-0.562);\fill[oiPurple!12] (0.810,-0.492) rectangle (0.933,-0.562);\fill[oiPurple] (0.945,-0.492) rectangle (1.068,-0.562);\fill[oiPurple!12] (1.080,-0.492) rectangle (1.203,-0.562);\fill[oiPurple!12] (1.215,-0.492) rectangle (1.338,-0.562);\fill[oiPurple!12] (1.350,-0.492) rectangle (1.473,-0.562);\fill[oiPurple!12] (1.485,-0.492) rectangle (1.608,-0.562);\fill[oiPurple] (0.000,-0.574) rectangle (0.123,-0.644);\fill[oiPurple] (0.135,-0.574) rectangle (0.258,-0.644);\fill[oiPurple!12] (0.270,-0.574) rectangle (0.393,-0.644);\fill[oiPurple] (0.405,-0.574) rectangle (0.528,-0.644);\fill[oiPurple!12] (0.540,-0.574) rectangle (0.663,-0.644);\fill[oiPurple!12] (0.675,-0.574) rectangle (0.798,-0.644);\fill[oiPurple] (0.810,-0.574) rectangle (0.933,-0.644);\fill[oiPurple] (0.945,-0.574) rectangle (1.068,-0.644);\fill[oiPurple!12] (1.080,-0.574) rectangle (1.203,-0.644);\fill[oiPurple!12] (1.215,-0.574) rectangle (1.338,-0.644);\fill[oiPurple!12] (1.350,-0.574) rectangle (1.473,-0.644);\fill[oiPurple!12] (1.485,-0.574) rectangle (1.608,-0.644);\fill[oiPurple] (0.000,-0.656) rectangle (0.123,-0.726);\fill[oiPurple] (0.135,-0.656) rectangle (0.258,-0.726);\fill[oiPurple!12] (0.270,-0.656) rectangle (0.393,-0.726);\fill[oiPurple!12] (0.405,-0.656) rectangle (0.528,-0.726);\fill[oiPurple] (0.540,-0.656) rectangle (0.663,-0.726);\fill[oiPurple] (0.675,-0.656) rectangle (0.798,-0.726);\fill[oiPurple!12] (0.810,-0.656) rectangle (0.933,-0.726);\fill[oiPurple] (0.945,-0.656) rectangle (1.068,-0.726);\fill[oiPurple!12] (1.080,-0.656) rectangle (1.203,-0.726);\fill[oiPurple!12] (1.215,-0.656) rectangle (1.338,-0.726);\fill[oiPurple!12] (1.350,-0.656) rectangle (1.473,-0.726);\fill[oiPurple!12] (1.485,-0.656) rectangle (1.608,-0.726);\fill[oiPurple] (0.000,-0.738) rectangle (0.123,-0.808);\fill[oiPurple] (0.135,-0.738) rectangle (0.258,-0.808);\fill[oiPurple!12] (0.270,-0.738) rectangle (0.393,-0.808);\fill[oiPurple!12] (0.405,-0.738) rectangle (0.528,-0.808);\fill[oiPurple!12] (0.540,-0.738) rectangle (0.663,-0.808);\fill[oiPurple] (0.675,-0.738) rectangle (0.798,-0.808);\fill[oiPurple] (0.810,-0.738) rectangle (0.933,-0.808);\fill[oiPurple] (0.945,-0.738) rectangle (1.068,-0.808);\fill[oiPurple!12] (1.080,-0.738) rectangle (1.203,-0.808);\fill[oiPurple!12] (1.215,-0.738) rectangle (1.338,-0.808);\fill[oiPurple!12] (1.350,-0.738) rectangle (1.473,-0.808);\fill[oiPurple!12] (1.485,-0.738) rectangle (1.608,-0.808);\fill[oiPurple] (0.000,-0.820) rectangle (0.123,-0.890);\fill[oiPurple] (0.135,-0.820) rectangle (0.258,-0.890);\fill[oiPurple] (0.270,-0.820) rectangle (0.393,-0.890);\fill[oiPurple] (0.405,-0.820) rectangle (0.528,-0.890);\fill[oiPurple!12] (0.540,-0.820) rectangle (0.663,-0.890);\fill[oiPurple!12] (0.675,-0.820) rectangle (0.798,-0.890);\fill[oiPurple!12] (0.810,-0.820) rectangle (0.933,-0.890);\fill[oiPurple] (0.945,-0.820) rectangle (1.068,-0.890);\fill[oiPurple!12] (1.080,-0.820) rectangle (1.203,-0.890);\fill[oiPurple!12] (1.215,-0.820) rectangle (1.338,-0.890);\fill[oiPurple!12] (1.350,-0.820) rectangle (1.473,-0.890);\fill[oiPurple!12] (1.485,-0.820) rectangle (1.608,-0.890);\fill[oiPurple] (0.000,-0.902) rectangle (0.123,-0.972);\fill[oiPurple] (0.135,-0.902) rectangle (0.258,-0.972);\fill[oiPurple] (0.270,-0.902) rectangle (0.393,-0.972);\fill[oiPurple!12] (0.405,-0.902) rectangle (0.528,-0.972);\fill[oiPurple] (0.540,-0.902) rectangle (0.663,-0.972);\fill[oiPurple!12] (0.675,-0.902) rectangle (0.798,-0.972);\fill[oiPurple!12] (0.810,-0.902) rectangle (0.933,-0.972);\fill[oiPurple] (0.945,-0.902) rectangle (1.068,-0.972);\fill[oiPurple!12] (1.080,-0.902) rectangle (1.203,-0.972);\fill[oiPurple!12] (1.215,-0.902) rectangle (1.338,-0.972);\fill[oiPurple!12] (1.350,-0.902) rectangle (1.473,-0.972);\fill[oiPurple!12] (1.485,-0.902) rectangle (1.608,-0.972);\fill[oiPurple] (0.000,-0.984) rectangle (0.123,-1.054);\fill[oiPurple] (0.135,-0.984) rectangle (0.258,-1.054);\fill[oiPurple!12] (0.270,-0.984) rectangle (0.393,-1.054);\fill[oiPurple] (0.405,-0.984) rectangle (0.528,-1.054);\fill[oiPurple] (0.540,-0.984) rectangle (0.663,-1.054);\fill[oiPurple] (0.675,-0.984) rectangle (0.798,-1.054);\fill[oiPurple!12] (0.810,-0.984) rectangle (0.933,-1.054);\fill[oiPurple!12] (0.945,-0.984) rectangle (1.068,-1.054);\fill[oiPurple!12] (1.080,-0.984) rectangle (1.203,-1.054);\fill[oiPurple!12] (1.215,-0.984) rectangle (1.338,-1.054);\fill[oiPurple!12] (1.350,-0.984) rectangle (1.473,-1.054);\fill[oiPurple!12] (1.485,-0.984) rectangle (1.608,-1.054);\fill[oiPurple] (0.000,-1.066) rectangle (0.123,-1.136);\fill[oiPurple] (0.135,-1.066) rectangle (0.258,-1.136);\fill[oiPurple!12] (0.270,-1.066) rectangle (0.393,-1.136);\fill[oiPurple] (0.405,-1.066) rectangle (0.528,-1.136);\fill[oiPurple] (0.540,-1.066) rectangle (0.663,-1.136);\fill[oiPurple!12] (0.675,-1.066) rectangle (0.798,-1.136);\fill[oiPurple] (0.810,-1.066) rectangle (0.933,-1.136);\fill[oiPurple!12] (0.945,-1.066) rectangle (1.068,-1.136);\fill[oiPurple!12] (1.080,-1.066) rectangle (1.203,-1.136);\fill[oiPurple!12] (1.215,-1.066) rectangle (1.338,-1.136);\fill[oiPurple!12] (1.350,-1.066) rectangle (1.473,-1.136);\fill[oiPurple!12] (1.485,-1.066) rectangle (1.608,-1.136);\fill[oiPurple] (0.000,-1.148) rectangle (0.123,-1.218);\fill[oiPurple] (0.135,-1.148) rectangle (0.258,-1.218);\fill[oiPurple] (0.270,-1.148) rectangle (0.393,-1.218);\fill[oiPurple!12] (0.405,-1.148) rectangle (0.528,-1.218);\fill[oiPurple!12] (0.540,-1.148) rectangle (0.663,-1.218);\fill[oiPurple] (0.675,-1.148) rectangle (0.798,-1.218);\fill[oiPurple!12] (0.810,-1.148) rectangle (0.933,-1.218);\fill[oiPurple] (0.945,-1.148) rectangle (1.068,-1.218);\fill[oiPurple!12] (1.080,-1.148) rectangle (1.203,-1.218);\fill[oiPurple!12] (1.215,-1.148) rectangle (1.338,-1.218);\fill[oiPurple!12] (1.350,-1.148) rectangle (1.473,-1.218);\fill[oiPurple!12] (1.485,-1.148) rectangle (1.608,-1.218);\fill[oiPurple] (0.000,-1.230) rectangle (0.123,-1.300);\fill[oiPurple] (0.135,-1.230) rectangle (0.258,-1.300);\fill[oiPurple] (0.270,-1.230) rectangle (0.393,-1.300);\fill[oiPurple!12] (0.405,-1.230) rectangle (0.528,-1.300);\fill[oiPurple!12] (0.540,-1.230) rectangle (0.663,-1.300);\fill[oiPurple!12] (0.675,-1.230) rectangle (0.798,-1.300);\fill[oiPurple] (0.810,-1.230) rectangle (0.933,-1.300);\fill[oiPurple] (0.945,-1.230) rectangle (1.068,-1.300);\fill[oiPurple!12] (1.080,-1.230) rectangle (1.203,-1.300);\fill[oiPurple!12] (1.215,-1.230) rectangle (1.338,-1.300);\fill[oiPurple!12] (1.350,-1.230) rectangle (1.473,-1.300);\fill[oiPurple!12] (1.485,-1.230) rectangle (1.608,-1.300);\fill[oiPurple] (0.000,-1.312) rectangle (0.123,-1.382);\fill[oiPurple] (0.135,-1.312) rectangle (0.258,-1.382);\fill[oiPurple!12] (0.270,-1.312) rectangle (0.393,-1.382);\fill[oiPurple!12] (0.405,-1.312) rectangle (0.528,-1.382);\fill[oiPurple] (0.540,-1.312) rectangle (0.663,-1.382);\fill[oiPurple!12] (0.675,-1.312) rectangle (0.798,-1.382);\fill[oiPurple] (0.810,-1.312) rectangle (0.933,-1.382);\fill[oiPurple] (0.945,-1.312) rectangle (1.068,-1.382);\fill[oiPurple!12] (1.080,-1.312) rectangle (1.203,-1.382);\fill[oiPurple!12] (1.215,-1.312) rectangle (1.338,-1.382);\fill[oiPurple!12] (1.350,-1.312) rectangle (1.473,-1.382);\fill[oiPurple!12] (1.485,-1.312) rectangle (1.608,-1.382);\fill[oiPurple] (0.000,-1.394) rectangle (0.123,-1.464);\fill[oiPurple] (0.135,-1.394) rectangle (0.258,-1.464);\fill[oiPurple!12] (0.270,-1.394) rectangle (0.393,-1.464);\fill[oiPurple] (0.405,-1.394) rectangle (0.528,-1.464);\fill[oiPurple!12] (0.540,-1.394) rectangle (0.663,-1.464);\fill[oiPurple] (0.675,-1.394) rectangle (0.798,-1.464);\fill[oiPurple!12] (0.810,-1.394) rectangle (0.933,-1.464);\fill[oiPurple] (0.945,-1.394) rectangle (1.068,-1.464);\fill[oiPurple!12] (1.080,-1.394) rectangle (1.203,-1.464);\fill[oiPurple!12] (1.215,-1.394) rectangle (1.338,-1.464);\fill[oiPurple!12] (1.350,-1.394) rectangle (1.473,-1.464);\fill[oiPurple!12] (1.485,-1.394) rectangle (1.608,-1.464);\fill[oiPurple] (0.000,-1.476) rectangle (0.123,-1.546);\fill[oiPurple] (0.135,-1.476) rectangle (0.258,-1.546);\fill[oiPurple!12] (0.270,-1.476) rectangle (0.393,-1.546);\fill[oiPurple] (0.405,-1.476) rectangle (0.528,-1.546);\fill[oiPurple!12] (0.540,-1.476) rectangle (0.663,-1.546);\fill[oiPurple] (0.675,-1.476) rectangle (0.798,-1.546);\fill[oiPurple] (0.810,-1.476) rectangle (0.933,-1.546);\fill[oiPurple!12] (0.945,-1.476) rectangle (1.068,-1.546);\fill[oiPurple!12] (1.080,-1.476) rectangle (1.203,-1.546);\fill[oiPurple!12] (1.215,-1.476) rectangle (1.338,-1.546);\fill[oiPurple!12] (1.350,-1.476) rectangle (1.473,-1.546);\fill[oiPurple!12] (1.485,-1.476) rectangle (1.608,-1.546);\fill[oiPurple] (0.000,-1.558) rectangle (0.123,-1.628);\fill[oiPurple] (0.135,-1.558) rectangle (0.258,-1.628);\fill[oiPurple!12] (0.270,-1.558) rectangle (0.393,-1.628);\fill[oiPurple!12] (0.405,-1.558) rectangle (0.528,-1.628);\fill[oiPurple] (0.540,-1.558) rectangle (0.663,-1.628);\fill[oiPurple] (0.675,-1.558) rectangle (0.798,-1.628);\fill[oiPurple] (0.810,-1.558) rectangle (0.933,-1.628);\fill[oiPurple!12] (0.945,-1.558) rectangle (1.068,-1.628);\fill[oiPurple!12] (1.080,-1.558) rectangle (1.203,-1.628);\fill[oiPurple!12] (1.215,-1.558) rectangle (1.338,-1.628);\fill[oiPurple!12] (1.350,-1.558) rectangle (1.473,-1.628);\fill[oiPurple!12] (1.485,-1.558) rectangle (1.608,-1.628);\fill[oiPurple] (0.000,-1.640) rectangle (0.123,-1.710);\fill[oiPurple] (0.135,-1.640) rectangle (0.258,-1.710);\fill[oiPurple] (0.270,-1.640) rectangle (0.393,-1.710);\fill[oiPurple] (0.405,-1.640) rectangle (0.528,-1.710);\fill[oiPurple!12] (0.540,-1.640) rectangle (0.663,-1.710);\fill[oiPurple!12] (0.675,-1.640) rectangle (0.798,-1.710);\fill[oiPurple!12] (0.810,-1.640) rectangle (0.933,-1.710);\fill[oiPurple!12] (0.945,-1.640) rectangle (1.068,-1.710);\fill[oiPurple!12] (1.080,-1.640) rectangle (1.203,-1.710);\fill[oiPurple] (1.215,-1.640) rectangle (1.338,-1.710);\fill[oiPurple!12] (1.350,-1.640) rectangle (1.473,-1.710);\fill[oiPurple!12] (1.485,-1.640) rectangle (1.608,-1.710);\fill[oiPurple] (0.000,-1.722) rectangle (0.123,-1.792);\fill[oiPurple] (0.135,-1.722) rectangle (0.258,-1.792);\fill[oiPurple] (0.270,-1.722) rectangle (0.393,-1.792);\fill[oiPurple!12] (0.405,-1.722) rectangle (0.528,-1.792);\fill[oiPurple] (0.540,-1.722) rectangle (0.663,-1.792);\fill[oiPurple!12] (0.675,-1.722) rectangle (0.798,-1.792);\fill[oiPurple!12] (0.810,-1.722) rectangle (0.933,-1.792);\fill[oiPurple!12] (0.945,-1.722) rectangle (1.068,-1.792);\fill[oiPurple] (1.080,-1.722) rectangle (1.203,-1.792);\fill[oiPurple!12] (1.215,-1.722) rectangle (1.338,-1.792);\fill[oiPurple!12] (1.350,-1.722) rectangle (1.473,-1.792);\fill[oiPurple!12] (1.485,-1.722) rectangle (1.608,-1.792);\fill[oiPurple] (0.000,-1.804) rectangle (0.123,-1.874);\fill[oiPurple] (0.135,-1.804) rectangle (0.258,-1.874);\fill[oiPurple!12] (0.270,-1.804) rectangle (0.393,-1.874);\fill[oiPurple] (0.405,-1.804) rectangle (0.528,-1.874);\fill[oiPurple] (0.540,-1.804) rectangle (0.663,-1.874);\fill[oiPurple!12] (0.675,-1.804) rectangle (0.798,-1.874);\fill[oiPurple!12] (0.810,-1.804) rectangle (0.933,-1.874);\fill[oiPurple!12] (0.945,-1.804) rectangle (1.068,-1.874);\fill[oiPurple!12] (1.080,-1.804) rectangle (1.203,-1.874);\fill[oiPurple!12] (1.215,-1.804) rectangle (1.338,-1.874);\fill[oiPurple] (1.350,-1.804) rectangle (1.473,-1.874);\fill[oiPurple!12] (1.485,-1.804) rectangle (1.608,-1.874);\fill[oiPurple!12] (0.000,-1.886) rectangle (0.123,-1.956);\fill[oiPurple] (0.135,-1.886) rectangle (0.258,-1.956);\fill[oiPurple] (0.270,-1.886) rectangle (0.393,-1.956);\fill[oiPurple] (0.405,-1.886) rectangle (0.528,-1.956);\fill[oiPurple] (0.540,-1.886) rectangle (0.663,-1.956);\fill[oiPurple!12] (0.675,-1.886) rectangle (0.798,-1.956);\fill[oiPurple!12] (0.810,-1.886) rectangle (0.933,-1.956);\fill[oiPurple] (0.945,-1.886) rectangle (1.068,-1.956);\fill[oiPurple!12] (1.080,-1.886) rectangle (1.203,-1.956);\fill[oiPurple!12] (1.215,-1.886) rectangle (1.338,-1.956);\fill[oiPurple!12] (1.350,-1.886) rectangle (1.473,-1.956);\fill[oiPurple!12] (1.485,-1.886) rectangle (1.608,-1.956);\fill[oiPurple] (0.000,-1.968) rectangle (0.123,-2.038);\fill[oiPurple!12] (0.135,-1.968) rectangle (0.258,-2.038);\fill[oiPurple] (0.270,-1.968) rectangle (0.393,-2.038);\fill[oiPurple] (0.405,-1.968) rectangle (0.528,-2.038);\fill[oiPurple] (0.540,-1.968) rectangle (0.663,-2.038);\fill[oiPurple!12] (0.675,-1.968) rectangle (0.798,-2.038);\fill[oiPurple!12] (0.810,-1.968) rectangle (0.933,-2.038);\fill[oiPurple] (0.945,-1.968) rectangle (1.068,-2.038);\fill[oiPurple!12] (1.080,-1.968) rectangle (1.203,-2.038);\fill[oiPurple!12] (1.215,-1.968) rectangle (1.338,-2.038);\fill[oiPurple!12] (1.350,-1.968) rectangle (1.473,-2.038);\fill[oiPurple!12] (1.485,-1.968) rectangle (1.608,-2.038);\fill[oiPurple!12] (0.000,-2.050) rectangle (0.123,-2.120);\fill[oiPurple] (0.135,-2.050) rectangle (0.258,-2.120);\fill[oiPurple] (0.270,-2.050) rectangle (0.393,-2.120);\fill[oiPurple] (0.405,-2.050) rectangle (0.528,-2.120);\fill[oiPurple] (0.540,-2.050) rectangle (0.663,-2.120);\fill[oiPurple] (0.675,-2.050) rectangle (0.798,-2.120);\fill[oiPurple!12] (0.810,-2.050) rectangle (0.933,-2.120);\fill[oiPurple!12] (0.945,-2.050) rectangle (1.068,-2.120);\fill[oiPurple!12] (1.080,-2.050) rectangle (1.203,-2.120);\fill[oiPurple!12] (1.215,-2.050) rectangle (1.338,-2.120);\fill[oiPurple!12] (1.350,-2.050) rectangle (1.473,-2.120);\fill[oiPurple!12] (1.485,-2.050) rectangle (1.608,-2.120);\fill[oiPurple!12] (0.000,-2.132) rectangle (0.123,-2.202);\fill[oiPurple] (0.135,-2.132) rectangle (0.258,-2.202);\fill[oiPurple] (0.270,-2.132) rectangle (0.393,-2.202);\fill[oiPurple] (0.405,-2.132) rectangle (0.528,-2.202);\fill[oiPurple] (0.540,-2.132) rectangle (0.663,-2.202);\fill[oiPurple!12] (0.675,-2.132) rectangle (0.798,-2.202);\fill[oiPurple] (0.810,-2.132) rectangle (0.933,-2.202);\fill[oiPurple!12] (0.945,-2.132) rectangle (1.068,-2.202);\fill[oiPurple!12] (1.080,-2.132) rectangle (1.203,-2.202);\fill[oiPurple!12] (1.215,-2.132) rectangle (1.338,-2.202);\fill[oiPurple!12] (1.350,-2.132) rectangle (1.473,-2.202);\fill[oiPurple!12] (1.485,-2.132) rectangle (1.608,-2.202);\fill[oiPurple] (0.000,-2.214) rectangle (0.123,-2.284);\fill[oiPurple] (0.135,-2.214) rectangle (0.258,-2.284);\fill[oiPurple!12] (0.270,-2.214) rectangle (0.393,-2.284);\fill[oiPurple] (0.405,-2.214) rectangle (0.528,-2.284);\fill[oiPurple!12] (0.540,-2.214) rectangle (0.663,-2.284);\fill[oiPurple!12] (0.675,-2.214) rectangle (0.798,-2.284);\fill[oiPurple] (0.810,-2.214) rectangle (0.933,-2.284);\fill[oiPurple!12] (0.945,-2.214) rectangle (1.068,-2.284);\fill[oiPurple!12] (1.080,-2.214) rectangle (1.203,-2.284);\fill[oiPurple!12] (1.215,-2.214) rectangle (1.338,-2.284);\fill[oiPurple] (1.350,-2.214) rectangle (1.473,-2.284);\fill[oiPurple!12] (1.485,-2.214) rectangle (1.608,-2.284);\node[font=\tiny,text=oiInk!80,anchor=north] at (0.81,-2.32) {orbital occupation (12 orbitals)};\end{tikzpicture}};
\node[anchor=center] at (2.75,0.80) {\begin{tikzpicture}\draw[oiInk!55,line width=0.5pt] (0.15,0.15)--(0.15,2.30);\draw[oiInk!45,line width=0.4pt] (0.05,0.15) node[anchor=east,font=\tiny,text=oiInk!80,xshift=-0.3mm]{0} --(0.15,0.15);\draw[oiInk!45,line width=0.4pt] (0.05,1.22) node[anchor=east,font=\tiny,text=oiInk!80,xshift=-0.3mm]{2} --(0.15,1.22);\draw[oiInk!45,line width=0.4pt] (0.05,2.30) node[anchor=east,font=\tiny,text=oiInk!80,xshift=-0.3mm]{4} --(0.15,2.30);\node[rotate=90,anchor=south,font=\tiny,text=oiInk] at (-0.42,1.22) {energy above $E_0$ (eV)};\draw[oiGreen,line width=1.6pt,line cap=round] (0.35,0.15)--(1.65,0.15);\node[anchor=west,font=\tiny,text=oiGreen] at (1.7,0.15) {$E_0$};\draw[oiGrey,line width=1.1pt,line cap=round] (0.5,0.59)--(1.5,0.59);\draw[oiGrey,line width=1.1pt,line cap=round] (0.5,1.74)--(1.5,1.74);\draw[oiGrey,line width=1.1pt,line cap=round] (0.5,1.87)--(1.5,1.87);\draw[oiGrey,line width=1.1pt,line cap=round] (0.5,2.26)--(1.5,2.26);\end{tikzpicture}};
\draw[-{Stealth[length=2.4mm]},oiInk!65,line width=1pt] (12.45,0.9)--(10.60,0.9);
\draw[-{Stealth[length=2.4mm]},oiInk!65,line width=1pt] (6.60,0.9)--(4.65,0.9);
\draw[-{Stealth[length=2.6mm]},oiVerm,line width=1.1pt] (14.65,4.7)--(14.65,3.5);
\node[anchor=west,font=\tiny,text=oiVerm,align=left] at (14.80,4.1) {recovered\\bitstrings};
\draw[oiInk!25,line width=0.5pt] (0,-2.0)--(17.4,-2.0);
\node[anchor=north,font=\footnotesize\bfseries,text=oiInk] at (8.70,-2.15) {Machine-learning entry points};
\node[circle,fill=oiOrange,text=white,font=\tiny\bfseries,inner sep=1.6pt] at (2.2,-2.95) {1};
\node[anchor=west,font=\scriptsize,text=oiInk] at (2.5,-2.95) {design the sampling circuit \emph{(GQE, stage 2)}};
\node[circle,fill=oiPurple,text=white,font=\tiny\bfseries,inner sep=1.6pt] at (9.6,-2.95) {2};
\node[anchor=west,font=\scriptsize,text=oiInk] at (9.9,-2.95) {generate / select configurations --- RBM $\cdot$ NQS $\cdot$ GFlowNet \emph{(stages 4--5)}};
\end{tikzpicture}
}
\caption{\textbf{The sample-based quantum diagonalization workflow} (N$_2$, CAS(10e,12o), exact-FCI-verifiable). A shallow LUCJ circuit on the quantum processor (orbital rotations $U(\theta)$ interleaved with diagonal-Coulomb Jastrow gates) is sampled in the computational basis; here $76\%$ of shots survive particle-number post-selection in the single-spin ($12$-qubit) half of the register at Heron-calibrated noise, and $57\%$ across the full $24$-qubit register (\texttt{calculations/supervivencia\_fig1.py}, $2\times10^{5}$ shots). The second figure is close to the square of the first because this noise model applies its depolarizing background per spin half, which makes the two halves very nearly independent under it; on a real device they need not be, and we give the full-register number as measured rather than inferred. The classical co-processor repairs the broken samples by occupation-weighted S-CORE, assembles the recovered determinants into a subspace $\mathcal{S}$, and diagonalizes the projected Hamiltonian, whose energy is a variational upper bound converging to the exact FCI value (panel~6). Machine learning augments either the sampler (\ent{1}, generative circuit design) or the generation/selection of configurations (\ent{2}). \emph{Panel~5 shows the target, not a sampled run:} each row is one of the 28 $\alpha$-strings carrying the largest exact-FCI amplitude, ordered by $|c|^2$ in the reproducible $D_{\infty h}$ symmetry-adapted basis of \S\ref{sec:2-1} --- the same gauge as the two insets of this figure and as Fig.~\ref{fig:coupon}.  The cut at 28 is illustrative rather than principled: ranks 28 and 29 are a degenerate pair, and 21 of the 28 rows have gauge-dependent weights, so membership near the boundary is defined only up to rotation within the degenerate $\pi$ shells --- the same oracle ordering that Fig.~\ref{fig:compact} plots as unreachable in practice. Panels 2--4 and 6 are sampled output of the accompanying calculations.}
\label{fig:loop}
\end{figure*}

The object shared by all methods in this review is a hybrid quantum--classical loop with four stages. (i) A parameterized quantum circuit prepares an approximate ground state. In the flagship demonstrations this is the \emph{local unitary cluster Jastrow} (LUCJ) ansatz \cite{motta2023}, a double-factorized, Trotterized representation of the unitary cluster Jastrow wavefunction --- each layer rotating into a new orbital basis, applying a diagonal Coulomb (Jastrow) factor there, and rotating back --- with entangling operations restricted to a physically motivated local connectivity, making it native to heavy-hex superconducting lattices and initializable from classical CCSD amplitudes. (ii) The circuit is measured in the computational basis; under the Jordan--Wigner mapping each length-$2M$ bitstring (for $M$ spatial orbitals) is an electronic configuration, split into an $\alpha$ and a $\beta$ half. Each measured bitstring is thus a \emph{proposed} Slater determinant, and the collection of distinct bitstrings, with their frequencies, is the raw material the rest of the pipeline processes. (iii) In SQD, noise corrupts a large fraction of these bitstrings to the wrong particle number or spin; a \emph{self-consistent configuration recovery} (S-CORE) step repairs them, flipping spin-orbitals toward the mean occupation of the current best eigenvector and iterating to self-consistency \cite{robledomoreno2025}; we make the mechanism precise below. (iv) The Hamiltonian is projected onto the recovered determinant subspace and diagonalized by a sparse (Davidson) eigensolver, distributed across high-performance classical hardware; the result is a variational upper bound and a sparse ground-state wavefunction. \textbf{What $D$ counts.} Throughout this paper $D$ denotes the number of distinct single-spin ($\alpha$) strings retained, and the subspace is the product $\mathcal{S}=\mathcal{S}_\alpha\otimes\mathcal{S}_\beta$ of that set with itself, so a quoted $D$ corresponds to a projected Hamiltonian of dimension $D^2$ within the $S_z$-conserving sector --- $D{=}120$, the value used throughout our own experiments, is a $14{,}400$-determinant subspace. Every method compared at a given $D$ is truncated to the same number of strings, so such comparisons are made at matched classical diagonalization cost.

\begin{figure*}[tbp]
\centering
\includegraphics[width=0.86\linewidth]{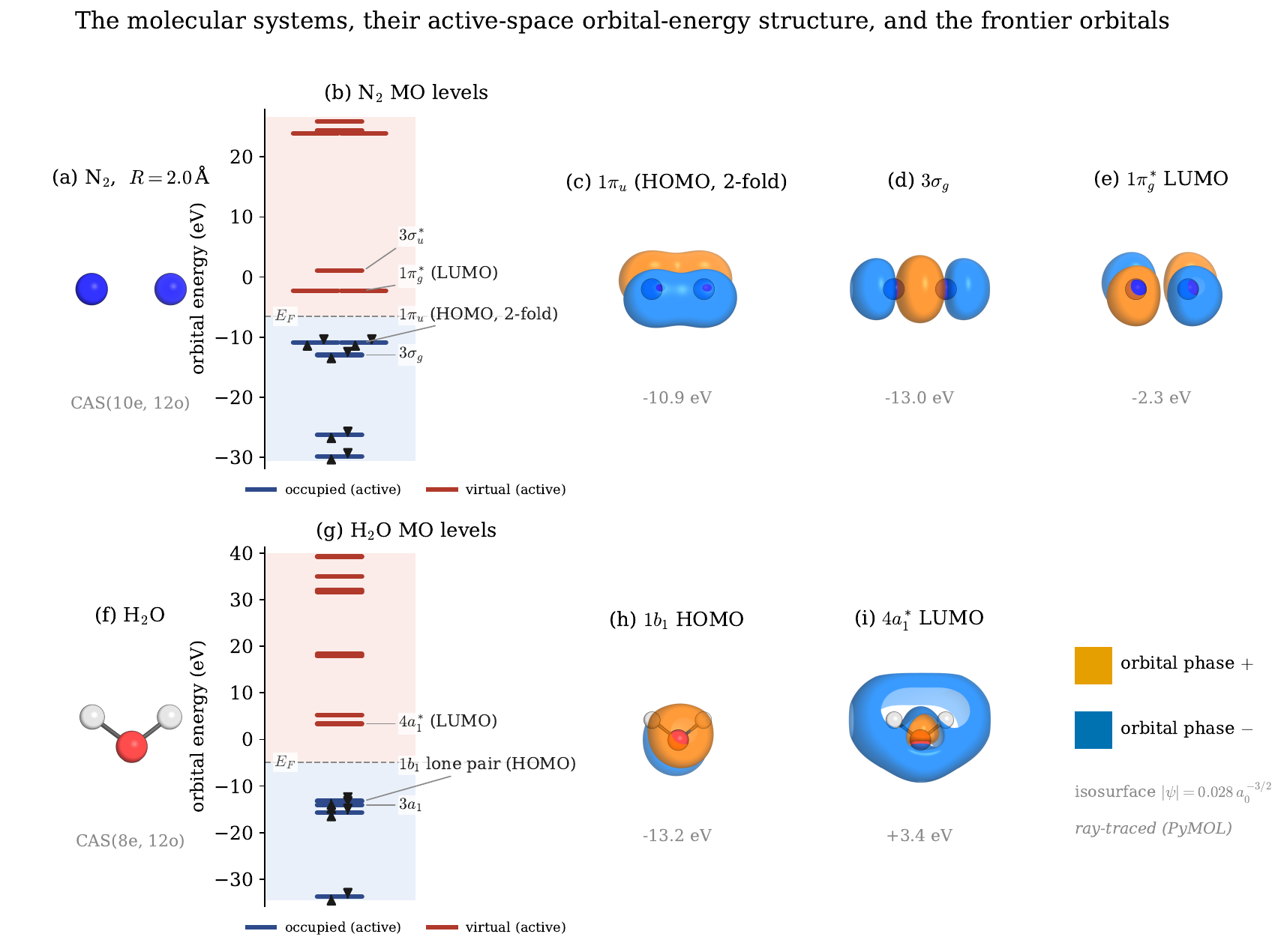}
\caption{\textbf{The molecular systems, their active-space orbital-energy structure, and the frontier orbitals.} \emph{(a,f)} Ray-traced ball-and-stick geometries of stretched N$_2$ ($R=2.0\,$Å, CAS(10e,12o)) and H$_2$O ($r_{\mathrm{OH}}=1.24\,$Å, $\angle\mathrm{HOH}=104.4^\circ$, CAS(8e,12o) --- also stretched, by $29\%$ against its $0.958\,$Å equilibrium, but far less so than N$_2$ and still dominated by its Hartree--Fock reference). \emph{(b,g)} Molecular-orbital energy-level diagrams (restricted Hartree--Fock, eV): occupied active levels (blue) carry their two electrons as up/down arrows, virtual active levels are red, the shaded band is the complete active space, and $E_F$ marks the HOMO--LUMO midpoint; the deep $1s$ core is frozen and off-scale. \emph{(c--e)} Ray-traced (PyMOL) isosurfaces of the N$_2$ frontier orbitals --- the two-fold-degenerate $1\pi_u$ HOMO ($-10.9$~eV; note the axis-containing nodal plane of $\pi$ symmetry), the non-degenerate $3\sigma_g$ level just below it ($-13.0$~eV; cylindrically symmetric about the bond axis), and the $1\pi_g^{*}$ LUMO ($-2.3$~eV) --- with their RHF energies. The crowding of these frontier levels as the triple bond stretches, and the collapse of the bonding--antibonding gaps ($1\pi_u/1\pi_g^{*}$, $3\sigma_g/3\sigma_u^{*}$) toward dissociation, is what makes stretched N$_2$ strongly multireference: at $R=2.0$~\AA{} the FCI natural-orbital occupations of the frontier pair are already $1.32/0.67$ (versus the closed-shell $2/0$) and the Hartree--Fock determinant carries only $34\%$ of the exact weight (\S\ref{sec:7}). \emph{(h,i)} The H$_2$O $1b_1$ lone-pair HOMO and $4a_1^{*}$ LUMO. Isosurfaces at $|\psi|=0.028\,a_0^{-3/2}$; orbital phases orange/blue.}
\label{fig:system}
\end{figure*}

Formally, the electronic Hamiltonian in second quantization is

\begin{equation}
\hat H \;=\; \sum_{pq} h_{pq}\, \hat a_p^{\dagger}\hat a_q \;+\; \tfrac{1}{2}\!\sum_{pqrs} (pq|rs)\, \hat a_p^{\dagger}\hat a_r^{\dagger}\hat a_s\hat a_q \;+\; E_{\mathrm{nuc}},
\end{equation}

whose exact diagonalization in the full configuration-interaction (FCI) basis of $M$ active spatial orbitals with $(n_\alpha,n_\beta)$ electrons has dimension $\dim\mathcal H_{\mathrm{FCI}} = \binom{M}{n_\alpha}\binom{M}{n_\beta}$, growing combinatorially. SQD replaces $\mathcal H_{\mathrm{FCI}}$ by a sampled subspace $\mathcal S = \mathrm{span}\{|D_x\rangle\}_{x\in\mathcal S}$ of Slater determinants and solves the projected eigenproblem for the projector $\hat P_{\mathcal S}=\sum_{x\in\mathcal S}|D_x\rangle\langle D_x|$,

\begin{equation}
\big(\hat P_{\mathcal S}\,\hat H\,\hat P_{\mathcal S}\big)\,\mathbf c \;=\; E_{\mathcal S}\,\mathbf c, \qquad E_{\mathcal S} \;\ge\; E_{\mathrm{FCI}},
\end{equation}

the inequality (the Rayleigh--Ritz variational principle; the Hylleraas--Undheim--MacDonald theorem is its interlacing extension to the excited roots) being the variational guarantee that makes every SQD energy a rigorous upper bound that improves monotonically as good determinants are added. This guarantee is a double-edged asset, and the edge it cuts against is the theme of \S\ref{sec:5}--\S\ref{sec:7}: the same projected-Hamiltonian structure that lets one \emph{certify} an SQD energy from the classical side is what lets one \emph{construct}, screen, and, for the shallow circuits actually run, \emph{simulate} the subspace classically. The variational principle that grounds the method's appeal is thus also the doorway through which its classical competitors enter. The self-consistent recovery (S-CORE) closes the loop: from the current subspace eigenvector it forms the mean spin-orbital occupations $\langle n_p\rangle = \sum_{x\in\mathcal S} |c_x|^2\, [x]_p$ and repairs each symmetry-broken bitstring by flipping spin-orbitals with probability increasing in the deviation $\big|[x]_p-\langle n_p\rangle\big|$, iterating recovery $\to$ diagonalization $\to$ occupation-update to self-consistency. Because the whole quantitative analysis of this review lives in an active space small enough for exact FCI (N$_2$ and H$_2$O in CAS(10e,12o)/(8e,12o), whose active-space orbitals are shown in Fig.~\ref{fig:system}), every energy we quote below is an \emph{error against exact truth} within that model space, not against another approximation. \textbf{Basis and gauge convention.} Every calculation we perform ourselves uses restricted Hartree--Fock orbitals in the cc-pVDZ basis, with frozen cores $n_{\mathrm{core}}{=}2$ for N$_2$ and C$_2$ and $n_{\mathrm{core}}{=}1$ for H$_2$O, giving CASCI(10e,12o) for N$_2$ and CASCI(8e,12o) for the other two. We state this with unusual precision because it is load-bearing, and we flag a trap that we fell into ourselves. The N$_2$ active space contains three two-fold degenerate $\pi$ shells, and the basis \emph{within} a degenerate shell is arbitrary: rotating inside one leaves $E_{\mathrm{FCI}}$ invariant to $4\times10^{-13}$~Ha while redistributing weight among the $\alpha$-strings, so only $84$ of the $792$ strings --- exactly those that wholly fill or wholly vacate every degenerate shell --- carry a gauge-invariant weight. \emph{And ``canonical Hartree--Fock orbitals'' does not by itself pin that choice.} Five runs of identical code at \texttt{scf.conv\_tol}${=}10^{-12}$ returned five \emph{different} $\pi$-shell orientations --- $85.5^\circ$, $29.2^\circ$, $169.8^\circ$, $92.3^\circ$ and $140.2^\circ$ in the run deposited with this version --- while agreeing on $E_{\mathrm{RHF}}$ to $1.9\times10^{-13}$~Ha. \emph{Those five particular angles are not reproducible, and that is the point rather than a caveat:} five further runs of \texttt{gauge\_study/determinismo.py} return five further angles, and an earlier version of this paper quoted a different five for the same reason. What reproduces is the phenomenon, not the draw. We traced this: it is an artefact of \emph{multithreaded} linear algebra, in which the reduction order inside the degenerate eigensolve varies between runs. Pinning the thread count restores bit-reproducibility --- four single-threaded runs returned $130.192^\circ$ with coefficients identical to nine decimals --- so the arbitrariness is not inherent to the method, but it is present in the default execution environment of every code we are aware of, and it is invisible in the energy. The same behaviour is documented in the PySCF issue tracker \cite{pyscf_degen}. \emph{$D_{\infty h}$ symmetry-adapted} orbitals fix the gauge and are reproducible --- the same five runs then return one orientation to nine decimals --- and we recommend them to anyone reporting determinant counts or compactness figures. The two files behind Fig.~\ref{fig:coupon} are computed in that gauge and carry the recipe, the active-space irreducible representations and the reference energy $E_{\mathrm{FCI}}$ in their headers, so that every number in that figure can be checked against the calculation that produced it; we adopted the practice only after an earlier version of those files, lacking both, could not be audited even by us, and we recommend it because the alternative cost us several days. The remaining deposited files are single coherent runs of the sampling experiments; the $\rho$ columns of \texttt{orderparam.dat} and \texttt{ladder.dat} are both computed in the symmetry-adapted, thresholded gauge and agree to four decimals, and \S\ref{sec:7} reports the residual gauge spread of $\rho$ measured across the three protocols we ran. Every per-string count, ranking, and rank correlation below is defined relative to an orbital basis in this sense, and we give the measured spread wherever it affects a stated figure (\S\ref{sec:3-1}, \S\ref{sec:7}). Quantities that are properties of the state rather than of the basis --- $E_{\mathrm{FCI}}$, natural-orbital occupations, the Hartree--Fock weight, and the excitation-level census --- are unaffected, and the paper's conclusions rest on those.

Two lineages coined the idea nearly simultaneously and use different names for it. \emph{Quantum-selected configuration interaction} (QSCI) was introduced by Kanno and co-workers \cite{kanno2023}, now published in \emph{Physical Review Research}, and framed the method as one in which the quantum device selects the important electronic configurations by sampling, and the classical machine then diagonalizes the Hamiltonian in that subspace. \emph{Sample-based quantum diagonalization} (SQD) is the IBM--RIKEN realization at quantum-centric-supercomputing scale \cite{robledomoreno2025}, adding the self-consistent recovery loop and the tight coupling to an HPC back end. We use SQD and QSCI interchangeably, noting the recovery step as the principal methodological difference and the natural home of the machine-learning contributions reviewed in \S\ref{sec:4}.

\subsection{A taxonomy of variants}\label{sec:2-2}

In two years the SQD/QSCI family has branched into some fifty distinct variants; we organize them by the stage of the loop they modify (Table~\ref{tab:sqd}). \emph{Input-state variants} replace the LUCJ sampler: ADAPT-QSCI grows the state one Pauli generator at a time without a variational optimization \cite{nakagawa2024}; time-evolved QSCI (TE-QSCI) and Hamiltonian-simulation QSCI (HSB-QSCI) generate multi-order excitations from real-time evolution rather than circuit optimization \cite{mikkelsen2025,sugisaki2025}; and generative circuit design learns the sampler itself (\S\ref{sec:4-5}). \emph{Krylov variants} (sample-based Krylov quantum diagonalization, SKQD, and its randomized provable-convergence descendant SqDRIFT) build a Krylov subspace from sampled time-evolved states, importing phase-estimation-like convergence guarantees at low depth \cite{yu2025,piccinelli2025}. \emph{Recovery and selection variants} attack the coupon-collector bottleneck directly: amplitude amplification of unseen bitstrings \cite{stockinger2026}, active-sampling by an Epstein--Nesbet acquisition function \cite{miura2026}, and (the focus of \S\ref{sec:4}) machine-learned recovery. \emph{Qubit-reduced variants} shrink the register by different factors: a half-qubit encoding combined with heat-bath-CI-inspired sample selection \cite{mcfarthing2026}, seniority-zero (DOCI) sampling \cite{yoshida2026}, and entanglement forging \cite{smith2025} each halve it, whereas neural fermionic compression (Lossy-QSCI) compresses further, to $\mathcal{O}(N\log M)$ qubits for $N$ electrons in $M$ spin-orbitals \cite{chen2025}, and a CI-matrix reformulation (CIM-QSCI) attains $\lceil \log_2 D_{\mathrm{CIM}} \rceil$ qubits, with $D_{\mathrm{CIM}}$ the CI-matrix dimension --- the same $\mathcal{O}(N \log M)$ order, reached by a different route \cite{graves2026}. \emph{Embedding variants} reach extended systems by using SQD as the impurity or fragment solver inside density-matrix embedding, localized-active-space SCF, or DFT embedding \cite{shajan2025,wang2026}. \emph{Hybrid post-processing variants} recover dynamical correlation missed by the finite subspace, using the SQD wavefunction as an AFQMC trial, a tailored-CC reference, or a multireference-PT2 anchor \cite{danilov2025,erhart2025}. Table~\ref{tab:sqd} is curated rather than exhaustive, and we say so rather than promising a completeness we do not ship: earlier drafts of this review referred to a full annotated tabulation in an accompanying dataset, and no such dataset is deposited.

\begin{table*}[t]\centering\footnotesize
\caption{The SQD/QSCI method family. \emph{Curated, not exhaustive:} the entries are those the argument of \S\ref{sec:4} turns on, chosen from the literature the search of \S\ref{sec:4-6} returned. ``Largest scale'' states hardware qubit counts or active spaces as reported.}
\label{tab:sqd}
\renewcommand{\arraystretch}{1.25}
\begin{tabularx}{\textwidth}{@{}>{\raggedright\arraybackslash}p{2.6cm} >{\raggedright\arraybackslash}p{2.0cm} >{\raggedright\arraybackslash}X >{\raggedright\arraybackslash}p{2.4cm} >{\raggedright\arraybackslash}p{1.8cm}@{}}
\toprule
\textbf{Category} & \textbf{Method} & \textbf{Key idea} & \textbf{Ref.} & \textbf{Largest scale} \\
\midrule
Core & QSCI & Quantum samples the subspace; classical diagonalization & \cite{kanno2023} & 8 qubits \\
 & SQD + S-CORE & Self-consistent configuration recovery; quantum-centric supercomputing & \cite{robledomoreno2025} & 77\,q / (54e,36o) \\
Sampler/input & ADAPT-QSCI & Iterative input-state growth, no VQE & \cite{nakagawa2024} & small molecules \\
 & TE-QSCI & Time-evolved sampling state & \cite{mikkelsen2025} & hydrogen chains, small molecules (numerical) \\
 & HSB-QSCI & Hamiltonian-simulation sampling state & \cite{sugisaki2025} & 36\,q (carbyne) \\
 & Generative-QSCI & Transformer (GQE) designs the sampling circuit & \cite{kemmoku2026} & 32\,q (N$_2$) \\
Krylov & SKQD / SqDRIFT & Sampled Krylov subspace; provable convergence & \cite{yu2025,piccinelli2025,piccinelli2026} & 100\,q (50 orb.) \\
Recovery/select & SQD-AA & Amplitude-amplify unseen bitstrings; quadratic advantage shown analytically for exponentially decaying distributions only & \cite{stockinger2026} & model systems \\
 & ML recovery & Learned generation/classification (see Table~\ref{tab:ml}) & --- & see Table~\ref{tab:ml} \\
Excited states & ext-SQD & Extra projection for low-lying excited states & \cite{barison2025} & N$_2$ 58\,q; [2Fe-2S] 45\,q \\
Orbital optimization & post-SQD oo & Orbital optimization applied \emph{after} the SQD solve (not self-consistent) & \cite{liepuoniute2025} & CH$_2$ 52\,q \\
Symmetry-adapted & SA-SQD & Space-group symmetry embedded in the sampled many-body subspace & \cite{nogaki2026} & two-leg Hubbard ladder (numerical) \\
Time-evolved compact & (unnamed) & Stochastic Hamiltonian time evolution to bias subspace expansion & \cite{weaving2025} & SiH$_4$ 42\,q (IQM) \\
Qubit-reduced & HCI-HSQD & Half-qubit encoding plus heat-bath-CI-inspired sample selection & \cite{mcfarthing2026} & Fe-S (54e,36o) \\
Embedding & DMET-SQD / LASSQD / HQC & SQD as fragment / impurity solver & \cite{shajan2025,wang2026} & protein--ligand 12{,}635 atoms \cite{merz2026} \\
Hybrid post-proc. & AFQMC-SQD; PT2 & Recover dynamical correlation on the SQD reference & \cite{danilov2025,erhart2025} & N$_2$; [2Fe-2S] \\
\bottomrule
\end{tabularx}
\end{table*}

\subsection{Hardware demonstrations and honest scale}\label{sec:2-3}

The method's hardware trajectory is genuinely impressive and worth stating precisely, because it is often conflated with an advantage claim it does not make. The founding demonstration reached N$_2$ at 58 qubits and the iron--sulfur clusters [2Fe-2S] and [4Fe-4S] on 45- and 77-qubit registers, with classical diagonalization distributed across up to 6,400 nodes of Fugaku \cite{robledomoreno2025}; the [4Fe-4S] run, a (54e,36o) active space (72 Jordan--Wigner spin-orbitals, the balance of the register serving the heavy-hex LUCJ mapping), was the largest quantum-computed at the time. A full-scale closed-loop follow-up used the entire 152,064-node Fugaku machine with an on-premises Heron processor and improved the [4Fe-4S] energy by 278 mHa over the 2024 result \cite{shirakawa2025}. The largest register reported in this family is larger still: a randomized sample-based Krylov (SqDRIFT) study of a half-M\"obius molecule reached $36$ orbitals on $72$ qubits and extended to $50$ orbitals on $100$ qubits on superconducting hardware, and is careful to claim only ``a promising path towards practically relevant quantum-assisted electronic-structure calculations'' rather than an advantage \cite{piccinelli2026}. Fragment-based embeddings then pushed the workflow, in 2025--2026, from a $\sim$300-atom miniprotein to two protein--ligand complexes of 11,608 and 12,635 atoms, sampled on two 156-qubit processors using up to 94 qubits --- 21,006 circuits, over 239 hours of device time, $3.0\times10^{9}$ measurement outcomes --- with the fragment diagonalizations distributed over Fugaku, Miyabi-G and ROQUO; the authors report more than a $40\times$ increase in system size over the previous state of the art \cite{merz2026}. These are landmark demonstrations of \emph{scale beyond exact diagonalization}. They are not, and are not claimed by their authors to be, demonstrations of an advantage over the best classical methods: in that same 12,635-atom study the quantum--classical fragment energies are reported as \emph{matching} CCSD, which is the accuracy target, not a method that CCSD cannot reach. The distinction is the one \S\ref{sec:5} makes central, and it is worth stating that the field's own flagship results state it too.

\subsection{Software and ecosystem}\label{sec:2-4}

The method is unusually well tooled for its age, which partly explains its rapid uptake. IBM maintains \texttt{qiskit-addon-sqd} (with an HPC C++ variant), while \texttt{ffsim}, used for fermionic-circuit simulation and LUCJ construction, is developed in the \texttt{qiskit-community} organization (Apache-2.0) rather than by IBM itself; QunaSys maintains \texttt{quri-parts-qsci} within the QURI SDK --- source-available under the QunaSys Source Available License 1.0, not open source --- and, notably, a continuously updated bibliography portal, \emph{Awesome-QSCI} (\texttt{awesome-qsci.qunasys.com}), curated by QunaSys rather than by the community, which functions as a living review of a field moving too fast for the conventional publication cycle. NVIDIA's CUDA-Q documents SKQD as a reference application and provides GPU primitives for the classical diagonalization back end. This tooling maturity is a mixed blessing: it lowers the barrier to method development --- including the machine-learning methods of \S\ref{sec:4} --- but it also lowers the barrier to producing demonstrations that are not benchmarked against the strong classical baselines of \S\ref{sec:5}.

\section{The configuration-recovery bottleneck: the machine-learning entry point}\label{sec:3}

\subsection{A coupon-collector law}\label{sec:3-1}

The efficiency of the entire loop reduces to a single question: how many distinct, \emph{important} determinants can be placed in the subspace, and at what cost? The answer is governed by the amplitude distribution of the exact ground state, which is heavy-tailed. In stretched N$_2$ in a (10e,12o) active space, for instance, 90\% of the exact FCI weight is carried by roughly 1.5\% of the single-spin strings (twelve of the 792), with per-configuration weights spanning many orders of magnitude. Sampling from such a distribution to \emph{discover} the low-amplitude but chemically decisive determinants is a coupon-collector problem with non-uniform probabilities: as sampling proceeds, an overwhelming fraction of draws re-select already-seen configurations, and the number of newly discovered determinants per shot collapses. Reinholdt and co-workers made this quantitative and damaging: for N$_2$ the discovery rate falls from $\sim$0.1 determinants per shot at $10^3$ shots to $\sim\!5\times10^{-4}$ at $10^9$, and reaching micro-hartree precision extrapolates to on the order of $10^{14}$ samples \cite{reinholdt2025}. This is the bottleneck that machine learning is invoked to break: if a learned model can \emph{generate} the important tail rather than wait to sample it, the coupon-collector wall might be circumvented. Whether it can --- and whether doing so beats simply running a classical selected-CI heuristic --- is the question of \S\ref{sec:4} and \S\ref{sec:5}.

\begin{figure*}[tbp]
\centering
\resizebox{\linewidth}{!}{%
\begin{tikzpicture}
\begin{groupplot}[group style={group size=2 by 1, horizontal sep=2.4cm}, paperplot, width=0.46\linewidth, height=0.40\linewidth]
\nextgroupplot[ymode=log, xlabel={single-spin string, ranked by weight}, ylabel={marginal weight $w_\alpha(i)$},
  xmin=0, xmax=792, ymin=1e-10, ymax=0.6, ytick={1e0,1e-2,1e-4,1e-6,1e-8}, ymajorgrids, clip=false,
  title={heavy-tailed determinant weights (N$_2$, exact FCI)}]
  \addplot[oiBlue, line width=1pt] table[x=rank,y=weight]{coupon_w.dat};
  \draw[oiOrange, densely dashed, line width=1pt] (axis cs:12,1e-10) -- (axis cs:12,0.55);
  \node[oiOrange!80!black, font=\scriptsize, anchor=west, align=left] at (axis cs:95,6e-2){90\% weight in\\12 strings};
  \draw[-{Stealth[length=1.7mm]}, oiOrange!80!black, line width=0.7pt] (axis cs:90,7e-2) -- (axis cs:12,2.4e-2);
  \node[oiOrange!65!black, font=\scriptsize, anchor=west] at (axis cs:140,2.4e-8){correlation-energy tail};
  \node[font=\bfseries\large, anchor=south east] at (rel axis cs:-0.14,1.0){a};
\nextgroupplot[xlabel={number of strings included}, ylabel={cumulative ground-state weight},
  xmin=0, xmax=260, ymin=0, ymax=1.04, ytick={0,0.2,0.4,0.6,0.8,1.0}, title={the coupon-collector signature}, clip=false]
  \addplot[oiGreen, line width=1.4pt, restrict x to domain=0:260] table[x=n,y=cum]{coupon_cum.dat};
  \addplot[only marks, mark=*, mark size=3pt, color=oiOrange, forget plot] coordinates {(12,0.9172)};
  \addplot[only marks, mark=*, mark size=3pt, color=oiVerm, forget plot] coordinates {(46,0.9902)};
  \addplot[only marks, mark=*, mark size=3pt, color=oiPurple, forget plot] coordinates {(186,0.9990)};
  \node[oiOrange!88!black, font=\scriptsize\bfseries, anchor=west] at (axis cs:20,0.46){90\% of weight $\rightarrow$ 12 strings (1.5\%)};
  \node[oiVerm, font=\scriptsize\bfseries, anchor=west] at (axis cs:20,0.33){99.0\% $\rightarrow$ 46 strings (5.8\%)};
  \node[oiPurple, font=\scriptsize\bfseries, anchor=west] at (axis cs:20,0.20){99.9\% $\rightarrow$ 186 strings};
  \node[font=\bfseries\large, anchor=south east] at (rel axis cs:-0.14,1.0){b};
\end{groupplot}
\end{tikzpicture}
}
\caption{\textbf{The coupon-collector bottleneck, computed exactly.} N$_2$ CAS(10e,12o), from the exact FCI wavefunction. \emph{(a)} The marginal single-spin-string weight $w_\alpha(i)=\sum_\beta|c_{i\beta}|^2$ falls across roughly nine orders of magnitude, from $0.35$ down to ${\sim}3\times10^{-10}$. \emph{(b)} Cumulative weight against the number of strings included: $90\%$ is carried by twelve of the $792$ strings ($1.5\%$).  The counts shown are basis-dependent; the analysis is in the text.}
\label{fig:coupon}
\end{figure*}
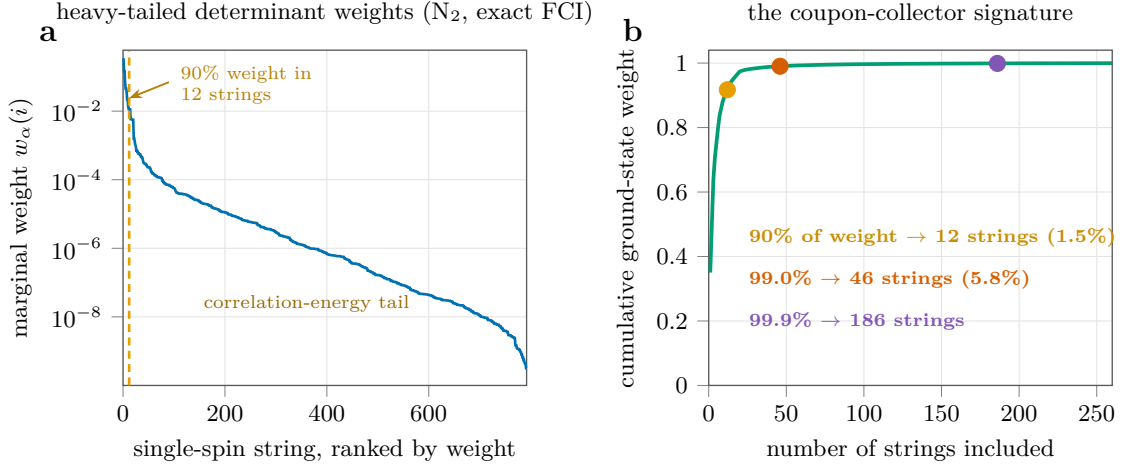

The effect is exact and computable. Writing the marginal weight of a single-spin string $i$ as $w_\alpha(i)=\sum_\beta|c_{i\beta}|^2$ and its sampling probability as $p_i$, the expected number of shots required to first observe every string in a target set $\mathcal T$ is bounded above by the coupon-collector sum (the leading inclusion--exclusion term, and a loose bound exactly where the argument invokes it: evaluated on this system's exact weights it overstates the true expectation by $2.5\times$ for the $90\%$ target set, $5.8\times$ at $99\%$ and $16\times$ at $99.9\%$ (\texttt{calculations/cota\_coleccionista.py}, $4000$ collection simulations per threshold), so it fixes the scaling, not a shot count)

\begin{equation}
\mathbb E\big[\,\text{shots to collect } \mathcal T\,\big] \;\le\; \sum_{i\in\mathcal T}\frac{1}{p_i} \;\xrightarrow[\;p_i\to 0\;]{}\; \infty,
\end{equation}

which upper-bounds the exact expectation (the leading term of an inclusion--exclusion series) and diverges as the heavy tail is approached. Fig.~\ref{fig:coupon} computes this directly from the exact FCI wavefunction of N$_2$ CAS(10e,12o): the same handful of dominant strings carry almost all the weight, but the correlation-energy tail that decides chemical accuracy is spread over hundreds of rare strings --- precisely the configurations whose $p_i$ make the sum above large, and precisely where a learned generator is supposed to earn its keep.

The counts in Fig.~\ref{fig:coupon} are not properties of the state \emph{alone}. \emph{String counts are properties of the state \textbf{and} the orbital basis, not of the state alone,} and we report them accordingly. The active space holds three two-fold degenerate $\pi$ shells, so only $84$ of the $792$ strings --- exactly those that wholly fill or wholly vacate every shell --- carry a gauge-invariant weight; a rotation inside a shell leaves $E_{\mathrm{FCI}}$ unchanged to $4\times10^{-13}$~Ha while redistributing the rest. Here the $90\%$ figure survives because the threshold falls inside an exactly degenerate pair ($w_\alpha(11)=w_\alpha(12)$ to $1.1\times10^{-9}$, the convergence floor) and a multiplet cannot be split. \emph{Recognising the pair is not a judgement call:} at \texttt{scf.conv\_tol}${=}10^{-12}$ and \texttt{fci.conv\_tol}${=}10^{-13}$ the partners within a degenerate $\pi$ pair differ by $4\times10^{-10}$ to $1.3\times10^{-9}$ while genuinely distinct strings are separated by at least $4.0\times10^{-4}$, so any tolerance across five orders of magnitude returns the same count: completing it returns \emph{twelve} in every one of $40$ random intra-shell rotations, though the raw crossing index alternates between $11$ and $12$. That stabilisation is a property of this system and not a general rule --- in C$_2$ at $1.24$~\AA{} multiplet completion \emph{widens} the $90\%$ range, from $4$--$5$ to $4$--$6$. The $99.9\%$ count is not robust here either, ranging over $185$--$196$, so we quote it as ${\approx}190$. \emph{How much does this matter?} Across the systems and thresholds we examined, the worst-case ratio between the largest and smallest count under intra-shell rotation is $1.25$ --- $1.09$ for N$_2$ and $1.25$ for C$_2$ over $40$ intra-shell rotations each, the two systems with degenerate shells that the deposited generator runs (\texttt{gauge\_study/bloque\_conteos.py}). An earlier version of this paper quoted $1.29$ from a wider set of runs we did not deposit, and we replace it with the figure the deposit can support. That is enough to make a factor-of-two compactness comparison meaningless without a stated gauge, and \emph{not} enough to disturb the order-of-magnitude claims on which \S\ref{sec:5} rests --- a distinction we draw explicitly rather than leave to the reader. As a control, H$_2$O in $C_{2v}$ has no degenerate shell at all, so there is no intra-shell rotation to perform and all $495$ strings are invariant by construction --- a structural statement rather than a measurement, and we give it as one. On enlarging the active space the fraction of gauge-invariant strings falls --- $84$ of $792$ ($10.6\%$) at twelve orbitals and $122$ of $2002$ ($6.1\%$) at fourteen, both pure combinatorics and both counted by \texttt{gauge\_study/bloque\_conteos.py} --- so the effect becomes structurally more pervasive as the active space grows. Whether the count \emph{spread} grows with it we do not claim: measuring that at fourteen orbitals means a four-million-determinant FCI we did not run, and we would rather leave the question open than answer it from a calculation we cannot deposit. Three quantities are exactly gauge-invariant and we give them as the reproducible anchors. The largest single-string weight, $0.350238$. The number of \emph{orbits} --- strings grouped into the sets a shell rotation mixes among themselves --- which had zero range over every gauge over $40$ intra-shell rotations of N$_2$: $6$, $23$ and $82$ orbits ($13$, $54$ and $208$ strings) at the three thresholds, against raw counts that wobble by $11$--$12$, $46$--$50$ and $185$--$196$ (\texttt{gauge\_study/bloque\_conteos.py}, which reproduces all six figures exactly; the same script finds $3$, $14$ and $46$ orbits for C$_2$, also with zero range). And the $\alpha|\beta$ Schmidt rank, $9$ at $90\%$ and $60$ at $99.9\%$, which majorization makes a lower bound on the string count in \emph{any} spin-collinear, $S_z$-conserving orbital basis (it does not extend to Bogoliubov or generalised-Hartree--Fock transformations, which mix the two spin sectors and so are not local to this cut). None of this is new in kind: that determinant weights and counts depend on the orbital basis is long established \cite{bytautas2003,abrams2005,alcoba2016,izsak2023}, was quantified this year at five orders of magnitude on [Fe$_4$S$_4$] \cite{zhangotten2026}, and the orbital-rotation invariance of the singular values of $C$ is stated by Stair and Evangelista, who use it to build a method rather than a metric \cite{stair2020}; the $\alpha|\beta$ Schmidt spectrum is itself the cost measure of entanglement forging \cite{eddins2022}. What we add here is the measurement, in the SQD setting, of how far the reported quantity actually moves --- and the observation that no study in this literature accompanies its subspace dimension with a sensitivity estimate, so counts are silently incommensurable across the corpus. The companion study \cite{bonilla2026spectral} reaches the same conclusion from the dynamical side. The curve plotted here is computed in the reproducible $D_{\infty h}$ symmetry-adapted gauge of \S\ref{sec:2-1}; a single-threaded canonical draw moves the same three counts to $12$, $49$ and $194$ --- that difference is the size of the gauge effect, not of an error. We say \emph{a} draw rather than \emph{the} draw deliberately: \S\ref{sec:2-1} establishes that canonical orbitals are not reproducible run to run in the presence of degenerate shells, so there is no single canonical answer to quote, only a family. Pinning the thread count fixes one member of it, and that is the member reported here.

\subsection{Recovery as denoising, and the peculiar role of noise}\label{sec:3-2}

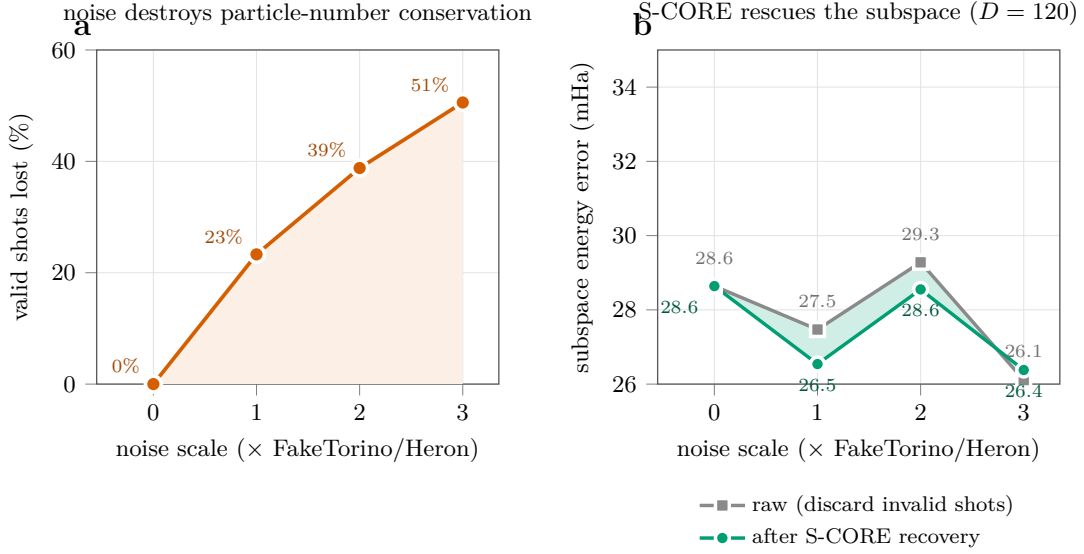
\begin{figure*}[tbp]
\centering
\begin{tikzpicture}
\begin{groupplot}[group style={group size=2 by 1, horizontal sep=2.1cm}, paperplot, width=0.46\linewidth, height=0.40\linewidth]
\nextgroupplot[xlabel={noise scale ($\times$ FakeTorino/Heron)}, ylabel={valid shots lost (\%)},
  xmin=-0.55, xmax=3.35, ymin=0, ymax=60, xtick={0,1,2,3}, title={noise destroys particle-number conservation},
  clip=false]
  \addplot[draw=none, fill=oiVerm!10, forget plot] table[x=scale,y=lost]{recovery.dat} \closedcycle;
  \addplot[oiVerm, line width=1.4pt, mark=*, mark size=3pt, mark options={fill=oiVerm,draw=white},
    nodes near coords={\pgfmathprintnumber[precision=0]{\pgfplotspointmeta}\%},
    nodes near coords style={font=\scriptsize, text=oiVerm!75!black, anchor=south east, xshift=-1pt}] table[x=scale,y=lost]{recovery.dat};
  \node[font=\bfseries\large, anchor=south east, yshift=2pt] at (rel axis cs:0.01,1.0){a};
\nextgroupplot[xlabel={noise scale ($\times$ FakeTorino/Heron)}, ylabel={subspace energy error (mHa)},
  xmin=-0.55, xmax=3.35, ymin=26, ymax=35, xtick={0,1,2,3}, ytick={26,28,30,32,34}, title={S-CORE rescues the subspace ($D=120$)},
  legend style={at={(0.5,-0.30)}, anchor=north, legend columns=1}, clip=false]
  \addplot[name path=raw, oiGrey, line width=1.3pt, mark=square*, mark size=2.6pt, mark options={fill=oiGrey,draw=white},
    nodes near coords, nodes near coords style={font=\scriptsize, text=oiInk!65, anchor=south, yshift=4.5pt, /pgf/number format/fixed, /pgf/number format/fixed zerofill, /pgf/number format/precision=1}]
    table[x=scale,y=eraw]{recovery.dat}; \addlegendentry{raw (discard invalid shots)}
  \addplot[name path=rec, oiGreen, line width=1.3pt, mark=*, mark size=2.6pt, mark options={fill=oiGreen,draw=white},
    nodes near coords, nodes near coords style={font=\scriptsize, text=oiGreen!55!black, anchor=north, yshift=-1.5pt, /pgf/number format/fixed, /pgf/number format/fixed zerofill, /pgf/number format/precision=1,
      /utils/exec={\ifnum\coordindex=0\relax\pgfkeysalso{anchor=north east, xshift=-2pt}\fi}}]
    table[x=scale,y=erec]{recovery.dat}; \addlegendentry{after S-CORE recovery}
  \addplot[oiGreen!18, forget plot] fill between[of=raw and rec];
  \node[font=\bfseries\large, anchor=south east, yshift=2pt] at (rel axis cs:0.01,1.0){b};
\end{groupplot}
\end{tikzpicture}
\caption{\textbf{Configuration recovery under backend-calibrated noise} (FakeTorino / Heron r1 rates; asymmetric readout $+$ depolarizing). \emph{(a)} The share of shots lost to particle-number violation climbs from $0$ at zero noise to ${\sim}51\%$ at $3\times$ --- equivalently, the particle-number-valid fraction falls from $100\%$ to ${\sim}49\%$. \emph{(b)} At fixed $D=120$ retained $\alpha$-strings, S-CORE recovery (green) lowers the subspace energy error relative to discarding the invalid shots (grey) --- most at the device-calibrated scale ($0.9$~mHa at $1\times$), shrinking monotonically to $-0.3$~mHa at $3\times$, where the repair is marginally worse than discarding --- while the raw curve (grey) drifts downward and the recovered curve (green) stays within a narrow band ($26.4$--$28.6$~mHa, not monotonically) as noise broadens Hilbert-space exploration, both consistent with the ``noise can help'' effect. Note the recovered subspace stays well above chemical accuracy here ($\sim$26--29~mHa at this fixed, illustrative $D$): recovery salvages otherwise-discarded samples, it does not by itself reach chemical precision. At zero noise no shots are invalid, so the two series coincide exactly and the grey marker at $0\times$ is hidden beneath the green one. \emph{What this figure's calculation implements is a surrogate, not the published S-CORE algorithm described in \S\ref{sec:2-1}:} it is deterministic (a broken bitstring is repaired by ranking orbitals on occupation and flipping the extreme ones, not by flipping with a probability increasing in the deviation), single-pass (there is no recovery $\to$ diagonalization $\to$ occupation-update loop), and it takes its occupation profile from the raw particle-number-valid shots rather than from a subspace eigenvector. It therefore illustrates the recovery mechanism and, if anything, understates it; it is not a benchmark of S-CORE as published.}
\label{fig:score}
\end{figure*}

On real hardware the bottleneck is compounded by noise, and here the picture is subtle in a way that is central to the field's clear-eyed self-assessment. Under readout and gate noise, the majority of measured bitstrings violate particle-number conservation and would be discarded by naive post-selection: for N$_2$ in a (10e,\,26o) active space on hardware, only about 0.17\% of sampled configurations carry the correct electron number \cite{robledomoreno2025}, a fraction quoted from that hardware run by \cite{vaquerosabater2026}. This survival fraction falls steeply with register width, since each additional qubit is another channel for a particle-number-violating readout error: our illustrative $12$-orbital N$_2$ CAS retains ${\sim}76\%$ of shots at $1\times$ Heron noise in the single-spin ($12$-qubit) half of the register, which is what Fig.~\ref{fig:loop} plots, and ${\sim}57\%$ across the full $24$-qubit register, whereas a large hardware active space over many more qubits retains a fraction of a percent --- the same mechanism, compared like for like on the full register, some two and a half orders of magnitude apart in scale. The self-consistent configuration recovery step is precisely a statistical imputation model that rescues these corrupted samples, and it is the true workhorse of SQD's noise tolerance. That it is a model is also where its bias enters. Because S-CORE repairs every broken bitstring against a \emph{single, global} mean-occupation vector, that reference becomes mixture-averaged wherever the dominant determinants are spread across several regions of determinant space, and the repair is pulled toward mean occupations that no dominant configuration actually has. Park and co-workers' cluster-adaptive SQD (CSQD, March 2026) clusters the pooled single-spin strings and recovers against \emph{cluster-specific} reference occupancies, reporting, at matched variational budget, energies below SQD by up to $15.95$~mHa for stretched N$_2$ in $(10e,26o)$ and $57.82$~mHa for [2Fe-2S] in $(30e,20o)$ \cite{park2026csqd}. We note what such a comparison does and does not settle: it is recovery measured against recovery, on active spaces beyond exact FCI and against SQD rather than against the classical selectors of \S\ref{sec:5-1}, so it says nothing about whether the quantum samples were needed. What it does establish is that the global-reference assumption is a real and fixable defect of the published algorithm --- and a $57.82$~mHa gap left on the table by a heuristic repair is the margin any learned recovery model of \S\ref{sec:4} would first have to capture before claiming anything beyond it. Strikingly, Vaquero-Sabater and co-workers showed that noise can \emph{help}: by scattering samples beyond the biased support of the ideal ansatz, moderate noise broadens Hilbert-space exploration, and recovery starting from \emph{uniformly random} bitstrings reaches accurate energies within a couple of iterations. The implication, that the classical recovery loop (not the quantum sampler) is responsible for the accuracy, is one of the sharpest results in the literature and is examined in \S\ref{sec:5}. For the present section the point is narrower: the recovery step is a learned/statistical model, and it is the natural site for the generative methods of \S\ref{sec:4}, whose promise is to replace an occupancy-weighted heuristic repair with a model that has learned the structure of the important subspace. Fig.~\ref{fig:score} makes the mechanism concrete on backend-calibrated noise: as the device-noise scale grows the fraction of particle-number-valid shots collapses, yet S-CORE recovery still lowers the fixed-dimension subspace error by imputing the broken samples back onto valid configurations, with its largest benefit at the device-calibrated scale itself and shrinking as noise grows further.

\subsection{Why the standard error-mitigation toolkit does not transfer}\label{sec:3-3}

A methodological subtlety that the field has had to internalize, and that a review should make explicit, is that SQD is a \emph{per-sample} post-selection pipeline, not an expectation-value pipeline. This distinction determines which of the near-term error-mitigation techniques transfer. Post-selection on particle number and spin, self-consistent recovery, and spin projection operate on individual bitstrings and are central to the method. In contrast, the techniques that dominate IBM's utility-scale expectation-value experiments (twirled readout error extinction, TREX; zero-noise extrapolation, ZNE; and probabilistic error cancellation, PEC) mitigate errors \emph{in expectation values}, not in individual samples, and so do not transfer. Dynamical decoupling and Pauli twirling give only marginal, inconsistent benefit once the recovery loop has converged \cite{bhuiyan2026}. The practical consequence, relevant to \S\ref{sec:6}, is that a claim about SQD's hardware performance must be benchmarked with the mitigation stack it actually uses (recovery + symmetry post-selection), not the expectation-value stack, and that the shot budget saturates surprisingly early, optimal at $10^3$--$10^4$ shots, with $10^5$ performing \emph{worse} \cite{bhuiyan2026}. This early hardware saturation does not contradict the ${\sim}10^{14}$-sample extrapolation of \S\ref{sec:3-1}: that figure is the noiseless discovery limit for the exact heavy tail, whereas this one is where, on noisy hardware with recovery already converged, additional shots mostly re-collect configurations recovery has restored.

\section{Machine learning and generative methods for determinant and subspace selection}\label{sec:4}

The proposition uniting this section is simple to state: if the important determinants are heavy-tailed and expensive to discover by sampling (\S\ref{sec:3-1}), replace or augment the raw sampler with a \emph{model} that has learned where they are. The proposition is also older than SQD: Coe's machine-learning configuration interaction (MLCI) trained a neural network on-the-fly to predict which unseen determinants a classical selected-CI loop should add, on CO and stretched water, as early as 2018 \cite{coe2018,coe2019}. What SQD added is a reason to revisit the idea urgently and at scale, and a quantum sampler to compete with or complement. The result is a fast-moving ecosystem that, to our knowledge, has not been surveyed. We organize it along two axes, the \emph{object} a method generates (a determinant, a subspace, a circuit) and the \emph{importance signal} it is trained on (CI-coefficient magnitude, a perturbative estimate, quantum-sample frequency, or the diagonalized energy itself), and we treat the five families in turn, closing with the conspicuous gap the taxonomy exposes.

\subsection{A taxonomy by generated object and importance signal}\label{sec:4-1}

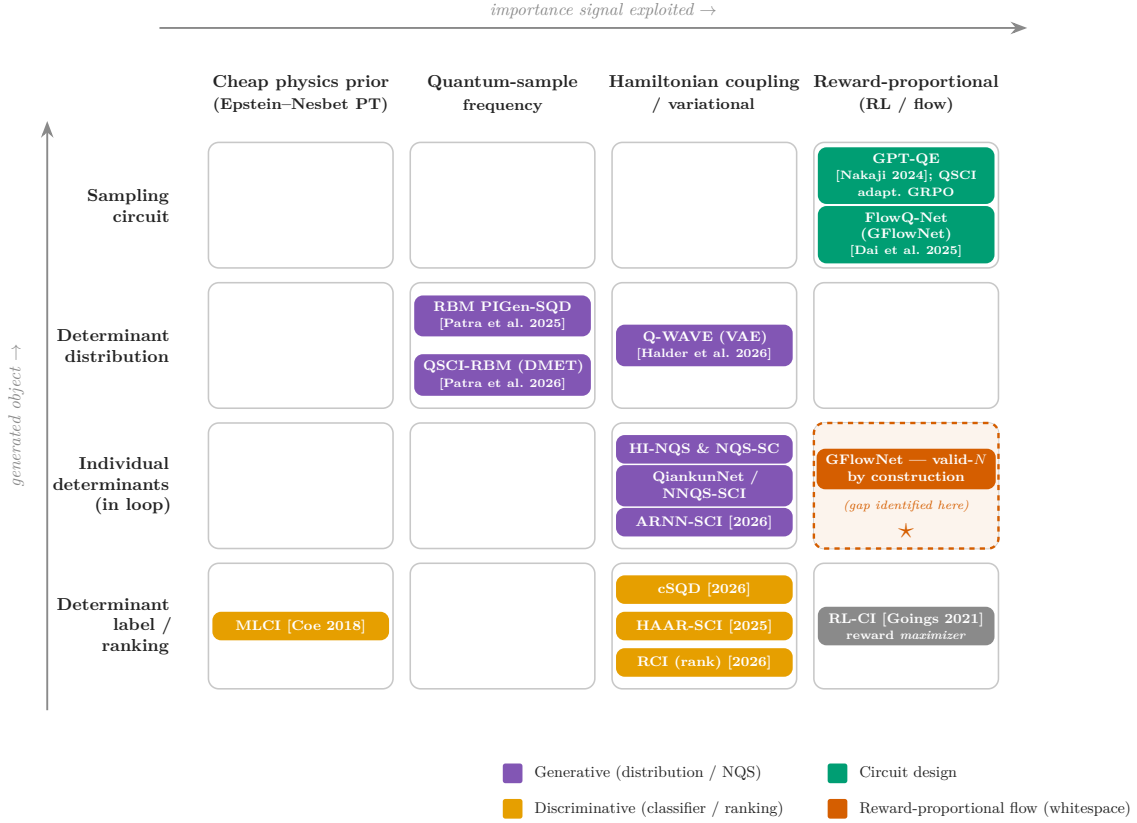
\begin{figure*}[tbp]
\centering
\resizebox{\linewidth}{!}{%
\begin{tikzpicture}[font=\sffamily,
  cell/.style={draw=oiInk!25, rounded corners=6pt, line width=0.9pt, minimum width=3.75cm, minimum height=2.55cm},
  wscell/.style={draw=oiVerm, dashed, line width=1.4pt, fill=oiVerm!7, rounded corners=6pt, minimum width=3.75cm, minimum height=2.55cm},
  chip/.style={rounded corners=5pt, text=white, font=\footnotesize\bfseries, align=center, inner sep=3.4pt, minimum height=0.58cm, text width=3.35cm},
  chipT/.style={rounded corners=5pt, text=white, font=\footnotesize\bfseries, align=center, inner sep=3.4pt, minimum height=0.78cm, text width=3.35cm},
  colh/.style={font=\bfseries, text=oiInk, align=center, text width=4.0cm, anchor=south},
  rowh/.style={font=\bfseries, text=oiInk, align=right, text width=2.5cm, anchor=east},
]
\def\cx{4.1}\def\cy{2.85}
\foreach \c in {0,1,2,3}{\foreach \r in {0,1,2,3}{\node[cell] at (\c*\cx,\r*\cy){};}}
\node[wscell] at (3*\cx,1*\cy){};
\def\hy{10.30}
\node[colh] at (0*\cx,\hy){Cheap physics prior\\[1pt]{\small (Epstein--Nesbet PT)}};
\node[colh] at (1*\cx,\hy){Quantum-sample\\[1pt]{\small frequency}};
\node[colh] at (2*\cx,\hy){Hamiltonian coupling\\[1pt]{\small / variational}};
\node[colh] at (3*\cx,\hy){Reward-proportional\\[1pt]{\small (RL / flow)}};
\node[rowh] at (-2.55,3*\cy){Sampling\\circuit};
\node[rowh] at (-2.55,2*\cy){Determinant\\distribution};
\node[rowh] at (-2.55,1*\cy){Individual\\determinants\\(in loop)};
\node[rowh] at (-2.55,0*\cy){Determinant\\label / ranking};
\node[chipT,fill=oiGreen] at (3*\cx,3*\cy+0.60){GPT-QE\\{\scriptsize [Nakaji 2024]; QSCI adapt.\ GRPO}};
\node[chipT,fill=oiGreen] at (3*\cx,3*\cy-0.60){FlowQ-Net (GFlowNet)\\{\scriptsize [Dai et al.\ 2025]}};
\node[chipT,fill=oiPurple] at (1*\cx,2*\cy+0.60){RBM PIGen-SQD\\{\scriptsize [Patra et al. 2025]}};
\node[chipT,fill=oiPurple] at (1*\cx,2*\cy-0.60){QSCI-RBM (DMET)\\{\scriptsize [Patra et al. 2026]}};
\node[chipT,fill=oiPurple] at (2*\cx,2*\cy){Q-WAVE (VAE)\\{\scriptsize [Halder et al.\ 2026]}};
\node[chip,fill=oiPurple] at (2*\cx,1*\cy+0.74){HI-NQS \& NQS-SC};
\node[chip,fill=oiPurple] at (2*\cx,1*\cy){QiankunNet / NNQS-SCI};
\node[chip,fill=oiPurple] at (2*\cx,1*\cy-0.74){ARNN-SCI [2026]};
\node[chip,fill=oiOrange] at (0*\cx,0*\cy){MLCI [Coe 2018]};
\node[chip,fill=oiOrange] at (2*\cx,0*\cy+0.74){cSQD [2026]};
\node[chip,fill=oiOrange] at (2*\cx,0*\cy){HAAR-SCI [2025]};
\node[chip,fill=oiOrange] at (2*\cx,0*\cy-0.74){RCI (rank) [2026]};
\node[chipT,fill=oiGrey] at (3*\cx,0*\cy){RL-CI [Goings 2021]\\{\scriptsize reward \emph{maximizer}}};
\node[chipT,fill=oiVerm,text width=3.4cm] at (3*\cx,1*\cy+0.34){GFlowNet --- valid-$N$\\by construction};
\node[oiVerm,font=\itshape\scriptsize] at (3*\cx,1*\cy-0.42){(gap identified here)};
\node[text=oiVerm] at (3*\cx,1*\cy-0.92){\fontsize{18}{18}\selectfont$\star$};
\draw[-{Stealth[length=3.4mm]}, oiGrey, line width=1.1pt] (-0.7*\cx,12.15) -- (3.6*\cx,12.15);
\node[oiGrey, font=\itshape, anchor=south] at (1.5*\cx,12.23){importance signal exploited $\rightarrow$};
\draw[-{Stealth[length=3.4mm]}, oiGrey, line width=1.1pt] (-5.15,-0.6*\cy) -- (-5.15,3.6*\cy);
\node[oiGrey, font=\itshape, rotate=90, anchor=south] at (-5.5,1.5*\cy){generated object $\rightarrow$};
\begin{scope}[shift={(1.05*\cx,-1.05*\cy)}]
  \foreach \i/\col/\lab in {0/oiPurple/{Generative (distribution / NQS)}, 1/oiGreen/{Circuit design}, 2/oiOrange/{Discriminative (classifier / ranking)}, 3/oiVerm/{Reward-proportional flow (whitespace)}}{
    \pgfmathsetmacro\lx{mod(\i,2)*6.6}\pgfmathsetmacro\ly{-int(\i/2)*0.72}
    \node[fill=\col, minimum size=4mm, rounded corners=1.5pt] at (\lx,\ly){};
    \node[anchor=west, font=\small, text=oiInk] at (\lx+0.32,\ly){\lab};
  }
\end{scope}
\end{tikzpicture}%
}
\caption{\textbf{A taxonomy of machine-learning methods for determinant and subspace selection in SQD}, organized by the object each method generates (rows) and the importance signal it exploits (columns). Generative models (purple) learn a distribution over determinants; discriminative models (orange) classify or rank them; circuit-design methods (green) learn the sampler itself. The reward-driven column (right) holds three occupants, and none of them \emph{samples determinants reward-proportionally}: two circuit designers (green) --- GPT-QE, whose logit-matching objective is genuinely reward-proportional over circuits, and FlowQ-Net [Dai et al.\ 2025], a true GFlowNet over gate sequences --- and the reinforcement-learning \emph{maximizer} RL-CI [Goings 2021] (grey), which does act on determinants, but by maximization rather than proportional sampling. The reward-\emph{proportional} flow cell over \emph{determinants} (dashed) that diverse tail discovery calls for --- a generative-flow network that constructs valid-$N$ determinants in the loop --- is the methodological whitespace this review identifies, unoccupied as of September 2026.}
\label{fig:taxonomy}
\end{figure*}

Table~\ref{tab:ml} and Fig.~\ref{fig:taxonomy} arrange the methods on the two axes. Two observations organize the rest of the section. First, the field is converging, from several directions, on a single algorithmic skeleton: \emph{propose configurations, diagonalize in their span, use the result to improve the proposer, iterate}. This is not a coincidence but an attractor: the variational guarantee (\S\ref{sec:2-1}) makes any proposal cheap to \emph{score} by exact diagonalization, and a cheap exact score is precisely what turns configuration selection into a learning problem, pulling classical and quantum approaches alike toward the same loop. The consequence reframes the field: the SQD loop is \emph{proposer-agnostic}, the quantum sampler one proposer among many (heat-bath screening, an RBM, an autoregressive network, a diffusion model, a GFlowNet), and the scientific question reduces to \emph{which proposer wins in which regime}. Read this way, the GFlowNet whitespace of \S\ref{sec:4-6} is not one more idea but the one systematic gap in the space of proposers the attractor admits. The attractor also pulls in proposers that do no learning at all, and the most recent of them is instructive about labelling as much as about method. Park, Kwon and Lee (31 August 2026) represent determinants as the nodes of a graph and the non-zero off-diagonal Hamiltonian elements as its edges, read the resulting block structure as a quadratic unconstrained binary optimization problem, and solve it by annealing over randomly drawn subsets of binary variables, accumulating the determinants the optimizer retains \cite{park2026fa}. The importance signal is the Hamiltonian coupling itself and the generated object an individual determinant, and because a QUBO solver is a reward \emph{maximizer} it belongs on the maximizer side with RL-CI, and --- its importance signal being the Hamiltonian coupling rather than a learned reward --- it falls outside the reward-proportional cell of Fig.~\ref{fig:taxonomy}, which it leaves empty. Its label deserves the discipline \S\ref{sec:6} asks of everyone: the work is \emph{quantum-assisted} only in that its QUBO is, in the authors' own words, ``directly compatible with quantum annealing hardware, while the numerical results presented in this work are obtained using simulated annealing'' --- run, they state, on a laptop. As of September 2026 it is therefore a classical combinatorial selector, and the baseline it owes a comparison to is SHCI, not SQD. Second, almost every method to date is a \emph{distribution-matching} or \emph{classification} model (it learns to reproduce, or to label as important, the determinants it has already seen). There is one true reward-proportional amortized sampler in the table, and we state it plainly because the gap claim depends on it: GPT-QE (\S\ref{sec:4-5}) is fitted by a logit-matching loss and therefore samples with probability proportional to $e^{-\beta E}$ --- but over \emph{circuits}, not determinants. Over \emph{determinants}, none, before the gap discussed in \S\ref{sec:4-6}, is a \emph{reward-proportional amortized sampler} explicitly built to place mass on diverse, high-importance determinants it has \emph{not} yet seen, which is precisely the tail-discovery problem the coupon-collector law makes hard.

\begin{table*}[t]\centering\footnotesize
\caption{Machine-learning and generative methods for determinant/subspace selection, organized by the object generated and the importance signal exploited. HW? indicates whether a quantum-hardware demonstration was reported.}
\label{tab:ml}
\renewcommand{\arraystretch}{1.25}
\begin{tabularx}{\textwidth}{@{}>{\raggedright\arraybackslash}p{2.0cm} >{\raggedright\arraybackslash}p{2.5cm} >{\raggedright\arraybackslash}X >{\raggedright\arraybackslash}p{1.55cm} >{\raggedright\arraybackslash}p{2.5cm}@{}}
\toprule
\textbf{Method} & \textbf{Generated object} & \textbf{Importance signal} & \textbf{HW?} & \textbf{Ref.} \\
\midrule
MLCI & determinant (add/skip) & on-the-fly NN prediction & classical & \cite{coe2018} \\
PIGen-SQD & determinants (recovery) & quantum-sample freq.\ + physics prior & IBM Heron & \cite{patra2025} \\
QSCI-RBM (DMET) & determinant distribution & learned Born distribution (RBM) & classical & \cite{patra2026} \\
HI-NQS & determinants (in loop) & PT score + distilled eigenvector & classical GPU & \cite{chang2026} \\
NQS-SC & selected-config energy & variational, on selected set & classical & \cite{solanki2026} \\
NNQS-SCI / QiankunNet & determinants & autoregressive Born + de-dup & classical HPC & \cite{shang2025,sun2026} \\
ARNN-SCI & determinants (in loop) & autoregressive NQS sampling & classical & \cite{thompson2026} \\
Q-WAVE & determinants (latent VAE) & variational energy on the expanded basis; SqDRIFT + CISD seed & IBM Heron r2/r3 & \cite{qwave2026} \\
cSQD & determinant label & binary classifier + active learning & classical (+SQD) & \cite{zeni2026} \\
HAAR-SCI & determinants & Hamiltonian coupling (gated transformer) & classical GPU & \cite{zhanghaar2025} \\
RCI (learning-to-rank) & determinant ranking & pairwise rank (transformer) & classical & \cite{nie2026} \\
GPT-QE (GQE) & \textbf{circuit} (not determinants) & circuit energy, logit matching ($\propto e^{-\beta E}$) & IBM 127\,q (H$_2$) & \cite{nakaji2024} \\
Generative-QSCI & \textbf{circuit} (not determinants) & QSCI subspace energy (RL policy, GRPO) & classical sim.\ & \cite{kemmoku2026} \\
RL-CI & determinant (add/skip) & RL policy, reward-\emph{maximizing} & classical & \cite{goings2021} \\
\textbf{GFlowNet (proposed)} & \textbf{determinant, valid-$N$ by construction} & \textbf{reward-\emph{proportional}: fused EN + sample freq.} & --- (whitespace) & \emph{this work, \S\ref{sec:4-6}} \\
\bottomrule
\end{tabularx}
\end{table*}

\subsection{Restricted Boltzmann machines for generative recovery}\label{sec:4-2}

The most direct instantiation of ``learn the distribution of important determinants and resample it'' uses a restricted Boltzmann machine (RBM). The Maitra group at IIT Bombay is the most active locus here. Their PIGen-SQD couples a physics-informed generative model to the SQD recovery step, using implicit low-rank tensor decompositions to steer the generator into the dominant sector, and, notably among the ML-for-SQD works, is validated on real IBM Heron hardware \cite{patra2025} (the transformer-QSCI line of \cite{zeng2026} is the other hardware-executed example, on Zuchongzhi 3.1). Its density-matrix-embedding successor, QSCI-RBM, learns the distribution of dominant determinants from quantum samples and generates high-probability configurations, reaching chemical accuracy on a SARS-CoV-2 main-protease--inhibitor complex while accessing only $\sim$4\% of the configuration subspace, against $\sim$20\% for standard SQD \cite{patra2026}. An independent lineage, building on generative-ML-in-configuration-space \cite{herzog2022}, uses an interpretable RBM to solve the CI problem directly, recovering 99.99\% of the correlation energy with orders of magnitude fewer determinants \cite{hernandezmartinez2025}. The strength of the RBM route is that it is a genuine generative model of the determinant distribution; its limitations, relevant to \S\ref{sec:4-6}, are that Gibbs sampling is iterative and mode-seeking, and that particle-number and spin symmetry are enforced by post-hoc filtering rather than by construction. (We note, for completeness and because the review record should be accurate, that an early QSCI-RBM preprint was withdrawn by arXiv administrators for a licensing-rights reason (not a scientific retraction) and its content re-appeared, expanded, in the density-matrix-embedding paper above.)

\subsection{Autoregressive and transformer neural quantum states in the loop}\label{sec:4-3}

A second, and currently the most vigorous, family replaces the quantum sampler with an autoregressive or transformer \emph{neural quantum state} (NQS) \cite{carleo2017} that generates occupation-number bitstrings by construction with fixed particle number and exact ancestral sampling. The striking development of 2025--2026 is that this classical community has converged, independently, onto the SQD paradigm. HI-NQS embeds a dual-channel transformer (with explicit spin-up/spin-down cross-attention encoding fermionic structure) inside a sample--diagonalize--distil loop that its authors explicitly call sample-based quantum diagonalization, driven by a neural network rather than a circuit, and reports roughly half the determinant count of CIPSI on an N$_2$ active-space series \cite{chang2026}; a parallel autoregressive-NQS selected-CI, ARNN-SCI, uses the network to guide subspace expansion directly \cite{thompson2026}. Solanki, Ding, and Reiher's NQS-SC evaluates the energy directly on a selected-configuration set and argues, on grounds of systematic improvability, that it should replace variational NQS as the default for electronic structure \cite{solanki2026}. At the high-performance-computing end, the QiankunNet line scales a transformer NQS-driven selected-CI engine to Hilbert spaces beyond $10^{14}$, with GPU-accelerated de-duplication and coupled-configuration kernels \cite{shang2025,sun2026}; a transformer-refined QSCI in this line has now been executed on the Zuchongzhi 3.1 superconducting processor, reaching chemical accuracy on the $40$-qubit [2Fe-2S] ferredoxin cluster, and extending to a nitrogenase P-cluster model in a $(114e,73o)$ active space where it lands some $18$~mHa from the DMRG extrapolation --- an order of magnitude short of chemical accuracy \cite{zeng2026}. These methods are, in effect, a classical mirror of SQD: the same loop, the same subspace diagonalization, but with a trained network as the proposer. Their relevance to a review of \emph{quantum}-sampled diagonalization is twofold: they are the most natural classical baseline against which any quantum-sampling advantage must be measured (\S\ref{sec:5}), and their architectural choices (symmetry-by-construction, distillation of the diagonalized eigenvector back into the proposer) are the design vocabulary any quantum-integrated learned proposer will draw on.

\subsection{Neural-network classifiers and learning-to-rank selectors}\label{sec:4-4}

A third family declines to model the full distribution and instead learns only to \emph{rank} or \emph{classify} determinants by importance --- a lighter, and in practice highly effective, target. Zeni and co-workers recast determinant selection as binary classification inside an active-learning loop, with a classical variant (cHCI) cutting per-iteration memory roughly fivefold and a quantum--classical variant (cSQD) converging in markedly fewer SQD iterations \cite{zeni2026}. HAAR-SCI samples determinants autoregressively with a gated transformer and retains only those with the largest Hamiltonian couplings via GPU min-heap kernels, reaching a mean absolute error of 0.51 mHa across eighteen molecules and \emph{matching} heat-bath CI accuracy on an iron--sulfur cluster with a $72\%$ determinant reduction --- a compactness result against HCI, not an accuracy win over its semistochastic form --- and retaining under 0.01\% of the Hilbert space after the workflow's final compression step \cite{zhanghaar2025}. A learning-to-rank formulation with a transformer reaches chemical accuracy on iron--sulfur clusters using 12\% of the CI space and reports higher \emph{efficiency} than heat-bath CI --- efficiency, not accuracy, and against HCI rather than its semistochastic form --- on the notoriously multireference Cr$_2$ \cite{nie2026}; and a natural-orbital-based neural-network CI selector chooses determinants using approximate natural orbitals drawn from intermediate many-body eigenstates rather than Hartree--Fock orbitals \cite{thirion2025}, building on the same group's earlier neural-network selective CI, which recovered the N$_2$ correlation energy from ${\sim}4\times10^{5}$ rather than ${\sim}10^{10}$ determinants \cite{schmerwitz2024}. These results matter for the honest assessment of \S\ref{sec:5} in a specific way: several of them match or outperform \emph{classical} selected-CI --- but they do so with a \emph{classical} learned model, using no quantum hardware, which sharpens rather than resolves the question of what the quantum sampler adds.

\subsection{Generative circuit design}\label{sec:4-5}

A fourth family is generative in a different sense: rather than generate determinants, it generates the \emph{circuit} that produces them. The generative quantum eigensolver (GQE) trains a GPT-style transformer over circuit tokens to design ansätze with desired properties, sidestepping the barren-plateau pathology of gradient-based variational optimization \cite{nakaji2024}. Its training principle matters for the gap claimed in \S\ref{sec:4-6} and is worth stating exactly: GPT-QE is fitted by a \emph{logit-matching} loss that aligns the model's cumulative logits with the evaluated circuit energies, so it samples circuits with probability proportional to $e^{-\beta E}$. It is, in other words, itself a reward-proportional amortized sampler. Applied to QSCI, this becomes generative circuit design in which the transformer is trained on the QSCI subspace energy to produce the state-preparation circuit whose samples define the subspace; on N$_2$ up to 32 qubits it reaches chemical accuracy with 98\% fewer two-qubit gates than first-order Trotterization and yields subspaces roughly half the size of heat-bath CI in strongly correlated regimes \cite{kemmoku2026}. This is the closest existing work to ``a generative model that produces the SQD sampler,'' and its QSCI adaptation departs from GPT-QE's Boltzmann training in a way that should not be elided: there the transformer is trained as a reinforcement-learning policy by group-relative policy optimization (GRPO) on the subspace energy as reward, so it is a reward \emph{maximizer}. Neither the parent nor the adaptation generates determinants, which is the distinction \S\ref{sec:4-6} turns on.

\subsection{Latent-variable generative models: variational autoencoders}\label{sec:4-5b}

A fifth family, and the newest, is generative in the \emph{latent-variable} sense: instead of modelling the determinant distribution autoregressively or as a Boltzmann machine, it learns a continuous latent embedding of the wavefunction's support and decodes new determinants from it. Q-WAVE (quantum wavefunction augmentation via variational autoencoders) seeds a $\beta$-annealed variational autoencoder with determinants sampled from SqDRIFT Krylov circuits \cite{piccinelli2025} together with a classical CISD set, then iteratively expands that basis toward the variational ground state \cite{qwave2026}. Its distinctive property, and the reason it bears on the tail-discovery argument of \S\ref{sec:4-6}, is that the decoder generates dominant determinants \emph{beyond any fixed excitation hierarchy}: because the latent space is continuous, the proposals are not confined to the singles-and-doubles manifold the seed came from --- precisely the constraint that makes CISD-anchored expansions blind to the high-order tail. The authors report sub-millihartree agreement with FCI for H$_2$O and N$_2$ dissociation and, at scale, a 52-qubit ethylene system at sub-millihartree accuracy against CCSD(T) and a 60-qubit Cr$_2$ stress test reaching chemical accuracy after a final perturbative correction. Two cautions belong in the record. The method is a hybrid by construction: its seed is part hardware-sampled, part classical CISD, so credit assignment between the two halves is untested, and the standard of \S\ref{sec:6}(ii) would require the CISD-only seed as an explicit baseline. And, like the RBM route of \S\ref{sec:4-2}, the autoencoder is trained to reproduce the support it has been shown; it is a distribution-matching model with a continuous latent bottleneck, not a reward-proportional sampler, so it broadens the generative family without occupying the whitespace of \S\ref{sec:4-6}.

\subsection{The generative-flow-network gap, and the transferable toolbox}\label{sec:4-6}

Surveying the five families above against the machine-learning literature exposes a conspicuous and, we argue, consequential gap. Generative flow networks (GFlowNets) \cite{bengio2021,malkin2022} are amortized samplers purpose-built to generate compositional discrete objects with terminal probability proportional to a reward, favouring \emph{diverse} high-reward states rather than collapsing onto the mode --- an advantage that comes from their amenability to \emph{off-policy} training rather than from the choice of divergence \cite{malkin2023}. They have been applied across molecular graphs, crystals, biological sequences, combinatorial optimization, and Bayesian structure learning, and in the past year to adjacent quantum tasks --- though here the generative-sampler families must be kept distinct: a conditional \emph{normalizing} flow (a different generative paradigm, not a GFlowNet) warm-starts the variational quantum eigensolver in Flow-VQE \cite{zou2026}, whereas GFlowNets \emph{proper} have been applied to Pauli-measurement grouping \cite{gfnmeas2024,flowmeas2025} and, in FlowQ-Net, to automated quantum \emph{circuit} design, including variational ans\"atze for molecular ground-state estimation \cite{flowqnet2025}. FlowQ-Net is the closest counterexample to the gap we claim, and we therefore state the claim against it rather than around it: FlowQ-Net samples \emph{gate sequences} in proportion to a circuit-level reward that encodes performance, depth and gate count, which places it in the circuit-design row of Fig.~\ref{fig:taxonomy} beside GQE --- the object it generates is not a configuration and its reward is not a determinant importance. With that case admitted, an exhaustive search across every relevant term (GFlowNet with configuration interaction, determinant, selected-CI, Slater determinant, SQD, QSCI, Fock space) still returns, as of September 2026, \emph{no} application of a GFlowNet to determinant or subspace selection for configuration interaction. The archetypal reward-driven CI selector is the reinforcement-learning configuration interaction of \cite{goings2021}, which trains a policy to \emph{maximize} the value of the determinants it adds; this occupies the reinforcement-learning, mode-seeking half of the reward-driven column and sharpens rather than fills the gap, because tail discovery calls for a reward-\emph{proportional}, diversity-preserving sampler, not a maximizer. The nearest generative work is the generative-circuit QSCI of \S\ref{sec:4-5}, whose own training is a GRPO reinforcement-learning policy; its parent GPT-QE \emph{is} reward-proportional, but samples circuits rather than determinants. The whitespace is therefore over the generated object, not over the training principle. The gap is not an accident of terminology but a genuine methodological opening: a determinant is a fixed-particle-number occupation bitstring built one orbital at a time (precisely the compositional discrete object a GFlowNet constructs), and the correlation energy lives in the \emph{many small-weight determinants of the tail}, the regime where a reward-proportional sampler that resists mode collapse is exactly the right tool and a reward-maximizing search is exactly the wrong one. We are careful about the scope of the resulting novelty claim. That a \emph{generative} model helps SQD is not novel (\S\ref{sec:4-2}--\S\ref{sec:4-5b}); that noise can help recovery is not novel (\S\ref{sec:3-2}). What is unclaimed is a specific combination that must be stated precisely to survive the closest relative, the autoregressive NQS of \S\ref{sec:4-3}, which \emph{also} builds valid determinants one orbital at a time but samples \emph{on-policy}, proportional to its own $|\psi|^2$. The open element is an \emph{off-policy}, mode-covering sampler trained by trajectory balance on a \emph{cheap, FCI-free} reward (the Epstein--Nesbet prior fused with the recovered sample statistics --- not on $|\psi|^2$ itself), with the fixed-particle-number sector enforced by construction and a \emph{learnable reward temperature} that a mode-seeking $|\psi|^2$ sampler lacks --- together with the controlled, exact-ground-truth study (the standard of \S\ref{sec:6}, and the kind of experiment begun in \S\ref{sec:7}) that would establish \emph{when} such a proposer helps. The adjacent literature already supplies the toolbox for it: energy-based GFlowNets as amortized samplers of unnormalized discrete distributions \cite{zhangd2022}, ``inexpensive-reward'' pretraining that mirrors the cheap-perturbative-estimate-then-expensive-diagonalization structure of the problem \cite{pandey2024}, temperature-conditional reward tempering that is a learnable analogue of CIPSI's threshold schedule \cite{kim2024}, and discrete-diffusion set-selection models for which a determinant subspace is literally a binary selection mask \cite{sun2023}. Whether any of these transfers \emph{usefully}, and whether it survives the classical baselines of \S\ref{sec:5}, is, we argue, the central open experimental question of the subfield.

\begin{figure}[tbp]
\centering
\begin{tikzpicture}
\begin{axis}[paperplot, width=0.82\linewidth, height=0.47\linewidth, ymode=log,
  xlabel={subspace dimension (number of strings)},
  ylabel={energy error vs exact FCI (mHa)},
  xmin=15, xmax=255, ymin=1.3, ymax=2600, xtick={50,100,150,200,250},
  title={A cheap, FCI-free GFlowNet beats blind sampling but not the classical selector\\(N$_2$; 5-seed mean $\pm$ s.d., greedy top-$K$ deterministic)},
  legend style={at={(0.5,-0.20)}, anchor=north, legend columns=2, /tikz/column 2/.style={column sep=10pt}},
  clip=false]
  \addplot[x filter/.expression={x-4.2}, oiGreen, densely dashed, line width=1.2pt, mark=star, mark size=3.5pt, mark options={solid,fill=oiGreen},
    error bars/.cd, y dir=both, y explicit, error bar style={oiGreen}]
    coordinates {(30,61.1)+-(0,8.5) (60,33.0)+-(0,2.3) (120,16.1)+-(0,1.6)};
  \addlegendentry{$|c|^2$-proportional sampling}
  \addplot[x filter/.expression={x-1.4}, oiGrey, line width=1.4pt, mark=*, mark size=2.6pt, mark options={fill=oiGrey},
    error bars/.cd, y dir=both, y explicit, error bar style={oiGrey}]
    coordinates {(30,1317.4)+-(0,779.2) (60,1003.0)+-(0,783.1) (120,492.4)+-(0,158.7) (240,321.9)+-(0,85.6)};
  \addlegendentry{uniform (blind)}
  \addplot[x filter/.expression={x+1.4}, oiVerm, line width=1.4pt, mark=diamond*, mark size=3.2pt, mark options={fill=oiVerm}]
    coordinates {(30,65.5)(60,54.6)(120,46.4)(240,45.4)};
  \addlegendentry{greedy top-$K$ (static cheap reward)}
  \addplot[x filter/.expression={x+4.2}, oiPurple, line width=1.4pt, mark=triangle*, mark size=3.4pt, mark options={fill=oiPurple},
    error bars/.cd, y dir=both, y explicit, error bar style={oiPurple}]
    coordinates {(30,306.4)+-(0,106.7) (60,243.3)+-(0,129.9) (120,110.8)+-(0,11.0) (240,83.1)+-(0,13.8)};
  \addlegendentry{GFlowNet on cheap reward (ours)}
  \draw[oiInk, densely dotted, line width=0.9pt] (axis cs:15,1.6) -- (axis cs:255,1.6);
  \node[oiInk, font=\scriptsize, anchor=south east] at (axis cs:253,1.72){chemical accuracy};
\end{axis}
\end{tikzpicture}
\caption{\textbf{Compactness under a cheap, FCI-free reward} (N$_2$, exact FCI; 5-seed mean~$\pm$~s.d.). Energy
error vs subspace dimension for blind uniform sampling, deterministic top-$K$ selection by the static cheap Epstein--Nesbet
reward, the GFlowNet trained on the same (tempered, $\beta{=}0.5$) cheap reward, and the exact $\propto|c|^2$ oracle.
Averaged over five seeds, the ordering is unambiguous at $D\ge120$, and it is the ordering the paper's verdict
predicts; two pairs are not resolved at the two smallest dimensions --- at $D{=}30$ the oracle's one-s.d.\ interval, $[52.6,69.7]$~mHa, still contains the deterministic greedy
point ($65.5$~mHa), and at $D{=}60$ the blind-sampling and generative intervals overlap, as the error bars show.
Markers are staggered horizontally for legibility; every series is evaluated at exactly $D{=}30$,
$60$, $120$ and $240$. (The $|c|^2$-proportional series stops at $D{=}120$, and that is a property of the \emph{method},
not a limit of the calculation: $90\%$ of the exact weight sits on twelve strings, so $5\times10^{4}$ i.i.d.\ draws
from it do not yield $240$ distinct ones. Sampling a concentrated distribution cannot fill a large subspace at
any budget --- which is the strongest thing the generative arm has going for it, since a flow network trained to
a tempered reward can.) In detail: the \emph{deterministic classical} greedy selector ($46.4$~mHa at $D{=}120$)
beats the GFlowNet ($111\pm11$~mHa), which in turn beats only blind uniform sampling ($492\pm159$~mHa); the exact
oracle ($16.1\pm1.6$~mHa) is out of reach of all three. The greedy curve flattens beyond $D{\approx}91$ --- the
number of strings the cheap reward scores above zero --- because past that point it is appending strings it cannot
rank. The GFlowNet's diversity buys a large margin over blind sampling but none over the cheap classical selector
it shares a reward with, consistent with the Spearman rank correlation of only ${\approx}0.41$ between that cheap
reward and the true weights (\S\ref{sec:7}).  The decisive comparison against the stronger \emph{iterative} EN-selected
CI is Fig.~\ref{fig:hci}.}
\label{fig:compact}
\end{figure}
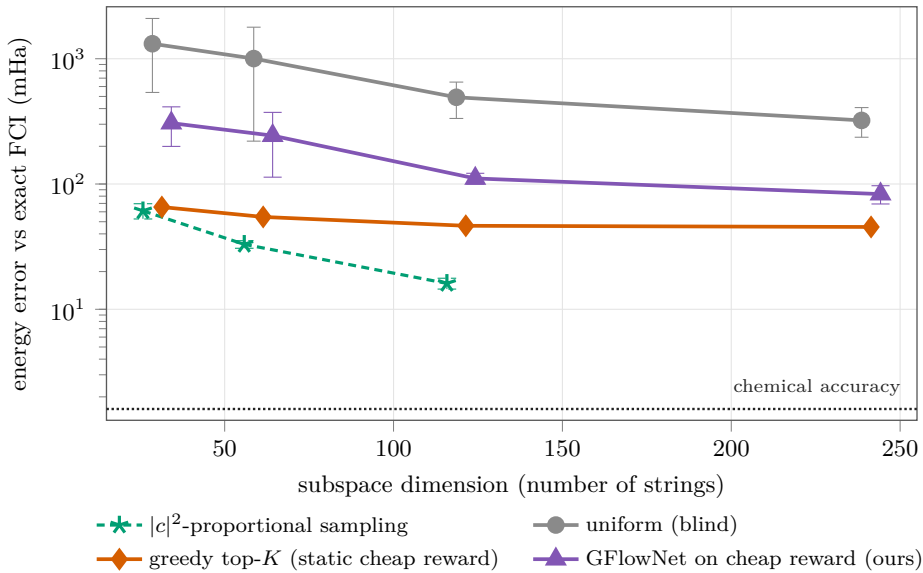

Concretely, such a proposer needs a reward that is cheap and FCI-free. The natural choice is the Epstein--Nesbet first-order coefficient of each determinant relative to the Hartree--Fock reference,

\begin{equation}
c^{(1)}_i \;=\; \frac{\langle D_i|\hat H|D_{\mathrm{HF}}\rangle}{E_{\mathrm{HF}} - \langle D_i|\hat H|D_i\rangle}, \qquad R(i) \;\propto\; \big(c^{(1)}_i\big)^2,
\end{equation}

computable with a single $\hat H|D_{\mathrm{HF}}\rangle$ matrix--vector product and the Hamiltonian diagonal. This is the Epstein--Nesbet functional form that CIPSI and heat-bath CI use, but frozen at the Hartree--Fock reference; those methods re-score every candidate against the \emph{current} correlated wavefunction at each iteration, and that asymmetry --- a static single-reference prior against an adaptively re-scored one --- is what the comparisons below turn on. By the Slater--Condon rules $\langle D_i|\hat H|D_{\mathrm{HF}}\rangle$ is nonzero only for determinants differing from the reference by at most two spin-orbitals, so this single-reference prior is structurally blind to the triple and higher excitations that carry weight in a strongly multireference wavefunction --- the physical origin of the imperfect ($\rho{\approx}0.41$, \S\ref{sec:7}) rank correlation, and a ceiling that no amount of proportional sampling of the prior can lift. A GFlowNet with forward policy $P_F(a_t\,|\,s_{t-1};\theta)$ builds a determinant as a trajectory $\tau: s_0\to s_1\to\cdots\to x$ that places electrons one orbital at a time --- masked so that $x$ is always a valid $n_\alpha$-electron string --- and is trained by the trajectory-balance objective

\begin{equation}
\mathcal L_{\mathrm{TB}}(\tau) \;=\; \Big(\log Z_\theta + \sum_{t}\log P_F(a_t\,|\,s_{t-1};\theta) - \log R(x)\Big)^{2},
\end{equation}

whose global minimizer samples terminal states in exact proportion to reward, $P^{\top}_\theta(x)\propto R(x)$, the diverse, tail-covering behaviour the coupon-collector regime demands. Fig.~\ref{fig:compact} evaluates exactly this construction on the FCI-verifiable N$_2$ system, over five seeds. Given only the cheap, FCI-free reward, the GFlowNet's diversity lets it beat blind uniform sampling by a wide margin; but, averaged over seeds, it does \emph{not} track the exact oracle, and it does not beat the \emph{deterministic} greedy top-$K$ selection by the \emph{same} cheap reward ($111\pm11$ vs $46.4$~mHa at $D{=}120$), an ordering the floor sweep of \S\ref{sec:4-6} holds at every exploration setting that can fill the subspace.

\textbf{That last comparison is not like-for-like, and we say so before a reader does.} A sampler must \emph{collect} $D$ distinct strings; a ranker merely \emph{takes} them, and \S\ref{sec:3-1} is this paper's own statement that the first is strictly the harder task. Charging the proposer for being a sampler at all confuses two effects, so we ran the control that separates them: i.i.d.\ sampling with probability exactly proportional to the same cheap reward, and to the exact weights, at the same draw budget and in the same gauge, five seeds (\texttt{figures/control\_iid.py} $\to$ \texttt{results/control\_iid.json}). Two things follow, and the first was not visible before.

\emph{Sampling a concentrated reward cannot fill the subspace at all.} I.i.d.\ draws proportional to the cheap reward reach $71.8\pm3.2$~mHa at $D{=}30$ and then stop: at $5\times10^{4}$ draws they never accumulate $60$ distinct strings, let alone $240$. The GFlowNet trained on that same reward reaches $D{=}240$. Whatever else the flow network fails to do, it converts a signal that proportional sampling cannot exploit at any practical budget into one that fills a subspace --- and that, not a compactness win, is the concrete thing it buys.

\emph{And the sampler--ranker penalty accounts for most of the remaining gap.} With a \emph{perfect} reward, sampling costs a factor $1.62$ against ranking the same reward ($15.7\pm1.5$ against $9.7$~mHa at $D{=}120$). The GFlowNet's shortfall against greedy top-$K$ at its best floor is a factor $1.25$ ($58.2\pm6.6$ against $46.4$) --- \emph{smaller} than the penalty an oracle-quality sampler pays. Read on the like-for-like axis, therefore, the proposer is not obviously inheriting the ceiling of its reward; what Fig.~\ref{fig:compact} mostly measures is the cost of sampling rather than selecting. The honest statement is the narrower one: \emph{a learned proposer trained on this cheap reward does not turn sampling into selection}, and against a classical method that never samples at all it loses --- which is the comparison \S\ref{sec:5} makes and the one the paper's verdict rests on. That ceiling is close to absolute for this proposer, and we say so on the strength of a sweep rather than a single setting: across the three reward floors of \S\ref{sec:4-6} the proposer overtakes the deterministic ranking in exactly one cell --- $D{=}60$ at a floor of $10^{-12}$, $51.6$ against $54.6$~mHa, unresolved at five seeds --- and that is a floor at which $D{=}120$ cannot be filled at all. (An earlier single-seed run of this figure suggested a near-oracle GFlowNet; the multi-seed standard of \S\ref{sec:6}(v), applied to our own figure, corrected it.) The lesson is not that the GFlowNet is worthless but that a learned proposer trained on a cheap classical reward tends to inherit that reward's ceiling; the case for it therefore rests on the two regimes where that ceiling might lift (the fused reward under noise --- which \S\ref{sec:7} finds gives a generic constraint-satisfaction gain rather than a reward-quality one --- and genuinely multireference systems, which \S\ref{sec:7} leaves untested), not on this noise-free, cheap-reward illustration, and certainly not against the strong \emph{iterative} EN-selected CI of Fig.~\ref{fig:hci} in \S\ref{sec:5}, where the classical selector wins at every FCI-verifiable scale.

\textbf{One hyperparameter decides Fig.~\ref{fig:compact}, and it is worth naming.} The GFlowNet there is trained on the cheap reward \emph{floored} at $10^{-3}$ of its maximum. That floor is not an implementation detail. With $701$ of the $792$ strings carrying identically zero reward --- the $546$ triples-and-higher of the Slater--Condon argument, plus $155$ doubles that spatial symmetry annihilates in this gauge --- and $774$ of $792$ ($97.7\%$) lying at or below the floor itself, the floor alone fixes how much probability mass the sampler places on strings the cheap reward cannot rank. We therefore ran the floor, rather than choosing it: three settings ($10^{-3}$, $10^{-6}$, $10^{-12}$ of the maximum), five seeds each, everything else held fixed. At $D{=}120$ the GFlowNet returns $110.8\pm11.0$, $58.2\pm6.6$, and --- at $10^{-12}$ --- no value at all, against a deterministic greedy selector at $46.4$~mHa. Lowering the floor does improve the proposer monotonically, but it never overtakes the static ranking \emph{at this dimension}, and below some point it stops being able to \emph{fill} the subspace: only $91$ of the $792$ strings carry non-zero cheap reward, and at a floor of $10^{-12}$ the tempered reward leaves the other $701$ six orders of magnitude beneath them, so in the $5\times10^{4}$ draws each arm is allowed the sampler never accumulates $120$ distinct strings. We put that as a budget and not an impossibility, because the floor is applied as $\max(w,\text{floor})$ and the floored reward therefore has full support: $D{=}120$ is reachable in principle, and not at any budget under which this would still be the same comparison. The same count explains why the greedy curve flattens beyond $D{\approx}91$ ($46.4$ at $D{=}120$, $45.4$ at $D{=}240$): past that point it is appending strings the reward scores at exactly zero. The floor is thus not a tuning knob with a lucky setting but the exploration--exploitation trade-off of this proposer made explicit, and the comparison of Fig.~\ref{fig:compact} is robust across it. What Fig.~\ref{fig:compact} tests is not whether generative models help, but whether \emph{this} proposer, at \emph{this} exploration setting, beats a static ranking of the same reward --- which is precisely why the gap of \S\ref{sec:4-6} is a question about sampler design rather than about generative modelling in general.

\subsection{A note on real-space neural wavefunctions}\label{sec:4-7}

For completeness and to forestall a common conflation, we distinguish the methods above from the real-space neural-wavefunction programme (FermiNet, PauliNet, Psiformer, and the emerging foundation models Neural Pfaffians and Orbformer) \cite{pfau2020,hermann2020,vonglehn2023,gao2024,foster2025}. These are variational Monte Carlo ans\"atze in continuous three-dimensional space, not determinant selectors in a fixed orbital basis; they are among the strongest classical methods for small strongly correlated molecules and thus appear in \S\ref{sec:5} as competitors. The discrete-orbital neural-network backflow of Liu and Clark reaches state-of-the-art molecular energies among them \cite{liuclark2024}; its optimized variant now matches heat-bath CI, ASCI, FCIQMC and DMRG and surpasses coupled cluster, making it the strongest single learned proposer to date \cite{liuclark2025} (a related neural ansatz reaches high accuracy with only a handful of Slater determinants \cite{fewdet2026}), and the programme is surveyed by \cite{hermann2023}. But these methods do not propose configurations for a diagonalization and are outside the loop this review concerns. The one bridge worth flagging is the discrete-basis autoregressive NQS (NAQS) \cite{barrett2022}, which \emph{does} sample determinants proportional to $|\psi|^2$ and is therefore the closest classical relative, and the most natural benchmark, for any learned determinant proposer, GFlowNet-based or otherwise.

\section{The critical assessment: does the quantum sampler, or its ML augmentation, beat classical selected CI?}\label{sec:5}

Every method in \S\ref{sec:4} is ultimately in service of an SQD subspace, and every SQD subspace inherits a single, unavoidable question. Because the final step is an exact diagonalization in a determinant subspace, SQD \emph{is} a selected configuration interaction whose determinants happen to be chosen by a quantum sampler. Its natural competitors are therefore the classical selected-CI heuristics --- and those are neither weak nor slow. A review that omitted this comparison would be advocacy, not assessment; we make it the centre of the paper.

\subsection{The classical baselines and the accuracy bar}\label{sec:5-1}

The relevant classical methods fall into two groups. The direct competitors are the classical determinant selectors: CIPSI, with its Epstein--Nesbet perturbative selection and extrapolation to the FCI limit \cite{huron1973,garniron2019,loos2024}; heat-bath CI (HCI) and its semistochastic form (SHCI), whose $|H_{ai}c_i|$ screening makes selection near-linear and which is generally the fastest route to sub-milli-hartree accuracy \cite{holmes2016,sharma2017}; and adaptive sampling CI (ASCI) \cite{tubman2016}. The best-in-class solvers for strong correlation are DMRG \cite{white1992,chan2002}, phaseless AFQMC \cite{zhangs2003}, full configuration-interaction quantum Monte Carlo \cite{booth2009}, and, for weak-to-moderate correlation, coupled cluster. Each of these baselines has a canonical, openly available implementation against which any claim can be reproduced (Dice for semistochastic heat-bath CI, Quantum Package for CIPSI, \texttt{block2} for DMRG, \texttt{ipie} for AFQMC, and NECI for FCIQMC, all interoperable through PySCF), and we take it as part of the standard below that a benchmark name the baseline code it was run against. The accuracy these methods \emph{bracket} sets the bar any quantum result must clear to matter, and that bar is now quantitative: in the blind benzene ground-state challenge (a (30e,108o) space, a multi-method community effort), DMRG, AFQMC, FCIQMC, SHCI and their relatives \emph{qualitatively} agree, clustering near $-863$ mHa of correlation energy, but the root-mean-square deviation among them is, in the authors' own word, ``considerable'' at 1.3 mHa \cite{eriksen2020} --- that is, the best classical solvers disagree with one another by very nearly the whole of chemical accuracy (1.6 mHa). The Simons hydrogen-chain benchmark reports the same structure of near-agreement-with-residual-spread in the thermodynamic limit \cite{motta2017}. The same collaboration's transition-metal capstone (twenty first-principles methods on 3$d$ atoms, ions, and monoxides, with experiment-free reference values \cite{williams2020}) fixes the classical bar in exactly the strongly-correlated, 3$d$/iron--sulfur arena that SQD most wants to claim, and a curated open hierarchy of chemically decisive multireference systems (N$_2$, FeS, [2Fe-2S], U$_2$) has since been assembled to standardize such comparisons \cite{sundar2026}. The operative standard, then, is not ``reaches chemical accuracy'' but ``lands inside the ${\sim}1.3$ mHa spread of the best classical solvers at \emph{lower cost} than they do, or in a regime where they all degrade.''

\textbf{Where our own experiments sit against that bar, stated here rather than buried in a limitation.} Every calculation we perform ourselves retains $D{=}120$ $\alpha$-strings, and on stretched N$_2$ that dimension is outside the bar for \emph{every} arm we have, including the exact-weight one: ranking the $120$ strings of largest exact FCI weight still leaves $9.71$~mHa, six times chemical accuracy and seven times the classical spread. A reader is entitled to ask what an ordering read off at that dimension is worth, and we give three measurements rather than an argument. \emph{First}, the bar is reachable in this paper: H$_2$O at the same $D{=}120$ puts both the classical selector and the exact-weight ranking inside it, at $0.56$ and $0.58$~mHa. \emph{Second}, we can now say \emph{where} N$_2$ reaches the bar rather than bound it: sweeping the grid out to $D{=}400$, both deterministic arms enter the $1.3$~mHa band at $D{=}300$ --- the iterative selector passing from $1.481$~mHa at $D{=}280$ to $1.168$ at $300$, the exact-weight ranking from $1.516$ to $1.170$ over the same step (\texttt{results/rejilla\_D.json}). That is $38\%$ of the $792$ $\alpha$-strings and $14\%$ of the $627\,264$ determinants, and we give it as measured rather than framed: a subspace that large is not what anyone means by a compact one. The bar is reachable on this system by classical selection alone, and we have now put a dimension on it. \emph{Third, and the reason we raise this at all:} the exact-weight ranking is not the ceiling, and which method wins depends on the system rather than on the dimension. Sweeping both deterministic arms over fourteen points in two molecules (\texttt{calculations/rejilla\_D.py} $\to$ \texttt{results/rejilla\_D.json}), the iterative selector beats the exact-weight top-$D$ ranking at \emph{all eight} N$_2$ dimensions --- $21.67$ against $23.68$ at $D{=}60$, $9.13$ against $9.71$ at $D{=}120$, $4.62$ against $4.82$ at $D{=}180$~mHa, and the margin holds at every point out to $D{=}400$ --- and at only two of six on H$_2$O, where the ranking takes three, one point is an exact tie, and no difference anywhere exceeds $0.14$~mHa: noise, not signal. The asymmetry is the physically meaningful part: an iterative selector scores candidates against the correlated wavefunction it has built so far, whereas a ranking of \emph{marginal} string weights cannot see that structure at all, and stretched N$_2$ has far more of it than equilibrium H$_2$O. Where correlation structure exists the selector exploits it; where it does not, the two arms tie. Orderings read at $D{=}120$ on N$_2$ are therefore illustrations of a mechanism rather than measurements of which method wins at the bar, and we use them only as the former. The verdict of \S\ref{sec:5-5} does not rest on them; it rests on the published comparisons of \S\ref{sec:5-2}. The spread is what makes the bar meaningful in both directions: a quantum result that misses it is not competitive, and a quantum result that hits it has matched the classical state of the art to within the state of the art's own internal disagreement --- which is the most any method can currently be asked to demonstrate, and which is why cost, not accuracy alone, is the discriminating axis. No SQD result has yet been measured against this bar and passed; indeed the largest systematic hardware benchmark to date (745 reactions on a superconducting processor) finds that SQD attains CCSD-level accuracy only after energy extrapolation, not directly \cite{raisuddin2025}.

\subsection{The critique literature}\label{sec:5-2}

Two peer-reviewed-track critiques frame the negative case. Reinholdt and co-workers, granting SQD an idealized noiseless sampler, show on N$_2$ and [2Fe-2S] (the very systems used to showcase the method) that its expansions are roughly an order of magnitude less compact than heat-bath CI (for [2Fe-2S], $50{,}551$ determinants against HCI's $4{,}260$ at a matched energy error of $0.1$~Ha --- a threshold worth stating, since it is some sixty times looser than chemical accuracy, and the order-of-magnitude gap is what survives it), and that the coupon-collector cost to reach micro-hartree precision extrapolates to $\sim10^{14}$ samples; their conclusion is that ``QSCI falls behind more efficient classical counterparts'' \cite{reinholdt2025}. Gaberle and Jattana sharpen the claim for strongly correlated lattice models: the number of configurations required for a fixed accuracy grows \emph{exponentially with system size even under optimal, decreasing-probability inclusion}: even a \emph{perfect} sampler fails, so the limitation is intrinsic wavefunction delocalization, not a sampling artefact \cite{gaberle2026}. This second result is the more fundamental, and to date it has no proposed remedy. A third line reaches this review's central negative independently, and from the benchmarking side. Van Rossum and co-workers show that SQD performance is strongly influenced by uncontrolled growth of the classical diagonalization subspace, with the consequence that when classical resources are not explicitly constrained, \emph{classical uniform random sampling can reproduce SQD benchmarks as noise increases the diversity of sampled configurations} --- an external replication, by a different group and a different route, of the recovery-does-the-work finding of \cite{vaquerosabater2026}. They draw the methodological conclusion directly: any fair benchmarking protocol for SQD must explicitly control the diagonalization size over \emph{unique} samples, a requirement we adopt into the standard of \S\ref{sec:6}. They then propose a non-orthogonal-CI measurement protocol that distributes measurements over Hamiltonian-optimized orbital bases and improves sample efficiency over Hartree--Fock-basis measurement, with the gain persisting at fixed classical cost \cite{vanrossum2026}.

\subsection{The 2026 dequantization}\label{sec:5-3}

The critique became concrete in 2026. Belagali and co-workers gave a polynomial-time classical algorithm for the energy of \emph{any} single-layer unitary cluster Jastrow circuit --- the exact ansatz SQD samples, and, in the authors' words, ``independent of locality constraints for quantum hardware,'' so the heavy-hex restriction that makes LUCJ hardware-native buys no protection, exploiting a structural feature of that layer rather than any Gaussian-state property. Their algorithm, in their words, ``is based on Heisenberg evolution of the fermionic Hamiltonian, as well as L\"owdin's formula for computing the energy'': the Hamiltonian is propagated through the circuit in the Heisenberg picture, and the energy of the Jastrow-dressed reference is then evaluated in closed form by L\"owdin's formula, which applies because the Jastrow factor is diagonal in the occupation-number basis. That diagonal structure is precisely what a second layer destroys --- the result ``does not hold for $L\ge2$ layers because orbital rotations between Jastrow operations perturb the diagonal structure that enables L\"owdin's formula'' --- which is exactly why the single-layer result does not extend to deep, multi-layer circuits. They reproduced the flagship 77-qubit [4Fe-4S] experiment on a laptop in under a minute, finding, through classical circuit optimization, an energy \emph{below} the hardware result \cite{belagali2026}. The authors are careful that this is weak (energy) rather than strong (bitstring) simulation, leaving a formal gap; but about the specific published experiments it is decisive. Independently, a classical iterative refinement seeded from random determinants (TrimCI) reached the [4Fe-4S] ground state with on the order of $10^6$-fold fewer determinants than the quantum run \cite{zhangotten2025}, and the noise-and-recovery study of \S\ref{sec:3-2} showed random determinants plus classical recovery reproducing the accuracy \cite{vaquerosabater2026}. Taken together, these results locate the accuracy not in the quantum sampler but in the classical recovery-and-diagonalization back end. The trend has since consolidated. A fermionic-linear-optical (extended-matchgate) simulation not only reproduces the LUCJ circuit classically but \emph{improves} on the noisy-hardware SQD energy \cite{hassman2025}; this is unsurprising given device noise, the substantive point being that no quantum resource was needed, not that a new accuracy record was set. And a classical polynomial-time exact-diagonalization result comes from an author group overlapping the one that reported the 49-qubit claim (W.\ Kirby and A.\ Mezzacapo appear on both): sparse guided eigenwalks establish that if a $\mathrm{poly}(n)$-sparse Hamiltonian has an $\mathcal{O}(1)$-sparse extremal eigenvector and a single element of its support is known, an exact eigenvector can be computed classically in $\mathrm{poly}(n)$ time \cite{chin2026}. That is a strict sub-regime of the sparse-ground-state setting to which the 49-qubit claim is scoped~\cite{kirby2026}---strict because it additionally assumes an $\mathcal{O}(1)$-sparse extremal eigenvector and a known element of its support. The eigenwalk paper reports no re-solution of those particular instances, so what it establishes is a tightening of the classical baseline inside the claim's own regime, not a refutation of it. In parallel, classical coupled-cluster and DMRG reached chemical accuracy on the FeMo-cofactor model long held up as the canonical quantum target \cite{zhai2026}, and a transformer neural-network backflow matched DMRG to within chemical accuracy on [2Fe-2S], the very iron--sulfur system SQD showcases, while predicting its magnetic exchange coupling constant closer to experiment than DMRG, CCSD(T), or prior neural-network methods \cite{ma2025}.

\subsection{The pro-quantum case, weighed on its own terms}\label{sec:5-4}

The defence is real but narrow. Hafid and co-workers prove that UCJ circuits can encode arbitrary instantaneous-quantum-polynomial (IQP) computations, so \emph{some} such circuits are classically hard to sample unless the polynomial hierarchy collapses \cite{hafid2025} --- a worst-case statement that does not describe the single-layer, chemically-structured LUCJ circuits of the flagship demonstrations, which \S\ref{sec:5-3} shows are classically tractable at the energy level; it does not dispose of the multi-layer case. Weaving and co-workers report, on a 42-qubit demonstration on an IQM superconducting device for stretched SiH$_4$ in a 6-31G basis, a configuration space more than two hundred times smaller than a conventional selected-CI criterion at comparable energies --- scoped, in their own words, to ``large separations, where static correlation dominates'' --- and, against best-in-class heat-bath CI, ``similar wavefunction compactness at convergence'' \cite{weaving2025}. They read this as ``early evidence for quantum utility in the chemical domain,'' while conceding the critique of \S\ref{sec:5-2} rather than answering it: ``While the criticisms of Reinholdt \emph{et al.} still stand, namely that the compactness of Heatbath Configuration Interaction (HCI) is challenging to improve upon, we make considerable progress towards this goal.'' We note that an earlier preprint version instead claimed evidence \emph{against} those criticisms; we quote the published text. That is the strongest form of the pro-quantum case at the subspace-selection stage, and it deserves to be weighed as what it is: a compactness result at convergence, on one molecule, rather than a cost comparison against a tuned classical selector --- and one whose advantage language (``the quantum computer can provide computational advantage in the selection of high quality configuration subspaces'') belongs to the paper's framing of the approach rather than to its measured results. The two are placed on the same axis only in \S\ref{sec:5-5}. The strongest experimental claim is Kirby and co-workers' 49-qubit sample-based Krylov result, reported to outperform classical sparse-ground-state solvers \cite{kirby2026} --- but on \emph{sparse-ground-state local Hamiltonians}, not general chemistry, a scope that must be stated precisely and whose classical rebuttals should be watched. The one genuinely open crack is depth: the dequantization proofs are for single-layer LUCJ, and whether deep, multi-layer LUCJ evades classical simulation is unresolved. Necaise and co-workers give the sharpest map of that terrain to date, characterizing three regimes of hardness for distribution complexity and classical simulability in a double-factorized representation and tracing the entanglement patterns to the interplay of coherent Gaussian orbital rotations and disordered Coulomb interactions \cite{necaise2026} --- precisely the two ingredients whose \emph{single-layer} combination \S\ref{sec:5-3} shows to be classically tractable, making their multi-fragment regime map the natural place to look for where tractability ends. Depth is the only place a future \emph{quantum-sampling-hardness} advantage could currently hide --- distinct from the elevated-noise robustness effect of \S\ref{sec:7}  and from the multireference opening that same section leaves untested.

\subsection{Verdict}\label{sec:5-5}

\begin{figure}[tbp]
\centering
\begin{tikzpicture}
\begin{axis}[paperplot, ybar, width=0.72\linewidth, height=0.47\linewidth,
  bar width=17pt, ymin=0, ymax=30, xmin=-0.6, xmax=1.6,
  xtick={0,1}, xticklabels={H$_2$O, N$_2$}, xtick style={draw=none},
  ylabel={energy error vs exact FCI (mHa)}, ytick={0,5,10,15,20,25,30},
  title={Learned proposer vs.\ classical selected CI ($D=120$)},
  nodes near coords, nodes near coords style={font=\footnotesize\bfseries, /pgf/number format/fixed, /pgf/number format/precision=1, anchor=south, yshift=1pt},
  legend style={at={(0.5,-0.14)}, anchor=north, legend columns=1},
  enlarge x limits=0.5, clip=false]
  \addplot[fill=oiVerm, draw=white, line width=0.6pt,
    nodes near coords style={/utils/exec={\ifnum\coordindex=0\relax\pgfkeysalso{yshift=11pt}\fi}}] coordinates {(0,0.56) (1,9.13)}; \addlegendentry{EN-selected CI, CIPSI-style (classical, noise-free)}
  \addplot[fill=oiPurple, draw=white, line width=0.6pt] coordinates {(0,3.22) (1,19.06)}; \addlegendentry{GFlowNet-SQD proposer (classical, simulated Heron noise)}
  \draw[oiInk, densely dotted, line width=1pt] ({rel axis cs:0,0}|-{axis cs:0,1.6}) -- ({rel axis cs:1,0}|-{axis cs:0,1.6});
  \node[oiInk, font=\scriptsize, anchor=south west] at (axis cs:-1.54,1.75){chemical accuracy};
\end{axis}
\end{tikzpicture}
\caption{\textbf{Classical selected CI versus a learned proposer at FCI-verifiable scale} ($D=120$; dotted line:
chemical accuracy, $1.6$~mHa). An iterative EN-selected CI (CIPSI-style) baseline, grown from the Hartree--Fock
reference by Epstein--Nesbet perturbative selection with each iteration re-scored against the current correlated
wavefunction (never seeded from the exact wavefunction), reaches $0.6$~mHa on H$_2$O ($r_{\mathrm{OH}}{=}1.24\,$\AA)
and $9.1$~mHa on stretched N$_2$, below the fused GFlowNet-SQD subspace built under simulated Heron noise ($3.2\pm1.3$
and $19.1\pm1.7$~mHa over five seeds; \texttt{calculations/gfn\_ruidoso\_fig7.py}). \emph{Both generative bars
are corrections.} Earlier versions printed $2.1$ and $26.7$~mHa, hand-entered values no deposited script produced
--- in a figure whose purpose is to make this comparison falsifiable. Measured, the N$_2$ bar is $4.5$ standard
deviations below what we had printed and the H$_2$O bar about one above, and the verdict survives both: the classical
selector wins by $2.7$~mHa on H$_2$O and by $9.9$~mHa on N$_2$, the latter at ${\approx}5.8\sigma$. This pits a
noise-free classical selector against a noise-affected learned proposer, so it illustrates the gap at realistic
conditions rather than isolating the proposer from the noise. Run at the same noise, the same five seeds and
the same $D$, the classical recovery loop fed by the cheap prior gives $1.79\pm0.18$~mHa on H$_2$O against the
proposer's $3.22\pm1.26$, and $21.51\pm1.57$~mHa on N$_2$ against the proposer's $19.06\pm1.70$ --- that is, \emph{on
N$_2$ the generative proposer wins the matched-noise comparison}, by $2.4$~mHa. \emph{This is a correction.} Earlier versions gave $1.5\sigma$ \emph{on the paired difference}; the five pairs were not deposited, and $1.5$ is not attainable as a paired statistic for these two arms at all. They are deposited now, and paired seed by seed the difference is $2.45\pm1.80$~mHa, $t_{4}{=}3.0$, with all five agreeing in sign --- which at $n{=}5$ floors the distribution-free significance at $p{=}0.0625$, so we read the $t$ as a reproducibility diagnostic and not a discovery significance, exactly as \S\ref{sec:7} does for the ladder (\texttt{calculations/gfn\_ruidoso\_fig7.py} $\to$ \texttt{results/gfn\_ruidoso\_fig7.json}). It wins against \emph{that}
baseline and not against the one this figure is about: selected CI, run noise-free, is at $9.13$~mHa on the same
system and dimension, less than half the proposer's error. The distinction is the whole point of \S\ref{sec:5}
--- a noisy sampler can beat a noisy classical recovery loop and still lose to a classical selector that never
needed the sampler --- The
zero-noise arm of Fig.~\ref{fig:crossover} adds a third view, where the classical control is ahead at three of
the four geometries and significantly so at two ($R{=}1.1$ and $R{=}2.5$); the present figure adds the realistic-conditions
gap that arm cannot show. A hardware run of ours on the N$_2$ active space of this figure \cite{bonilla2026spectral}
uses iterative self-consistent recovery at an unreported subspace dimension, not the fixed dimension scored here;
\S\ref{sec:7} says why the two energies are not commensurable and why neither is an advantage claim.}
\label{fig:hci}
\end{figure}
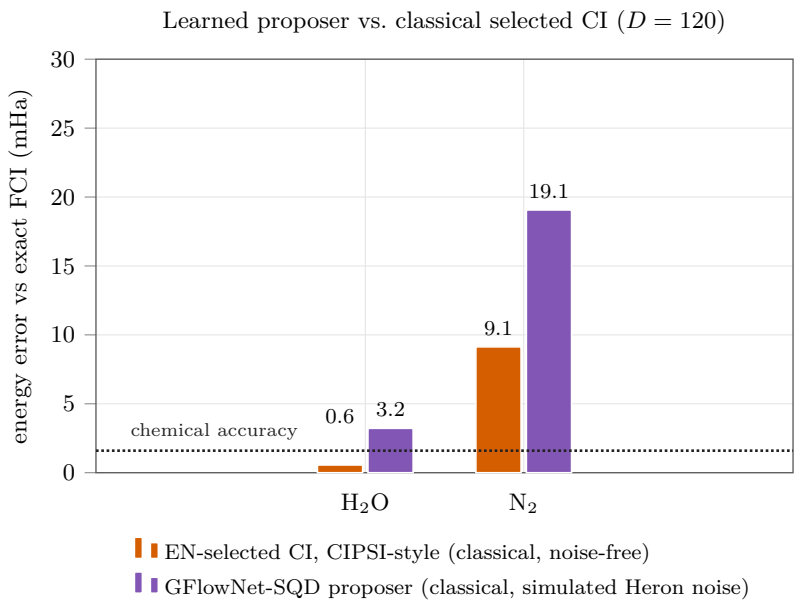

The current state of the field, as of September 2026, is that there is \emph{no reproducible, same-active-space demonstration that the SQD/QSCI quantum sampler beats SHCI, CIPSI, DMRG, or AFQMC (or even random determinants passed through classical recovery) on molecular electronic structure}, and that the flagship sampling circuits have been classically reproduced at the energy (weak-simulation) level. This is not a claim that SQD is useless: it is a genuinely useful, noise-tolerant, well-tooled workflow that has reached chemically relevant scale, and its configuration-recovery idea is a real contribution. It is a claim that the burden of proof for \emph{advantage} rests squarely on the quantum side, that this burden has not been met, and, the point of \S\ref{sec:6} and \S\ref{sec:7}, that meeting it, or ruling it out, requires a standard of benchmarking the field has not yet adopted. This verdict is not ours to claim: it was reached and published in 2025, on the same two systems, by Reinholdt and co-workers \cite{reinholdt2025}, and it rests on the independent critique and dequantization literature above (not on any single computation of ours) and now coincides with independent macro-assessments of quantum advantage in electronic structure \cite{lee2023,gundlach2025,legeza2026}. Fig.~\ref{fig:hci} merely illustrates it on two FCI-verifiable systems, where an iterative EN-selected CI (CIPSI-style) baseline --- grown from the Hartree--Fock reference by Epstein--Nesbet perturbative selection, each iteration re-scored against the current (increasingly correlated) wavefunction and never seeded from the exact solution --- reaches $0.6$~mHa on H$_2$O ($r_{\mathrm{OH}}{=}1.24\,$\AA) and $9.1$~mHa on stretched N$_2$, below both the noisy fused GFlowNet-SQD subspace and the noise-free cheap-reward GFlowNet subspace (Fig.~\ref{fig:compact}). \emph{We name our own baseline, since \S\ref{sec:5-1} asks everyone else to.} It is an in-house iterative Epstein--Nesbet selector (\texttt{calculations/hci\_baseline.py}), not Dice, Quantum Package, block2, ipie or NECI, and not the $|H_{ai}c_i|$ heat-bath screening that defines HCI proper --- the deposited file carried the name \texttt{hci} and we correct the label here rather than let it stand. A released semistochastic code would very likely beat ours, which strengthens the comparison rather than weakening it: the classical side already wins with the weaker implementation.

\section{A standard for honest benchmarking}\label{sec:6}

The pattern that recurs across \S\ref{sec:5} is not that quantum methods were shown to fail, but that they were often not measured against the baselines that would settle the question. We therefore propose, and argue the field should adopt, a benchmarking standard with ten elements; we state it as a checklist because its value is in being applied uniformly. Systematic hardware benchmarking of SQD has recently begun \cite{raisuddin2025}; our emphasis differs in insisting on exact-FCI ground truth and an explicit dequantization baseline, and the checklist is exactly the protocol any defensible study of a learned proposer --- ours or others' --- would follow. The Quantum Advantage Tracker is an open, publicly reviewed ledger of quantum-advantage candidates, each classified \emph{active}, \emph{baseline} or \emph{superseded} --- the last being instances where subsequent classical progress closed a gap a quantum computation once appeared to open \cite{qatracker}. Its variational track is this review's subject matter: as of September 2026 its five problems --- the N$_2$/cc-pVDZ dissociation series, [2Fe-2S] $(30e,20o)$, [4Fe-4S] $(54e,36o)$, an Anderson impurity model, and a guided sparse ground-state instance at $49$ qubits --- set SQD, SKQD and HI-VQE on Heron processors beside classical HCI, ASCI, CIPSI, DMRG and TrimCI \cite{zhangotten2025}. Two consequences follow, and they cut in opposite directions. The first is a verdict rather than infrastructure: that track lists no \emph{active} candidate at all --- all five are classified baseline benchmarks --- so the community's own open ledger records no standing quantum-advantage claim for ground-state energies, the conclusion of \S\ref{sec:5-5} arrived at independently and in public. The second is what is left for a checklist to add: the tracker adjudicates upper bounds on Hamiltonians chosen precisely because they lie beyond exact diagonalization, and requires neither exact-FCI ground truth (i), nor a dequantization baseline (iii), nor multi-seed error bars (v), nor a backend-calibrated asymmetric readout model (vi), nor orbital-gauge convergence (vii), nor pre-registration (x). The ten elements below are that tightening, and we would count the standard as having succeeded if it were adopted \emph{there} rather than alongside.

\textbf{(i) Exact ground truth at verifiable scale.} Claims about which determinants matter, and how compact a subspace is, are only meaningful where the exact answer is known. The controlled experiment should live in active spaces small enough for exact FCI (up to roughly a dozen orbitals) --- large enough for the coupon-collector effect to appear, small enough that every energy is an error against truth, not against another approximation. This is a deliberate inversion of the ``largest active space'' instinct: the point is not scale but falsifiability.

\textbf{(ii) The strong classical baseline, not a straw man.} The comparison must be against iterative heat-bath / semistochastic CI bootstrapped from the correlated wavefunction --- and, following \S\ref{sec:5-3}, against the two baselines the field has repeatedly found decisive: random determinants passed through the classical recovery loop, and the classical simulation of the sampling circuit itself. A method that beats uniform sampling but not SHCI has demonstrated nothing about quantum value.

\textbf{(iii) A dequantization baseline.} Given \cite{belagali2026}, any hardware claim should report whether the sampling circuit is classically simulable at the depth used, and if so, the energy that classical simulation attains. Five independent routes exist, to be reported against wherever they apply: the Heisenberg--L\"owdin route of \S\ref{sec:5-3} \cite{belagali2026}; matchgate simulation \cite{hassman2025}; sparse guided eigenwalks \cite{chin2026}; Pauli propagation \cite{shrikhande2025}; and direct matrix-product-state contraction of the circuit --- the last two being the depth-truncable routes that most directly probe the multi-layer-LUCJ crack of \S\ref{sec:5-4}.

\textbf{(iv) Full cost accounting in the currency that is scarce, with the diagonalization size explicitly controlled.} The scarce resource is quantum measurement (shots = QPU time), and the honest axis is energy error versus shots, or versus subspace dimension at matched classical diagonalization cost --- not energy error versus wall-clock on incomparable hardware. Reporting the shot budget honestly also surfaces the early saturation of \S\ref{sec:3-3}. To this we add a control that van Rossum and co-workers show to be decisive and that is routinely absent: the diagonalization size must be constrained over the \emph{unique} samples entering the subspace, not merely reported. Left free to grow, it lets noise inflate sample diversity until classical uniform random sampling can reproduce the SQD benchmark outright \cite{vanrossum2026} --- so an unconstrained subspace size makes the comparison uninterpretable in exactly the direction that flatters the quantum sampler.

\textbf{(v) Statistical error bars.} Sampling is stochastic and neural proposers are seed-dependent; a single run is an anecdote. Multi-seed means and standard deviations, and a significance statement on any claimed separation, should be mandatory. In our own work a single-seed run of the compactness study reported a \emph{near-oracle} generative value; the five-seed standard corrected it to one an order of magnitude above the oracle it had appeared to match (Fig.~\ref{fig:compact}). And with only five seeds even a real separation is fragile: the sharpest distribution-free statement is a sign test (floor $p{=}0.0625$ at $n{=}5$), so a controlled sweep should carry many seeds and report the test, not a parametric $\sigma$ extrapolated past what five points resolve (Fig.~\ref{fig:crossover}). Both are cautionary instances of exactly this point.

\textbf{(vi) Backend-calibrated, asymmetric noise.} Because SQD's noise sensitivity runs through particle-number-violating readout errors (\S\ref{sec:3-2}), a credible noise model must be backend-calibrated and must include the $T_1$-driven asymmetry of readout, which symmetric bit-flip models understate; and the analysis should include an ablation isolating that asymmetry. A readout-only model will, in turn, be criticized where gate errors dominate, so the strongest studies will move toward full noisy-circuit simulation or a confirmatory hardware point.

\textbf{(vii) Convergence of the reduction to the model problem.} Every SQD number is a claim about a chosen active space in a chosen basis; the benchmark should report active-space, basis-set, and orbital-choice (natural- versus canonical-orbital) convergence, so that an apparent advantage is not an artefact of a fortunate truncation. We add a requirement that our own analysis of \S\ref{sec:3-1} forced on us: because determinant counts and compactness figures are \emph{not} invariants of the state --- a rotation inside a degenerate shell can move them while leaving the energy untouched, and, we find, even a nominally deterministic selection is not bit-reproducible, since a subspace solver reused across calls carries state and returns energies differing by two to three per cent depending on how many diagonalizations preceded it --- a compactness claim should give the complete orbital specification, including how degeneracies were resolved, together with a sensitivity estimate wherever a degenerate shell is partly occupied. This bears directly on the compactness comparisons of \S\ref{sec:5-2}, whose determinant counts are quoted across studies that need not share a gauge.

\textbf{(viii) Energy differences and properties, not only total energies.} Chemistry is decided by relative quantities --- barriers, spin gaps, dissociation energies --- and a method can reach a good total energy while misranking them; a benchmark should report at least one chemically decisive energy difference, not a total energy alone.

\textbf{(ix) More than one geometry.} A single-point win can be a coincidence of the reference; several geometries along a dissociation or reaction coordinate test whether the advantage is systematic, and are the natural setting for the amortization question of \S7.

\textbf{(x) Pre-registered, commensurate total-resource accounting.} Beyond the shot axis of (iv), the true total cost is quantum and classical together --- QPU-seconds reported alongside the classical node-hours of the diagonalization and recovery, in the commensurate-resource spirit of \cite{raisuddin2025} --- and the seed count and significance threshold of (v) should be fixed before the runs, not chosen after.

\textbf{We ran the ten elements against our own deposit.} A checklist is worth what it catches, so on 11 September 2026 we applied this one to our own deposit --- its code, its data files and its figure generators --- rather than to anyone else's. It returned five defects. Three are reported where they arise and we do not restate them here: the undeclared threshold inside $\rho$ and the non-replicating $1\times$ ladder (\S\ref{sec:7}, elements (v) and (vii)), and the exploration hyperparameter that decides Fig.~\ref{fig:compact} (\S\ref{sec:4-6}). Two are new, and both are failures of the deposit rather than of the analysis. \emph{First}, the script we had deposited to regenerate Fig.~\ref{fig:compact} was not the script that made it: it floored the reward nine orders of magnitude lower, which reverses the figure's verdict. None of the ten elements as stated requires a deposit to regenerate the figures it accompanies, and no referee reading a manuscript could detect that it does not; we read that requirement into (x). \emph{Second}, three of seven deposited \texttt{.dat} files did not come from their declared recipes --- two split a degenerate $\pi$ pair that exact FCI leaves degenerate, and a third was not reproducible even by re-running its own original code, three runs returning three answers. Both trace to the unfixed orbital gauge of \S\ref{sec:2-1}, which is the failure (vii) exists to catch, and the first two shipped with the first public version of the deposit.

\emph{The standard kept returning defects after the audit date, and we report those too rather than quietly repairing them.} Preparing this version we found that several figures quoted in \S\ref{sec:3-1} and \S\ref{sec:7} had no deposited generator at all --- the same violation of (x) as the first item above, in the sections that argue for (x). Writing the generators changed four of those numbers: the inflation and gauge sensitivity of $\rho$ and the emptiness of the gap that fixes its threshold (\S\ref{sec:7}), and the share of reward mass carried by the singly excited strings (\S\ref{sec:7}). We also found that the deposit had shipped two irreconcilable versions of the controlled ladder, and that the instructions it gave for reading its own data files did not run.

\emph{And it kept returning them after that.} A second pass, run adversarially against the corrected manuscript rather than against the original, returned seven more, which we list because the list is the evidence. \emph{One:} the two shot-survival fractions in the caption of Fig.~\ref{fig:loop} --- the opening figure --- had no generator; measured, they are $76\%$ and $57\%$ rather than $77\%$ and $59\%$, and the second turned out to be the square of the first, an independence assumption typeset as a second measurement. \emph{Two:} the looseness of the coupon-collector bound in \S\ref{sec:3-1} had none either; measured, $2.5\times$ and $16\times$ rather than ${\approx}2.7\times$ and ${\approx}17\times$. \emph{Three:} a classical control measured under the same simulated device noise as Fig.~\ref{fig:hci}, on the same seeds and dimension, sat in the deposit unprinted --- and on N$_2$ it runs against the direction of this paper, which is precisely the asymmetry element (x) exists to remove. \emph{Four:} the data-availability statement claimed that every result we compute is reproduced in the companion notebook, which is false and was printed inside the journal-mandated statement itself. \emph{Five:} \S\ref{sec:2-2} and the caption of Table~\ref{tab:sqd} twice promised an annotated dataset the deposit does not contain. \emph{Six:} the caption of Fig.~\ref{fig:compact} contradicted itself in adjacent sentences about why one series stops, and gave a reason its own generator refutes. \emph{Seven:} the deposit records a negative energy error against ``exact'' truth for C$_2$ --- which the variational principle forbids, and which therefore says the FCI reference of that near-degenerate active space is not exact. That is a failure of element (i), found by our own deposit rather than by us, and it is why C$_2$ is absent from Fig.~\ref{fig:hci}; the deposited file now carries the flag and the warning. \emph{Eight, and the largest correction in this paper:} the order parameter of v1 is retracted outright (\S\ref{sec:7}).

A checklist whose authors fail it twice in one manuscript, in the sections defending it, and then fail it again on the corrected manuscript, is evidence about how hard element (x) is to satisfy. We would rather print that than a cleaner story, and we note what it implies: the defects stopped appearing not when the deposit became correct but when we stopped looking, and we have no way to show that those two moments coincided.

All of them were found by applying this standard to our own work rather than by a referee, several changed a number printed here, and one retracted a headline result of the previous version. We say \emph{applying} rather than \emph{the author noticing}: the later passes were adversarial audits we commissioned against our own manuscript and deposit, with the explicit instruction to find what the corrections had broken. That is a method, and we recommend it over re-reading. The claim we draw is narrow: the standard is testable, we tested it on the one deposit we control, and its yield on a twelve-orbital, FCI-exact study of our own design was five defects.

We contend that a study meeting all ten is publishable and valuable \emph{whatever its result} --- a rigorous negative result that maps where a generative or quantum proposer does and does not help is, in a field this contested and this prone to under-benchmarked demonstrations, more useful than another hopeful positive that a strong baseline would deflate.

\section{Outlook: where generative and quantum methods retain a defensible advantage}\label{sec:7}

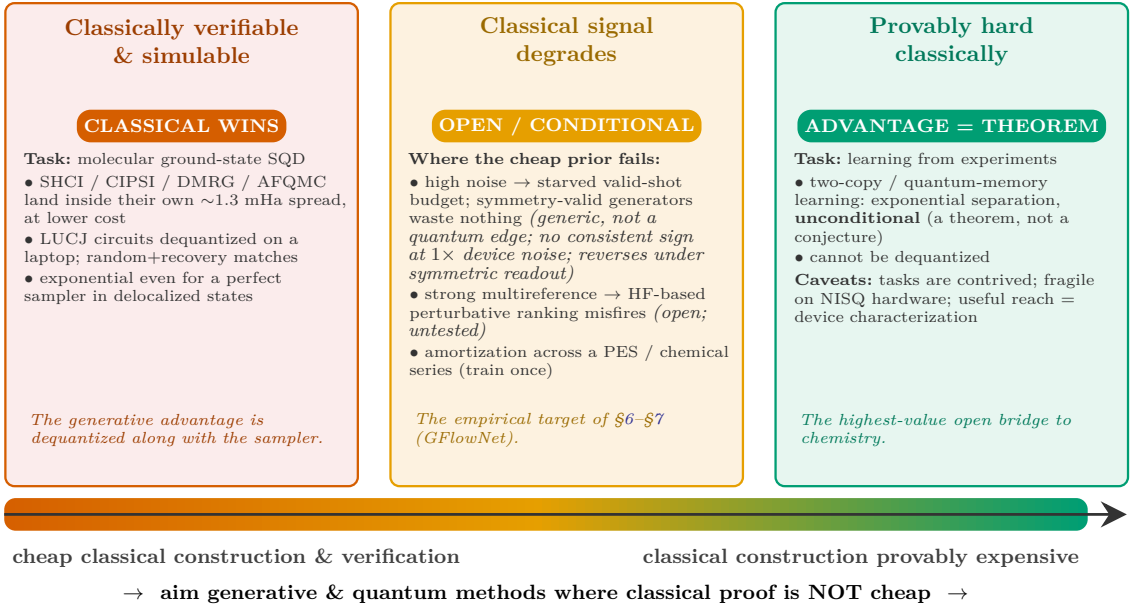
\begin{figure*}[tbp]
\centering
\resizebox{\linewidth}{!}{%
\begin{tikzpicture}[font=\sffamily,
  panel/.style={rounded corners=4pt, minimum width=5.4cm, minimum height=7.4cm, line width=1pt, anchor=north west},
  hdr/.style={font=\bfseries, align=center},
  chip/.style={rounded corners=6pt, text=white, font=\footnotesize\bfseries, inner sep=3pt, minimum height=0.5cm},
  body/.style={anchor=north west, align=left, font=\scriptsize, text width=4.9cm, inner sep=0pt},
  ital/.style={anchor=south west, align=left, font=\scriptsize\itshape, text width=4.9cm},
]
\definecolor{zRedF}{HTML}{FBECEC}\definecolor{zRedS}{HTML}{D55E00}\definecolor{zRedT}{HTML}{9C3D00}
\definecolor{zOrgF}{HTML}{FDF2E0}\definecolor{zOrgS}{HTML}{E69F00}\definecolor{zOrgT}{HTML}{9A6A00}
\definecolor{zGrnF}{HTML}{E7F6F0}\definecolor{zGrnS}{HTML}{009E73}\definecolor{zGrnT}{HTML}{00795A}
\node[anchor=north west, font=\large\bfseries, text=black] at (0,0.5) {Where a quantum or generative advantage can, and cannot, live};
\node[anchor=north west, font=\small, text=black!55, text width=17cm] at (0,-0.15)
  {The structure that makes SQD classically \emph{verifiable} is the structure that makes it classically \emph{constructible}. Advantage survives only where that equivalence breaks.};
\node[panel, fill=zRedF, draw=zRedS] (P1) at (0,-1.7) {};
\node[panel, fill=zOrgF, draw=zOrgS] (P2) at (5.9,-1.7) {};
\node[panel, fill=zGrnF, draw=zGrnS] (P3) at (11.8,-1.7) {};
\node[hdr, text=zRedT] at ([yshift=-6mm]P1.north) {Classically verifiable\\\& simulable};
\node[hdr, text=zOrgT] at ([yshift=-6mm]P2.north) {Classical signal\\degrades};
\node[hdr, text=zGrnT] at ([yshift=-6mm]P3.north) {Provably hard\\classically};
\node[chip, fill=zRedS] at ([yshift=-19mm]P1.north) {CLASSICAL WINS};
\node[chip, fill=zOrgS] at ([yshift=-19mm]P2.north) {OPEN / CONDITIONAL};
\node[chip, fill=zGrnS] at ([yshift=-19mm]P3.north) {ADVANTAGE = THEOREM};
\node[body, text=black!80] at ([xshift=3mm,yshift=-23mm]P1.north west)
  {\textbf{Task:} molecular ground-state SQD\\[2pt]
   $\bullet$ SHCI / CIPSI / DMRG / AFQMC land inside their own ${\sim}$1.3 mHa spread, at lower cost\\[1pt]
   $\bullet$ LUCJ circuits dequantized on a laptop; random+recovery matches\\[1pt]
   $\bullet$ exponential even for a perfect sampler in delocalized states};
\node[ital, text=zRedT] at ([xshift=3mm,yshift=5mm]P1.south west) {The generative advantage is dequantized along with the sampler.};
\node[body, text=black!80] at ([xshift=3mm,yshift=-23mm]P2.north west)
  {\textbf{Where the cheap prior fails:}\\[2pt]
   $\bullet$ high noise $\rightarrow$ starved valid-shot budget; symmetry-valid generators waste nothing {\footnotesize\itshape(generic, not a quantum edge; no consistent sign at $1\times$ device noise; reverses under symmetric readout)}\\[1pt]
   $\bullet$ strong multireference $\rightarrow$ HF-based perturbative ranking misfires {\footnotesize\itshape(open; untested)}\\[1pt]
   $\bullet$ amortization across a PES / chemical series (train once)};
\node[ital, text=zOrgT] at ([xshift=3mm,yshift=5mm]P2.south west) {The empirical target of \S\ref{sec:6}--\S\ref{sec:7} (GFlowNet).};
\node[body, text=black!80] at ([xshift=3mm,yshift=-23mm]P3.north west)
  {\textbf{Task:} learning from experiments\\[2pt]
   $\bullet$ two-copy / quantum-memory learning: exponential separation, \textbf{unconditional} (a theorem, not a conjecture)\\[1pt]
   $\bullet$ cannot be dequantized\\[2pt]
   \textbf{Caveats:} tasks are contrived; fragile on NISQ hardware; useful reach = device characterization};
\node[ital, text=zGrnT] at ([xshift=3mm,yshift=5mm]P3.south west) {The highest-value open bridge to chemistry.};
\shade[left color=zRedS, right color=zGrnS, middle color=zOrgS, rounded corners=5pt] (0,-9.8) rectangle (16.6,-9.3);
\draw[-{Stealth[length=4mm,width=4mm]}, line width=1.2pt, black!80] (0,-9.55) -- (17.2,-9.55);
\node[anchor=north west, font=\small\bfseries, text=black!75] at (0,-9.95) {cheap classical construction \& verification};
\node[anchor=north east, font=\small\bfseries, text=black!75] at (16.6,-9.95) {classical construction provably expensive};
\node[anchor=north, font=\small\bfseries, text=black] at (8.3,-10.5) {$\rightarrow$\ \ aim generative \& quantum methods where classical proof is NOT cheap\ \ $\rightarrow$};
\end{tikzpicture}
}
\caption{\textbf{The advantage landscape of \S5--\S7.} Colour-blind-safe (Okabe--Ito) palette; zones are also distinguished by header text and verdict chips, not colour alone.}
\label{fig:landscape}
\end{figure*}

The reason classical selected CI wins the SQD contest, on the systems studied, is structural: SQD's task is \emph{classically verifiable}, and, empirically on molecular systems, the same structure that lets one check a subspace (a projected Hamiltonian one can diagonalize) also makes a cheap classical \emph{importance heuristic} available with which to construct it; the dequantization of \S\ref{sec:5-3} is the extreme statement of the same fact, that even the sampler is classically reproducible. (This is an observation about the available classical signal, not a general implication of verifiability: verifiability alone need not imply cheap constructibility, as the P-versus-NP gap makes plain.) There is little quantum edge to be had where such a cheap classical construction happens to exist.

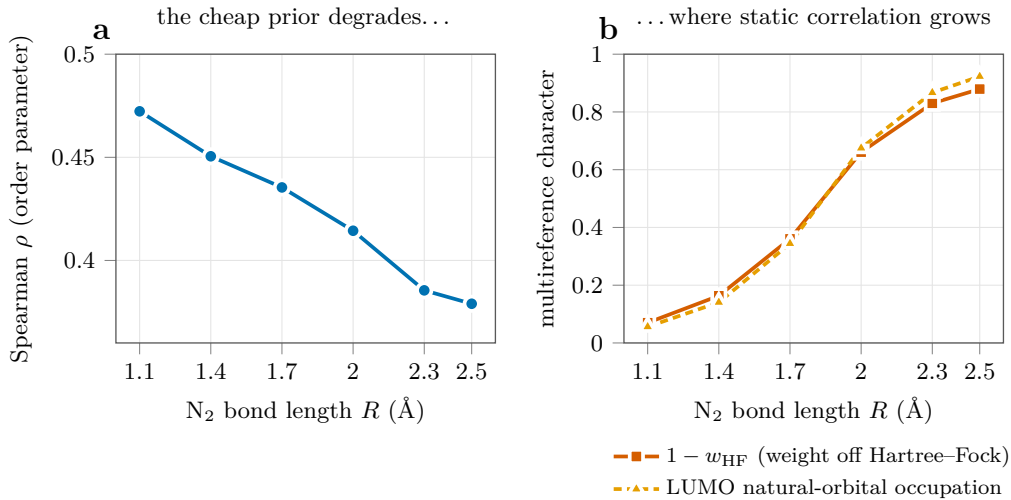
\begin{figure}[tbp]
\centering
\begin{tikzpicture}
\begin{groupplot}[group style={group size=2 by 1, horizontal sep=1.7cm}, paperplot,
  width=0.44\linewidth, height=0.36\linewidth, xmin=1.0, xmax=2.6, xtick={1.1,1.4,1.7,2.0,2.3,2.5}]
\nextgroupplot[clip=false, xlabel={N$_2$ bond length $R$ (\AA)}, ylabel={Spearman $\rho$ (order parameter)},
  ymin=0.36, ymax=0.50, title={the cheap prior degrades\dots}]
  \addplot[oiBlue, line width=1.4pt, mark=*, mark size=2.7pt, mark options={fill=oiBlue,draw=white}]
    table[x=R,y=spearman]{orderparam.dat};
  \node[font=\bfseries\large, anchor=south east, yshift=2pt] at (rel axis cs:0.01,1.0){a};
\nextgroupplot[clip=false, xlabel={N$_2$ bond length $R$ (\AA)}, ylabel={multireference character},
  ymin=0, ymax=1, title={\dots where static correlation grows},
  legend style={at={(0.5,-0.32)}, anchor=north, legend columns=1, draw=none, fill=none, font=\footnotesize}]
  \addplot[oiVerm, line width=1.4pt, mark=square*, mark size=2.5pt, mark options={fill=oiVerm,draw=white}]
    table[x=R,y=mrweight]{orderparam.dat}; \addlegendentry{$1-w_{\mathrm{HF}}$ (weight off Hartree--Fock)}
  \addplot[oiOrange, line width=1.4pt, densely dashed, mark=triangle*, mark size=3.2pt, mark options={fill=oiOrange,draw=white,solid}]
    table[x=R,y=lumo_noon]{orderparam.dat}; \addlegendentry{LUMO natural-orbital occupation}
  \node[font=\bfseries\large, anchor=south east, yshift=2pt] at (rel axis cs:0.01,1.0){b};
\end{groupplot}
\end{tikzpicture}
\caption{\textbf{The cheap prior degrades with multireference character (N$_2$ across geometries).} \emph{(a)} The Spearman rank correlation between the cheap Epstein--Nesbet reward and the exact FCI $|c|^2$ (over all 792 $\alpha$-strings) falls monotonically as the triple bond stretches, from $\rho{=}0.47$ near equilibrium to $0.38$ at $R{=}2.5$~\AA{} (symmetry-adapted gauge; see the header of \texttt{orderparam.dat} for the thresholding convention). \emph{(b)} Over the same interval the wavefunction becomes strongly multireference: the weight carried by determinants \emph{other} than Hartree--Fock rises from $0.07$ to $0.88$, and the frontier natural-orbital occupations open from the closed-shell $2/0$ toward the dissociation limit $1/1$ (LUMO occupation $0.06\to0.92$). The cheap classical importance signal that makes an SQD subspace \emph{verifiable} degrades exactly where static correlation makes it harder to \emph{construct} --- so we propose the single scalar $\rho$ as a low-cost \emph{coordinate}, computable system by system before any hardware is run, for where a learned or quantum proposer might have room to help. Whether it in fact predicts that is the open test set out in \S\ref{sec:7}, not a result of this figure.}
\label{fig:orderparam}
\end{figure}

This is not a wall but a compass (Fig.~\ref{fig:landscape}): it says to aim generative and quantum methods at the regimes where a strong classical baseline provably \emph{cannot} be cheap. This converts the compass from a metaphor into a falsifiable prediction the standard of \S\ref{sec:6} can test: if the locus of advantage is exactly where the cheap importance signal misranks the truly important determinants, then any genuine generative or quantum advantage on molecular electronic structure should appear first, and preferentially, where the \emph{rank correlation between the cheap prior (e.g. the Epstein--Nesbet coefficient) and the exact CI weights degrades most}. That rank correlation is directly measurable at FCI-verifiable scale, and our own minimal computational experiment both confirms it is imperfect and shows it behaving as a monotone multireference coordinate. Over the 792 $\alpha$-strings of the N$_2$(10e,12o) active space, the Spearman correlation between the cheap Epstein--Nesbet reward and the exact $|c|^2$ is $\rho{\approx}0.41$ at $R=2.0$~\AA{}; more tellingly, it \emph{declines as the wavefunction becomes more multireference}, falling from $\rho{=}0.47$ near equilibrium ($R=1.1$~\AA{}: Hartree--Fock weight $0.93$, frontier natural-orbital occupations $1.95/0.06$) through $0.41$ at $R=2.0$~\AA{} (HF weight $0.34$, occupations $1.32/0.67$) to $0.38$ at $R=2.5$~\AA{} (HF weight $0.12$, occupations $1.08/0.92$, the sum of fractional occupancy rising from $0.29$ to $5.10$). This is that coordinate, measured (Fig.~\ref{fig:orderparam}): the scalar declines monotonically where static correlation grows --- though it is the Hartree--Fock weight ($0.93\to0.12$) and the natural-orbital occupations, not $\rho$ itself, that genuinely collapse. Notably $\rho$ stays ${\approx}0.38$-informative even at maximal static correlation, so the cheap prior is \emph{degraded, not destroyed} by multireference character (consistent with Fig.~\ref{fig:compact}, where deterministic top-$K$ on that prior still beats the GFlowNet). Two caveats a specialist will demand: by the Slater--Condon rules the reward is exactly zero for the $546$ $\alpha$-strings that are triple-or-higher excitations of Hartree--Fock. They are not the only zeros: in the symmetry-adapted gauge a further $155$ of the $210$ doubly excited strings are annihilated by spatial symmetry, so $701$ of the $792$ strings ($88.5\%$) are tied at zero and only $91$ --- the Hartree--Fock string, all $35$ singles, and $55$ doubles --- carry rankable reward at all (\texttt{figures/cuenta\_rankeables.py}). $\rho$ over all strings is therefore an even blunter, more tie-laden statistic than the Slater--Condon count alone suggests. Brillouin's theorem annihilates the \emph{determinant}-level single excitations, but not the $\alpha$-string reward, which marginalizes over $\beta$: for a singly excited $\alpha$-string the reward is carried by the $(\alpha\text{-single}\otimes\beta\text{-single})$ \emph{double} excitations, which Brillouin does not touch. The $35$ singly excited $\alpha$-strings therefore carry $66.6\%$ of the total reward mass --- $83.6\%$ of it once the Hartree--Fock string's own $20.4\%$ is set aside --- and $47.5\%$ of the exact FCI $\alpha$-string weight, more than the Hartree--Fock string itself ($35.0\%$). (An earlier version of this paper reported $99.2\%$ here; the deposited census returns $66.6\%$, and we correct it. The two neighbouring figures, $47.5\%$ and $35.0\%$, reproduce exactly.) The blunt-statistic point stands on the $546$ exact ties alone; and its precise value carries a gauge dependence from the arbitrary rotation within the degenerate $\pi$ shells (\S\ref{sec:2-1}), and a subtler defect that we found only on auditing our own deposit, and that we state because it changes the number we report. The Epstein--Nesbet reward is exactly zero for the $546$ triple-or-higher $\alpha$-strings, but in floating point those matrix elements leave cancellation residue as small as $10^{-42}$, and a rank correlation orders that residue \emph{above} the exact zeros. The statistic is therefore only defined once a threshold is declared: we set rewards below $10^{-12}$ of the maximum to exactly zero. The cut is not a tuned knob, because the reward spectrum is bimodal: across $41$ gauges the genuine entries begin at $1.2\times10^{-12}$ of the maximum and the cancellation residue tops out around $10^{-34}$, and only in the worst gauge do as many as $2$ of the $792$ entries fall anywhere between $10^{-16}$ and $10^{-12}$. Without the cut, $\rho$ reads high by $0.085$ on average ($0.055$--$0.090$ across those gauges) and its gauge spread is $0.063$; with it the residual spread is $0.034$. \emph{These three figures are corrections, and they are not the largest one this paper has made.} \textbf{The order parameter itself is a retraction.} Version~1 of this preprint reported $\rho$ falling from $0.7245$ to $0.5982$ across the same ladder, where this version reports $0.4723$ to $0.3790$. Diffing the two deposited \texttt{orderparam.dat} files, \emph{only} the $\rho$ column moved --- the fractional-occupancy sum, the Hartree--Fock weight and both frontier occupations are identical to every digit --- by $0.2522$, $0.2508$, $0.2421$, $0.2277$, $0.2252$ and $0.2192$ at the six geometries. The declared threshold accounts for about a third of that ($0.085$ on average); the remainder is the orbital gauge of \S\ref{sec:2-1}, which v1 did not fix. What survives the correction is the decline and its ordering, which is what \S\ref{sec:5-5} and \S\ref{sec:8} rest on; what does not survive is any absolute value printed in v1, and readers who took one should replace it. Earlier versions of this paper also put the three sensitivity figures at ${\approx}0.2$, $0.18$ and $0.03$; only the third survived being measured by a script we then deposited (\texttt{gauge\_study/rho\_sensibilidad.py}, which refuses to report anything unless each rotation leaves $E_{\mathrm{FCI}}$ invariant, here to $2.3\times10^{-13}$~Ha). The direction is unchanged --- thresholding halves the gauge sensitivity --- and so is the reason for declaring the threshold, which does not rest on the size of either effect: a rank correlation that orders $10^{-42}$ cancellation residue above exact algebraic zeros is not a well-defined statistic at any spread. We quote $\rho$ thresholded, single-threaded, in the symmetry-adapted gauge; the \emph{decline} is then $0.09$, some three times the residual spread, and the monotone ordering survives in all three protocols we ran (raw, thresholded, and thresholded in the symmetry-adapted gauge). We rely on that monotone \emph{trend}, not on the absolute value, and propose $\rho$ as a low-cost coordinate for where to look --- computable system by system before any hardware is run, provided the gauge and threshold are declared with it. A method that helps where the prior already ranks well would falsify the picture; one that helps only as $\rho$ degrades would confirm it. Three such regimes are visible.

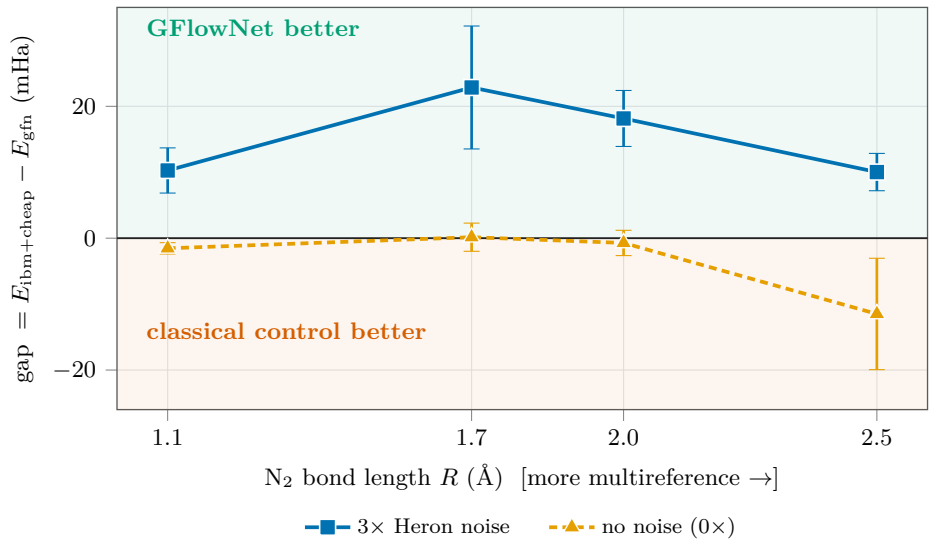
\begin{figure}[tbp]
\centering
\begin{tikzpicture}
\begin{axis}[paperplot, width=0.82\linewidth, height=0.46\linewidth,
  xlabel={N$_2$ bond length $R$ (\AA)\ \ [more multireference $\rightarrow$]},
  ylabel={gap $\;{=}\;E_{\mathrm{ibm+cheap}}-E_{\mathrm{gfn}}$\,\,(mHa)},
  xmin=1.0, xmax=2.6, ymin=-26, ymax=35, xtick={1.1,1.7,2.0,2.5}, xticklabels={1.1,1.7,2.0,2.5},
  title={Controlled single-molecule test: at $3\times$ noise the crossover is universal\\across N$_2$ dissociation --- driven by shot starvation, not multireference},
  legend style={at={(0.5,-0.24)}, anchor=north, legend columns=2, draw=none, fill=none, /tikz/every even column/.append style={column sep=12pt}}, clip=false]
  \addplot[draw=none, fill=oiGreen, fill opacity=0.06, forget plot] coordinates {(1.0,0)(2.6,0)(2.6,35)(1.0,35)}\closedcycle;
  \addplot[draw=none, fill=oiVerm, fill opacity=0.06, forget plot] coordinates {(1.0,-26)(2.6,-26)(2.6,0)(1.0,0)}\closedcycle;
  \draw[oiInk, line width=0.7pt] (axis cs:1.0,0) -- (axis cs:2.6,0);
  \addplot[oiBlue, line width=1.4pt, mark=square*, mark size=3pt, mark options={fill=oiBlue, draw=white, line width=0.6pt},
    error bars/.cd, y dir=both, y explicit, error bar style={oiBlue, line width=1pt}, error mark options={rotate=90, mark size=3pt}]
    table[x=R, y=gap3, y error=gap3sd]{ladder.dat};
  \addlegendentry{$3\times$ Heron noise}
  \addplot[oiOrange, line width=1.4pt, densely dashed, mark=triangle*, mark size=3.6pt, mark options={fill=oiOrange, draw=white, line width=0.5pt, solid},
    error bars/.cd, y dir=both, y explicit, error bar style={oiOrange, line width=1pt, solid}, error mark options={rotate=90, mark size=3pt}]
    table[x=R, y=gap0, y error=gap0sd]{ladder.dat};
  \addlegendentry{no noise ($0\times$)}
  \node[oiGreen, font=\small\bfseries, anchor=west] at (axis cs:1.04,32){GFlowNet better};
  \node[oiVerm, font=\small\bfseries, anchor=west] at (axis cs:1.04,-14){classical control better};
\end{axis}
\end{tikzpicture}
\caption{\textbf{A controlled single-molecule test refutes multireference-specificity of the noise crossover.}
Classical-minus-generative energy-error gap for N$_2$ at fixed active space (10e,12o), swept along the dissociation
coordinate so that \emph{only} multireference character varies (HF weight $0.93\to0.12$ as $R{=}1.1\to2.5$~\AA),
with molecule, electron count, and Hilbert dimension held constant; mean~$\pm$~s.d.\ over five seeds at $D{=}120$.
\emph{At $3\times$ Heron noise} (blue) the constraint-respecting generative proposer wins at \emph{every} geometry
($+10.0$ to $+22.9$~mHa, positive in all five seeds at every geometry), including near-single-reference $R{=}1.1$
--- so the crossover is driven by valid-shot starvation ($\sim$51\% lost at $3\times$), \emph{not} by multireference
character. \emph{Two bounds belong with this panel:} at $1\times$ Heron noise --- the device-calibrated scale ---
the comparison does not even keep its sign along the ladder, running from $-4.03\pm1.68$~mHa at $R{=}1.1$ (a significant
\emph{classical} win, $t_4{=}-5.4$) to $+3.01\pm1.25$ at $R{=}2.0$ ($t_4{=}5.4$) --- a sign reversal along the
ladder that an independent five-seed replication does not reproduce (\S\ref{sec:7}), and under a symmetric rather
than backend-calibrated readout the ordering reverses over five seeds (generative $47.9\pm3.2$~mHa vs.\ $31.1\pm3.9$
classical and $33.5\pm2.8$ raw, $t_4{=}10.2$). The effect shown here is a high-noise, asymmetric-readout effect,
and \S\ref{sec:6} asks that it be reported as such. \emph{At zero noise} (orange) the classical control is ahead
at three of the four geometries, significantly at two (deepest at $R{=}2.5$, $-11.5$~mHa, $t_4{=}-3.0$), with a
null $+0.15\pm2.14$~mHa at $R{=}1.7$ --- the opposite of a generative opening. The two effects are decoupled. With
five seeds the distribution-free significance floors at $p{=}0.0625$ (all seeds agree in sign); the parametric
$t_4$ is a reproducibility diagnostic only. The apparent multireference-specificity of an earlier stretched-N$_2$-versus-near-single-reference-H$_2$O
contrast was a subspace-coverage confound ($24\%$ vs $15\%$ of the target space at fixed $D$; \S\ref{sec:7}), which
this single-molecule design removes.}
\label{fig:crossover}
\end{figure}

\textbf{Noise-limited recovery, and a controlled test of the multireference hypothesis.} Within SQD itself the classical selectors win because a cheap classical importance signal (a Hartree--Fock-referenced perturbative estimate) is available and accurate. The one regime where a learned proposer has room is high noise: as the valid-shot budget is starved, a generator that emits only symmetry-valid configurations \emph{by construction} wastes nothing, whereas the classical control loses a growing fraction of its samples to particle-number violation. The fair baseline here is no straw man --- it is the noise-robust recovery loop of \S\ref{sec:3-2} itself (S-CORE plus the cheap prior that already reproduces the accuracy), so any generative gain is measured \emph{beyond} that classical recovery; and because the classical learned rankers of \S\ref{sec:4-4} already contest heat-bath CI with no quantum resource --- matching its accuracy on far fewer determinants, and reporting higher efficiency on Cr$_2$ --- the live question is the \emph{incremental} value of the quantum samples, not of learning over perturbation theory. Our first reading of this gain was that it is \emph{multireference-specific}: that it should appear only where a single-reference prior misranks the important determinants. \emph{A controlled experiment refutes that reading.} We ran the noise sweep along the N$_2$ dissociation coordinate at fixed active space (10e,12o), so that only multireference character varies --- from near-single-reference at $R{=}1.1$~\AA{} (HF weight $0.93$, $\rho{=}0.47$) to strongly multireference at $R{=}2.5$~\AA{} (HF weight $0.12$, $\rho{=}0.38$) --- with molecule, electron count, and Hilbert dimension held \emph{constant} (Fig.~\ref{fig:crossover}). The crossover is \emph{universal}: at $3\times$ Heron noise the generative proposer wins at \emph{every} geometry, by $+10.0$ to $+22.9$~mHa and positive in all five seeds at each point, including the near-single-reference $R{=}1.1$ case; an independent five-seed replication (seeds $5$--$9$) reproduces this arm in full --- four of four positive, all significant, $+7.0$ to $+14.3$~mHa. Two boundaries belong with that number, because they fix what it licenses. \emph{First, it is a high-noise effect, not a device-scale one.} We have now run the intermediate point on the ladder itself. At $1\times$ Heron noise --- the calibrated scale of the actual processor --- the effect does not even keep its sign across the ladder: $-4.03\pm1.68$~mHa at $R{=}1.1$, where the \emph{classical} control wins significantly ($t_4{=}-5.4$); $+0.34\pm2.51$ at $R{=}1.7$, a null ($t_4{=}0.3$); $+3.01\pm1.25$ at $R{=}2.0$ ($t_4{=}5.4$); and $+5.23\pm3.19$ at $R{=}2.5$ ($t_4{=}3.7$). At device noise there is no consistent effect in either direction along the coordinate, and every figure is a factor of two to twenty below the $3\times$ values. \emph{That is a statement about the sign along the ladder, not about the sign of every device-scale comparison we ran.} At the geometry and dimension of Fig.~\ref{fig:hci}, against a classical recovery loop under the same simulated device noise and the same five seeds, the proposer is ahead on N$_2$ by $2.4$~mHa ($19.06\pm1.70$ against $21.51\pm1.57$; paired seed by seed, $2.45\pm1.80$~mHa and $t_{4}{=}3.0$, with all five agreeing in sign, which five seeds floor at $p{=}0.0625$ distribution-free) and behind on H$_2$O by $1.4$. We report the positive because it exists and because withholding it would be the asymmetry element (x) exists to remove; we do not build on it, because it does not resolve at five seeds and because the same proposer loses to the noise-free classical selector of \S\ref{sec:5} by more than a factor of two. We can now put a number on how little this arm resolves, because we ran it twice: an independent five-seed replication (seeds $5$--$9$) agrees in \emph{sign} at only two of the four geometries and returns $-2.47$, $-0.14$, $+0.35$, $-1.83$~mHa, so the two $t_4$-significant positives above are not reproducible at this seed count. At $3\times$ both runs agree at every geometry; at $1\times$ neither run should be read as measuring anything but the absence of a stable effect. A monotone ordering of the $1\times$ gaps with $\rho$ appeared in the first run (Spearman $-1.000$ over four geometries, exact $p{=}0.083$) and did \emph{not} survive the replication ($-0.400$, $p{=}0.75$); we report it only to record that we looked and that it failed. Every substantial positive figure quoted here is at three times the noise of the device the model is calibrated from, and we claim no useful margin at $1\times$. \emph{Second, the sign is noise-model dependent:} replacing the asymmetric, backend-calibrated readout by a symmetric readout channel reverses the ordering, the generative proposer degrading to $47.9\pm3.2$~mHa against $31.1\pm3.9$~mHa for the classical recovery loop and $33.5\pm2.8$~mHa raw, over five seeds at $2000$ shots ($t_4{=}10.2$ against the recovery loop, all five seeds agreeing in sign; \texttt{calculations/lectura\_simetrica.py}) --- the proposer collapses once the corruption it was implicitly exploiting becomes unbiased. \emph{This is a correction, and it makes the caveat stronger.} Earlier versions of this paper quoted $40.9$, $32.4$ and $33.1$~mHa from a \emph{single} seed with the orbital gauge unpinned --- the precise failure element (v) exists to catch, applied to the one number that most limits our own positive result. Re-run to the standard we ask of others, the two classical arms land inside the spread of the values we had printed and the generative arm does not: the single seed had been a favourable draw for the proposer. We report both as a test of our own standard rather than an exception to it: a result whose sign is not even consistent at device noise, and which reverses under a different readout model, is exactly what the pre-registration and noise-model-reporting requirements of \S\ref{sec:6} are meant to force into the open, and the requirement is worth little if it is not applied to the one positive result in this paper. Nor does multireference character control the \emph{zero-noise} baseline, which we had expected it to. Regenerated in the fixed gauge, that arm reads $-1.54\pm0.86$~mHa at $R{=}1.1$ ($t_4{=}-4.0$), $+0.15\pm2.14$ at $R{=}1.7$ ($t_4{=}0.2$), $-0.72\pm1.92$ at $R{=}2.0$ ($t_4{=}-0.8$) and $-11.49\pm8.45$ at $R{=}2.5$ ($t_4{=}-3.0$): the classical control is ahead at three of the four geometries and significantly so at two, but the magnitude does \emph{not} increase with stretching --- it is smaller at $R{=}2.0$ than at near-equilibrium $R{=}1.1$, and the sign flips to a null positive at $R{=}1.7$. This is the third prediction of ours that the ladder refutes, and we report it as such. Either way it is the opposite of a generative opening. The two axes are decoupled: \emph{valid-shot starvation}, not multireference character, drives the crossover, and it does so across the whole dissociation curve.

This also exposes the confound in a naive stretched-N$_2$-versus-near-single-reference-H$_2$O contrast (H$_2$O --- here at its \emph{equilibrium} geometry, $r_{\mathrm{OH}}{=}0.958\,$\AA{}, not the stretched $1.24\,$\AA{} of Fig.~\ref{fig:hci} --- runs the classical control ahead at low and intermediate noise --- $-1.39\pm0.95$~mHa at $1\times$ ($t_4{=}-3.3$) and $-2.91\pm1.02$ at $2\times$ ($t_4{=}-6.4$) --- and shows a nominal sign flip at $3\times$, $+1.36\pm4.45$~mHa, which at five seeds is not resolved at all, $t_4{=}0.7$): at fixed $D{=}120$, H$_2$O covers $24\%$ of its 495-string space against N$_2$'s $15\%$ of 792, so its classical fill is simply more complete --- a coverage effect, not a chemistry effect, and not the multireference-specificity we first inferred from it. One disclosure is owed here under the standard \S\ref{sec:6} asks of others: the generalization script behind these runs covers a third system, C$_2$ at $1.40\,$Å in CAS(8e,12o) --- the same 495-string space as H$_2$O, and strongly multireference --- which is not reported. We claim no result for it. It is named so that its absence is a stated exclusion rather than a silent one, which is exactly the asymmetry a pre-registration requirement exists to remove.

\textbf{A hardware run of our own that reads as a counterexample.} A reader following our own citation \cite{bonilla2026spectral} finds --- in that paper's deposit, not ours --- a CCSD-seeded LUCJ circuit sampled on \texttt{ibm\_marrakesh} on the same active space as Fig.~\ref{fig:hci}, the same $E_{\mathrm{FCI}}$ to $1.7\times10^{-8}$~Ha, recovering to $0.59\pm0.14$~mHa over eight recovery seeds on one cached $10{,}000$-shot job --- against that figure's $9.13$~mHa classical selector. \emph{We state the resolution rather than leave it to be found.} Fig.~\ref{fig:hci} draws $1{,}000$ shots per seed from the exact FCI $|c|^2$ under simulated Heron noise, repairs them in a \emph{single} occupation-weighted pass, and scores a fixed $D{=}120$ subspace; the companion samples a physical circuit, runs \emph{five} self-consistent recovery iterations over five batches of $300$ and keeps the minimum, works in an unpinned orbital gauge where ours is symmetry-adapted (\S\ref{sec:2-1}), and never reports the diagonalization size that element (iv) asks be controlled --- which leaves it uninterpretable in exactly the direction that flatters the sampler. They were not built as a controlled comparison and we do not offer them as one. \emph{Neither number is an advantage:} this active space is diagonalized exactly on a laptop, which is how both papers obtain their reference. The verdict of \S\ref{sec:5-5} --- SHCI, CIPSI, DMRG, AFQMC, \emph{or even random determinants passed through classical recovery} --- is untouched, because none of those controls was run. One feature of that run does bear on this section. The \emph{noiseless} simulation of the same circuit lands at $29.5$~mHa: that is not the valid-shot starvation argued for here, which turns on a valid-by-construction generator and the LUCJ circuit is not one, but the unablated noise-assisted broadening of \S\ref{sec:3-2} --- and the spread across recovery seeds on those same device counts runs from $0.40$ to $0.85$~mHa.

The honest conclusion is narrower than our first one: the only generative advantage we can demonstrate is a \emph{noise-robustness} effect at elevated noise (valid-by-construction sampling under shot starvation) --- with no consistent sign at the device-calibrated scale, and reversing under a symmetric readout model --- available in principle across chemistry rather than confined to the multireference regime, and \emph{not} a quantum or chemistry advantage, since any constraint-respecting classical sampler shares the mechanism. The multireference \emph{prediction} of the order parameter above is left \emph{untested} by this experiment, which varies noise rather than prior quality at fixed noise; cleanly separating those two axes is the controlled study \S\ref{sec:6} still calls for (e.g.\ on Cr$_2$ or a metal dimer). Two statistical caveats belong with the result: with five seeds the sharpest distribution-free statement is that all five agree in sign (a sign-flip test floors at $p{=}0.0625$; the parametric $t_4$ values we quote are reproducibility diagnostics, not discovery significances), and each per-geometry test certifies seed reproducibility, not generalization across chemical space. That a controlled test of our own hypothesis returns a partial negative is, by the standard of \S\ref{sec:6}, exactly the kind of result this field needs more of.

\textbf{Amortization across chemical space.} A learned proposer offers something a per-instance classical selector does not: amortization. A model conditioned on molecular geometry and trained once can, in principle, generate compact subspaces across an entire dissociation curve or chemical series without re-running an expensive selection at each point --- the value proposition of the foundation-wavefunction programme \cite{foster2025}, transposed to the determinant-selection setting. Here the metric is not accuracy at one geometry but total cost across many, and it is a genuinely open, and genuinely ML-flavoured, question whether amortization beats re-running SHCI per point.

\textbf{Learning from experiments: where the advantage is a theorem.} The deepest reason for optimism lies just outside the SQD loop, and it inverts the logic of \S\ref{sec:5} exactly. The one place in this landscape with an \emph{unconditional, information-theoretic} quantum advantage is the quantum-data-native task of learning properties of an unknown quantum state or process: a learner with quantum memory that measures two coherent copies jointly needs exponentially fewer experiments than any single-copy strategy --- where the single-copy baseline is itself a strong one, randomized measurement protocols of the classical-shadow family \cite{huang2020} --- a separation proved without complexity assumptions \cite{chen2021} and demonstrated on superconducting and photonic hardware \cite{huang2022,liuphotonic2025}, and surveyed among the field's few unconditional advantages by \cite{huang2025}. Unlike SQD, this advantage cannot be dequantized, because its classical lower bound is a theorem, not a conjecture about circuit structure. This suggests a reorganizing principle for the enterprise as a whole: candidate quantum advantages should be sorted not by their promised magnitude but by the \emph{type} of their classical lower bound. Those resting on a complexity-theoretic conjecture (the presumed hardness of sampling a particular circuit family) are perpetually one clever classical algorithm away from collapse, as SQD's own two-year arc from flagship experiment to laptop reproduction (\S\ref{sec:5-3}) attests; those resting on an unconditional, information-theoretic separation are not, because no future algorithm can close a gap that is already a theorem. A field repeatedly surprised by dequantization would do well to spend its scarce hardware where the lower bound is a theorem, and to treat conjecture-backed advantages as demonstrations of scale, not of separation, until proven otherwise. Its caveats are equally sharp and belong in any forward-looking account: the demonstrated-advantage \emph{tasks} (predicting Pauli expectations, purity, and random-channel characteristic functions) are information-theoretically natural but scientifically contrived; the advantage is fragile --- in theory the two-copy purity-testing advantage collapses under local depolarizing noise \cite{cotler2025}, and on IBM's heavy-hex superconducting processors it survives only as the reduced, mitigation-propped separation actually measured \cite{seif2024}; and the one \emph{useful} application delivered with rigorous backing is the characterization of quantum devices themselves. The bridge from this provable-but-contrived frontier to genuinely useful chemistry is, we argue, the highest-value open problem adjacent to the subfield, and the natural destination for the generative machinery reviewed here. That bridge might be built by learning a device's effective error model (as machine-learning quantum error mitigation already does at utility scale \cite{liao2024}) to improve an SQD estimator, or by learning the properties of a molecular state prepared on hardware.

\section{Conclusions and open problems}\label{sec:8}

Machine learning entered sample-based quantum diagonalization for a sound reason (the configuration-recovery step is a selection-and-denoising problem that a learned model should do better than an occupancy-weighted heuristic), and in two years has produced a rich ecosystem of restricted Boltzmann machines, neural-quantum-state proposers, classifiers, ranking models, and generative circuit designers. This review has organized that ecosystem, mapped its methods and open-source tooling, and, most importantly, insisted on holding it to the question it exists to answer, and stated at falsifiable precision the verdict the critique literature had already reached \cite{reinholdt2025}. That verdict, in \S\ref{sec:5-5}, is negative, and it is scoped: no verifiable-scale demonstration that the quantum sampler or its machine-learning augmentation beats classical selected configuration interaction on molecular electronic structure, and flagship sampling circuits now classically reproducible at the energy (weak-simulation) level. We have argued that it is a compass rather than a counsel of futility: it points generative methods toward the noise-limited and multireference regimes where the cheap classical signal fails, toward amortization across chemical space, and, as an adjacent horizon, toward the quantum-data-native learning tasks where the advantage is provable but not yet connected to useful chemistry. A controlled FCI-exact experiment already sharpens the first of these into a candid partial negative: the prior's rank correlation with the truth declines with multireference character (a usable coordinate), but the generative noise-crossover turns out to be a generic shot-starvation effect, present across the N$_2$ dissociation curve at elevated noise rather than multireference-specific --- and, held to the same standard, without a consistent sign at device-calibrated noise, and reversing under a symmetric readout model (\S\ref{sec:7}).

Four open problems structure the road ahead. First, the \emph{intrinsic-delocalization} barrier of \cite{gaberle2026} (that some strongly correlated ground states require exponentially many configurations even under a perfect sampler) has no proposed remedy and may be fundamental. Second, the \emph{depth} question: whether deep, multi-layer LUCJ circuits evade the classical simulability that \cite{belagali2026} proved for the single-layer case is the one place a \emph{quantum-sampling-hardness} advantage could still live, and settling it is urgent. We flag a structural worry worth confronting rather than postponing. The VQE programme collapsed when it emerged that the conditions letting a circuit avoid barren plateaus tend also to render it classically simulable \cite{cerezo2025}; SQD may face an isomorphic dichotomy in a different dialect, since the shallow, structured, low-entanglement circuits whose samples are noise-tolerant and amenable to configuration recovery are precisely those whose Jastrow structure the dequantization of \cite{belagali2026} exploits, while the depth that would defeat classical simulation may also be the depth at which readout noise and the coupon-collector tail (\S\ref{sec:3-1}) overwhelm the loop. Whether a usable corridor exists between these two failure modes is, we conjecture, the question on which a near-term \emph{chemistry} advantage for SQD ultimately rests. Third, the \emph{useful-target} gap in learning-from-experiments (bridging provable-but-contrived tasks to science anyone needs) is the highest-value problem adjacent to the field. Fourth, and most immediately actionable, the \emph{GFlowNet gap}: the one generative paradigm built for diverse, reward-proportional sampling over exponential discrete spaces has not been applied to determinant selection, and whether it helps, measured (per \S\ref{sec:6}) against exact ground truth and strong classical baselines, is a clean, falsifiable, and currently unclaimed question. We have tried to write the review such that its most useful legacy would be not a catalogue but a standard: that the next claim in this field, positive or negative, arrives with exact ground truth, strong and dequantization baselines, full cost accounting, and error bars, so that the field can, at last, know what it knows.

\emph{Limitations of this review.} This survey is a snapshot of a subfield that is expanding on a timescale of weeks; a substantial fraction of the primary literature we cite is preprint and not yet peer-reviewed. We have verified the authors, venues, and arXiv/DOI identifiers of the works most central to the verdict, but preprint metadata can still shift, and readers should confirm against the published versions of record. Our coverage extends to September 2026 and will inevitably miss work that appears between writing and publication, including, very possibly, the first application of a generative-flow network to determinant selection whose absence we here record. We have tried to make the review robust to that churn by organizing it around durable questions (compactness, verifiability, the locus of advantage) and a benchmarking standard, rather than around any single result. The GFlowNet gap is presented as an open, falsifiable question (one that forthcoming study may well answer in the negative), not a solved one.

\section*{Data and code availability}
\small
Every quantitative result we compute ourselves in this review --- as distinct from the numbers we quote from the cited literature --- is reproduced from executable code in the deposit, and the README's generator table names the script behind each one. \emph{We are specific rather than sweeping here, because a blanket claim of this kind is one of the defects \S\ref{sec:6} caught us in.} The companion computational notebook \texttt{GFlowNet\_SQD\_calculations.ipynb} covers the workflow of \S\ref{sec:2}--\S\ref{sec:5}: the FCI reference, the coupon-collector statistics, S-CORE recovery, the cheap reward and the GFlowNet compactness study, the order parameter, the noise crossover and the selected-CI test. It does \emph{not} contain the gauge study, the reward-floor sweep, the symmetric-readout control, the $\rho$ sensitivity measurement, the reward-mass census or the independent five-seed replication; those live in \texttt{gauge\_study/} and \texttt{calculations/}, each with its own generator. Several notebook cells deliberately \emph{read} the deposited data files rather than recompute them, so that a number cannot drift between the notebook and the manuscript; the cells say so where they do it. The notebook runs in Google Colab or any Python~$\ge$3.10 with \texttt{pyscf}, \texttt{torch}, \texttt{scipy} and \texttt{matplotlib}, and the calculation scripts regenerate, from first
principles and with fixed random seeds, the exact FCI reference of the systems in Fig.~\ref{fig:system}, the coupon-collector
statistics of \S\ref{sec:3}, the S-CORE recovery of \S\ref{sec:3-2}, the Epstein--Nesbet reward and GFlowNet compactness study of
\S\ref{sec:4}, the order-parameter (Fig.~\ref{fig:orderparam}) and controlled noise-crossover (Fig.~\ref{fig:crossover}) sweeps of \S\ref{sec:7}, and the EN-selected-CI decisive test of \S\ref{sec:5}. The complete code --- the notebook,
the calculation scripts behind every reported number, one script per figure, the generated figures, and the
Docker build environments --- is openly available (MIT license) at \url{https://github.com/nicolasbonilla/ml-for-sqd-review},
with a one-click Google~Colab entry point and every notebook cell and figure rendered inline in the browser.
Fig.~\ref{fig:loop} was produced as native vector graphics; Fig.~\ref{fig:system} combines vector molecular-orbital diagrams with high-resolution ($\ge$300~dpi at print size) ray-traced (PyMOL) isosurface renders, and all figure sources are available with the code. No proprietary data or
hardware access is required to reproduce any figure or table; all ``real-noise'' results use local,
backend-calibrated noise models (\texttt{FakeTorino}, IBM Heron~r1). The one device measurement
this paper discusses (\S\ref{sec:7}) belongs to the companion study \cite{bonilla2026spectral} and is
cited rather than reproduced here; no figure, table or claim of this review depends on it.

\section*{Author contributions}
\small
N.B.V.\ conceived the review, performed the literature synthesis, designed and ran all computational
experiments, produced the figures, and wrote the manuscript.

\emph{Use of AI tools, disclosed at the level of detail \S\ref{sec:6} asks of others.} A large language model (Anthropic's Claude, used interactively as a writing and coding assistant) was used in producing this work. Against the categories a disclosure is asked to address, the scope was: \emph{text production} --- drafting and revision of the manuscript, including this statement; \emph{production of code and calculations} --- the companion notebook, the calculation scripts, and the figure sources deposited with them; \emph{bibliographic research} --- candidate-literature searches, every hit of which was then verified against its DOI or arXiv record before being cited, a step that rejected fabricated and misattributed identifiers on more than one occasion; and \emph{reformatting}. It was \emph{not} used for image production: the figures are authored as vector graphics in this document's own typesetting language, the exception being the ray-traced isosurfaces of Fig.~\ref{fig:system}, which are PyMOL renders. The author takes full responsibility for all of it, and the deposit is built so that a reader need not take that on trust: every number reported here is regenerated by a named script, and the figure data, the noise ladder, the shot-survival fractions, the subspace sweep and the paired statistic are in addition checked back against the typeset manuscript by \texttt{calculations/verifica\_deposito.py}, which fails if the two disagree. That check is not ceremonial. It, and the audit against \S\ref{sec:6} that accompanied it, are what found the errors this version corrects and the results it retracts --- several of them numbers that assistance had produced and that survived until something mechanical compared them against the data. We report this because a review that asks the field for pre-registration, deposited generators and honest negative results cannot decline to say how it was itself made.

\section*{Competing interests}
\small
The author declares no competing financial interests. Daita AI, listed among the author's affiliations, neither funded this review nor holds any commercial interest in sample-based quantum diagonalization or in the methods assessed here; no company resources were used, and every calculation reported runs on a laptop or in the freely available container images deposited with the code. As a non-financial interest, the author is developing a generative-flow-network determinant proposer for SQD of the kind this review identifies as a methodological gap (\S\ref{sec:4-6}, \S\ref{sec:8}). The review documents that gap from the published literature and rests its central verdict on independent results, not on the author's own illustrative computations.

\section*{Acknowledgements}
\small
The author thanks the IBM Quantum and Qiskit communities and the organizers of the Qiskit Global Summer School,
whose open materials seeded this line of work, and the developers of \texttt{PySCF} and the
open-source scientific-Python ecosystem.

\end{document}